\documentclass[aps,twocolumn,amssymb, amsmath,nobibnotes]{revtex4}

\usepackage{graphicx}
\usepackage{dcolumn}
\usepackage{bm}
\usepackage{epsf}

\begin{document}

\newcommand{\bk}{{\bf k}}
\newcommand{\bp}{{\bf p}}
\newcommand{\bv}{{\bf v}}
\newcommand{\bq}{{\bf q}}
\newcommand{\br}{{\bf r}}
\newcommand{\bR}{{\bf R}}
\newcommand{\bB}{{\bf B}}
\newcommand{\bA}{{\bf A}}
\newcommand{\ba}{{\bf a}}
\newcommand{\bK}{{\bf K}}
\newcommand{\vd}{{v_\Delta}}
\newcommand{\vf}{{v_F}}
\newcommand{\tr}{{\rm Tr}}
\newcommand{\kslash}{\not\!k}
\newcommand{\qslash}{\not\!q}
\newcommand{\pslash}{\not\!p}
\newcommand{\rslash}{\not\!r}
\newcommand{\phiup}{\varphi_A}
\newcommand{\phidown}{\varphi_B}
\newcommand{\g}{{\gamma}}
\newcommand{\gm}{{\gamma_{\mu}}}
\newcommand{\gn}{{\gamma_{\nu}}}
\newcommand{\gl}{{\gamma_{\lambda}}}
\newcommand{\gnm}{{\gamma^n_{\mu}}}
\newcommand{\gnn}{{\gamma^n_{\nu}}}
\newcommand{\gnl}{{\gamma^n_{\lambda}}}
\newcommand{\eps}{{\epsilon}}
\newcommand{\la}{{\lambda}}
\newcommand{\abs}[1]{\lvert#1\rvert}

\title{QED$_3$ theory of pairing pseudogap in cuprates:\\
From d-wave superconductor to antiferromagnet via
``algebraic'' Fermi liquid}

\author{M. Franz}
\affiliation{Department of Physics and Astronomy, University of British
Columbia, Vancouver, BC, Canada V6T 1Z1}
\author{Z. Te\v{s}anovi\'c and O. Vafek}
\affiliation{Department of Physics and Astronomy,~Johns Hopkins University,
Baltimore, MD 21218}

\date{\today}

\begin{abstract}
High-$T_c$ cuprates differ from conventional superconductors in three
crucial aspects: the superconducting state descends from a strongly
correlated Mott-Hubbard insulator (as opposed to a Fermi liquid), the
order parameter exhibits d-wave symmetry and fluctuations play an all
important role.  We formulate an effective theory of underdoped
cuprates within the pseudogap state 
by taking the advantage of these unusual features.
In particular, we introduce a concept of ``pairing protectorate"  and we
seek to describe various phases within this protectorate by
phase disordering a
d-wave superconductor. The elementary excitations of
the protectorate are the Bogoliubov-de Gennes quasiparticles and
topological defects in the phase of the pairing
field -- vortices and antivortices -- which appear as quantum and
thermal fluctuations. The effective low energy theory of these elementary
excitations is shown to be, apart from intrinsic anisotropy, equivalent to
the quantum electrodynamics in (2+1) space-time dimensions (QED$_3$). A
detailed derivation of this QED$_3$ theory is given and some of its main
physical consequences are inferred for the pseudogap state. As the
superconducting order is destroyed by underdoping two possible
outcomes emerge: (i) the system can go into
a symmetric normal state characterized
as an ``algebraic Fermi liquid" (AFL) before developing
antiferromagnetic (AF) order, or (ii) a direct transition into the
AF state can occur.
In both cases the AF order arises spontaneously, through an intrinsic
``chiral"  instability of QED$_3$/AFL. Here we focus
on the properties of AFL and propose that
inside the pairing protectorate it assumes
the role reminiscent of that played
by the Fermi liquid theory in conventional metals.
We construct a gauge invariant electron propagator of
AFL and show that within the $1/N$ expansion
it has a non-Fermi liquid, Luttinger-like form
with positive anomalous dimension $\eta'=16/3\pi^2N$, where $N$ denotes
the number of pairs of nodes. We investigate the effects of Dirac
anisotropy by the perturbative renormalization group analysis and find
that the theory flows into isotropic fixed point. We therefore conclude
that, at long lengthscales, the AFL is stable against
anisotropy.
\end{abstract}
\maketitle


\section{Introduction}

In the classic tradition of condensed matter physics,
the phenomenon of superconductivity is usually described as an
instability of a normal metal
towards a quantum state in which electrons bind into  Cooper
pairs \cite{bcs}. This traditional picture, further fortified
within the Eliashberg formalism \cite{eliashberg}, has met with
spectacular success when applied
to conventional, low $T_c$-superconductors.
This success is no accident: the BCS-Eliashberg theory uses
the Landau theory of Fermi liquids as the underlying
description of a normal metal. In turn, the Landau
theory of Fermi liquids is one of the most successful
theories in physics -- by starting from the free gas
of fermions it exploits the constraint on
the phase space imposed by the Pauli exclusion principle
to provide a comprehensive understanding of a low
energy behavior of an {\em interacting} system. Among
its many pleasing features, the Landau theory allows
for methodical understanding of the seeds of its own
destruction -- while Fermi liquid is hardly the
true ground state of any system,  \cite{kohn1} we routinely seek to
understand such ground states by investigating various
instabilities of the Fermi liquid (FL) description, either
in the particle-hole or
particle-particle channel, few notable examples being spin- and
charge-density waves, itinerant ferromagnetism and, of course,
superconductivity.
This remarkable success story of the BCS-Eliashberg-FL theory
gave birth to the traditional
paradigm: ``one must understand the normal state
before one can understand the superconductor''  \cite{anderson,pines}.

The discovery of high-$T_c$ superconductors (HTS) and the subsequent
efforts to decipher their mysteries altered the above
state of affairs dramatically. As experimental information
began to pour in, it became rapidly clear
that HTS depart qualitatively from the BCS-Eliashberg-FL
orthodoxy in at least three important ways. First,
the high-$T_c$ superconducting copper
oxides are strongly interacting systems in which correlations
play an essential role. Their parent compounds, believed to contain
one hole per copper within CuO$_2$ layers, are far from good metals
that the simple band-structure
theory would predict and are instead insulating Neel
antiferromagnets. Even when doped and the long range
antiferromagnetic order had subsided, these materials do not seem
to follow the dictates of the Fermi liquid theory, except
perhaps in the heavily overdoped regime. Rather, the ``normal''
metal state of the cuprates is anything but, and has been routinely
dubbed ``anomalous'' or ``strange''. In line with the traditional
paradigm, a major theoretical
effort, mainly inspired by Anderson \cite{anderson},
has been directed at first understanding this anomalous ``normal'' state
as arising from the Hubbard-Mott-Neel antiferromagnetic insulator
at half-filling once a small density of mobile dopant holes
has been induced in the CuO$_2$ layers. The physics of
such ``doped Mott insulator'' and
particularly the microscopic mechanism
through which the high temperature superconductivity itself
is generated from within such a ``normal'' state,
even after years of concentrated
theoretical and experimental onslaught, remain
a deep challenge to the condensed matter physics community.

The second departure from the BCS-Eliashberg-FL 
orthodoxy \cite{departure} has
by now been firmly established experimentally:
the superconducting state in cuprates is itself of unconventional
symmetry  \cite{kirtley}. Instead of a usual ``s-wave'' gap function whose
magnitude generally varies over the Fermi surface but whose
sign does not, the cuprates are ``d-wave'' superconductors.
It is believed at the present time that this d-wave symmetry
reflects the presence of correlations caused by strong on-site
repulsion and is thus a close dynamical relative of the
antiferromagnetic state at half-filling. 
Majority of authors have focused on such purely electronic
``mechanism'' of superconductivity arising from strong correlations
in sharp distinction to the phonon-mediated pairing of
the traditional BCS-Eliashberg approach. Such unconventional
symmetry of the superconducting state and the ensuing presence
of low-energy fermionic excitations near the gap nodes result
in a rich phenomenology of cuprates. Most remarkably, however,
this phenomenology, at low energies and temperatures and
deep inside the superconducting state, seems to fit within the
theoretical mold of a model BCS-like d-wave superconductor, quite
separately from the detailed microscopic mechanism that
is at its origin. We consider this an experimental fact of 
crucial significance which for the rest of this paper
we intend to fully exploit to our advantage.

Finally, more than the intrinsic gap symmetry distinguishes
HTS from their conventional kin. They are also strongly fluctuating
systems  \cite{emerykivelson}. While strong fluctuations away from
a simple mean-field description are familiar occurrence in
magnets or liquid crystals they are a relatively novel phenomenon
in superconductors. In most conventional superconductors, well-described
by the BCS-Eliashberg-FL theory, the fluctuations are simply not
an issue: the dimensionless parameter which controls the deviations
from the mean-field theory, the inverse of the product
of the BCS coherence length $\xi_0$ and the Fermi wavevector $k_F$,
is typically as small as $10^{-3}$ or $10^{-4}$ and the fluctuation effects
are rarely observable  \cite{tinkham}.
In contrast, early experiments on HTS clearly established
strong fluctuations in numerous physical quantities
and various estimates put $(1/\xi_0 k_F) \sim 10^{-1}$ or
larger. The strongly fluctuating nature of the
superconducting state is particularly pronounced in underdoped
cuprates and is rather vividly manifested in recent experiments on
Nernst effect by the Princeton group  \cite{ong}.
The Nernst effect is routinely used
to detect superconducting vortex fluctuations and a strong
signal observed up to $T_{\rm Nernst}\gg T_c$, with $T_{\rm Nernst}$
comparable to the pseudogap temperature $T^*$, is most
naturally interpreted in those terms. Other
experiments have also provided both direct  \cite{corson}
and indirect  \cite{fischer} support for the pairing fluctuations
far above $T_c$. The experimental evidence for such fluctuations
has been compiled in Ref.  \cite{oda}.

Recently, a different path toward the theory of HTS was
proposed in Refs.  \cite{nodalliquid,ftqed}.
A key aspect of this ``inverted'' approach is
the observation that, when compared to its neighbors
in the phase diagram of cuprates, the superconducting
state appears to be the ``least correlated'' and its fermionic
excitations best defined  \cite{ftqed}. A tantalizing
theoretical feature of this approach is that it holds promise to
turn the above triple predicament of strong correlations, fluctuations
and unconventional symmetry in HTS into an advantage
by using the superconducting state as its departure point. The
first step in constructing a theory based on this ``inverted''
philosophy is to
assume that the most important effect of strong correlations at
the basic microscopic level is to build a large pseudogap of
d-wave symmetry, which is predominantly pairing in origin, i.e.
arises form the particle-particle (p-p) channel. In this regard,
we are following the wisdom of the FL theory but are
replacing the free electron gas starting point with
the free Bogoliubov-deGennes (BdG) d-wave quasiparticles.
Under the umbrella of this large d-wave pseudogap, the low energy
fermionic excitations enjoy remarkably sheltered existence.
They are completely impervious to weak residual short range interactions
left over after the effect of the pseudogap had been built in.

There is an important new element in this parallel with the FL.
Our reference state being a superconductor, we must also consider
interactions of BdG quasiparticles with relevant collective
modes of the pairing pseudogap, i.e.  fluctuating
thermal and quantum vortex-antivortex pairs. Within
the theory of Ref.  \cite{ftqed} this interaction
is represented by two U(1) gauge fields, $v_\mu$ and $a_\mu$. $v_\mu$
describes Doppler shift in quasiparticle energies  \cite{volovik}
and has been studied in Refs.  \cite{franzmillis,dorsey}.
Its effect on low energy fermions is rather modest since
$v_\mu$ gains mass from fermions
both in a superconductor and in a
phase-incoherent pseudogap state.
In contrast, the Berry gauge field $a_\mu$, minimally
coupled to fermions and encoding the
topological frustration inflicted upon BdG quasiparticles
by fluctuating vortices, is massless in the pseudogap state
and is the main source of strong
scattering at low energies  \cite{ftqed}.
The effective
theory was found to take the form equivalent to (2+1)-dimensional
(anisotropic) quantum electrodynamics (QED$_3$)  \cite{ftqed}.
In its symmetric phase, QED$_3$ is governed by the
interacting critical point leading to a non-Fermi
liquid behavior for its fermionic excitations.
This ``algebraic'' Fermi liquid (AFL)  \cite{ftqed} displaces conventional FL
as the underlying theory of the pseudogap state.

The AFL (symmetric QED$_3$) suffers an intrinsic instability
when vortex-antivortex fluctuations and residual interactions
become {\em too strong}. The topological frustration
is relieved by the spontaneous generation of mass for
fermions, while the Berry gauge field remains
massless. In the field theory literature on QED$_3$ this
instability is known as the dynamical chiral symmetry
breaking (CSB) and is a well-studied and established
phenomenon  \cite{csb}, although few clouds of uncertainty still hover
over its more quantitative aspects  \cite{appelquist}. In cuprates,
the region of such strong vortex fluctuations corresponds
to heavily underdoped samples and the CSB leads to
spontaneous creation of a whole plethora of nearly degenerate
ordered and gapped states from within the AFL  \cite{tvfqed}.
An important check on the internal consistency of the ``inverted'' approach
is that the manifold of CSB states contains an incommensurate
antiferromagnetic insulator (spin-density wave (SDW))
\cite{herbut1,tvfqed}.
Remarkably, both the ``algebraic'' Fermi liquid and
the SDW and other CSB insulating states arise from the one
and the same QED$_3$ theory  \cite{ftqed}, echoing
the satisfying features of Fermi liquid theory in conventional metals.
It therefore appears that the ``inverted'' approach
can be used to advance along the
doping axis of the HTS phase diagram (Fig. 1) in the ``opposite''
direction, from a d-wave superconductor all the way to
an antiferromagnetic insulator at very low doping, the low
energy physics held under overall
control of the symmetric QED$_3$ (AFL).

In this paper we first present a detailed derivation of the
QED$_3$ theory of the pairing pseudogap state in underdoped
cuprates previously
introduced in Ref.  \cite{ftqed} and then embark on a systematic exploration
of its fermionic excitation spectrum and other related
properties. To keep the
paper at manageable length we confine ourselves to the
the chirally {\em symmetric} phase of QED$_3$, i.e. to
the algebraic Fermi liquid (AFL) and its main properties. The
chiral symmetry broken phase is discussed separately,
in Part II of this paper.
Section II and Appendix A contain a step-by-step manual
on vortex-quasiparticle interactions and how the low-energy physics
of such interactions can be given its mathematical formulation
in the language of the QED$_3$
effective theory. In Sections III and IV we focus on
the AFL, give a detailed accounting
of vortex-quasiparticle interactions within QED$_3$ and compute
quasiparticle spectral properties within the pseudogap symmetric
phase. In Section V we then discuss the effects of Dirac anisotropy
on the QED$_3$ infrared fixed point and demonstrate that, to leading order
in $1/N$-expansion, the anisotropic  QED$_3$ scales to an isotropic
limit. Finally, we present a brief summary of our
results and conclusions in Section VI.

In Part II of the paper, to appear separately, we introduce the
concept of chiral symmetry within the d-wave pairing pseudogap
state \cite{herbut1,tvfqed}. We enumerate
different physical states within the chiral manifold
and discuss in depth various patterns of chiral symmetry breaking (CSB)
(Fig. 1) and fermion
mass generation within our QED$_3$ theory. This is followed by
the applications to the phase diagram of cuprates within the
pseudogap state.


\section{Preliminaries and vortex-quasiparticle interactions}

\subsection{Protectorate of the pairing pseudogap}

Our starting point is the assumption that the pseudogap is predominantly
particle-particle or pairing (p-p) in origin and
that it has a d$_{x^2-y^2}$ symmetry.
This assumption is given mathematical expression in the partition function:
\begin{eqnarray}
Z&=&\int{\cal D}\Psi^{\dag}({\bf r},\tau)
\int{\cal D}\Psi({\bf r},\tau)\int{\cal D}\varphi({\bf r},\tau)
\exp{[-S]},\nonumber \\
S&=&\int d\tau\int d^2r\{\Psi^{\dag}\partial_{\tau}
\Psi + \Psi^{\dag}{\cal H}\Psi + (1/g)\Delta^*\Delta \},
\label{action0}
\end{eqnarray}
where $\tau$ is the imaginary time, ${\bf r}=(x,y)$,
$g$ is an effective coupling constant in
the  d$_{x^2-y^2}$ channel, and
$\Psi^{\dag}=(\bar\psi_{\uparrow},\psi_{\downarrow})$
are the standard Grassmann variables. The effective
Hamiltonian
${\cal H}$ is given by:
\begin{equation}
{\cal H} =\left( \begin{array}{cc}
\hat{\cal H}_e  & \hat{\Delta} \\
\hat{\Delta}^*  & -\hat{\cal H}_e^*
\end{array} \right) + {\cal H}_{\rm res}
\label{h1}
\end{equation}
with
$\hat{\cal H}_e={1\over 2m}(\hat{\bf p}-{e\over c}{\bf A})^2-\epsilon_F$,
$\hat{\bf p}=-i\nabla$ (we take $\hbar=1$),
and $\hat\Delta$ the $d$-wave pairing operator \cite{ft,ft2},
\begin{equation}
\hat{\Delta} =
{1\over k_F^{2}}\{\hat{p}_x,\{\hat{p}_y,\Delta\}\}
-\frac{i}{4k_F^2}\Delta
\bigl (\partial_x\partial_y\varphi\bigr ),
\label{gauge}
\end{equation}
where $\Delta ({\bf r},\tau) =
|\Delta |\exp(i\varphi ({\bf r},\tau))$ is the center-of-mass
gap function and $\{a,b\}\equiv (ab+ba)/2$.
As discussed in Ref.\ \cite{ft2} the second term in Eq.\ (\ref{gauge}) is
necessary to preserve the overall gauge invariance.
Notice that we have rotated $\hat{\Delta}$ from
d$_{x^2-y^2}$ to d$_{xy}$ to simplify the continuum limit.
$\int{\cal D}\varphi ({\bf r},\tau)$ denotes the integral
over smooth (``spin wave") and singular (vortex) phase
fluctuations. Amplitude fluctuations of
$\hat{\Delta}$ are suppressed at or just below $T^*$ and the amplitude
itself is frozen at $2\Delta \sim 3.56 T^*$ for $T\ll T^*$.

The fermion fields $\psi_{\uparrow}$ and
$\psi_{\downarrow}$ appearing in Eqs. (\ref{action0},\ref{h1})
do not necessarily refer to the bare
electrons. Rather, they represent some {\em effective} low-energy
fermions of the theory, already fully renormalized by high-energy
interactions, expected to
be strong in cuprates due to Mott-Hubbard
correlations near half-filling  \cite{anderson}. The precise structure of
such fermionic effective fields follows from a more microscopic
theory and is not of our immediate concern here -- we are
only relying on the {\em absence} of {\em true} spin-charge
separation which allows us to
write the effective pairing term (\ref{h1})
in the BCS-like form. The experimental evidence that supports
this reasoning, at least within the superconducting state and
its immediate vicinity,
is rather overwhelming  \cite{ong,corson,fischer,oda}. Furthermore,
${\cal H}_{\rm res}$  represents the ``residual''
interactions, i.e. the part dominated by the effective
interactions in the p-h channel.
Our main assumption is equivalent to stating
that such interactions are in a certain sense ``weak'' and
less important part of the effective Hamiltonian
${\cal H}$ than the large pairing interactions already
incorporated through $\hat{\Delta}$. As we progress, this notion
of ``weakness'' of ${\cal H}_{\rm res}$ will be defined more precisely
and with it the region of validity of our theory.

The above discussion reveals
that our theory is at least partly of phenomenological character
and it must rely on a more microscopic description to fully define
its basic form specified by Eqs. (\ref{action0},\ref{h1}).
While it is still far from established just what such more microscopic
description might be in the case of cuprates, various gauge theories
of the $t-J$ and related models  \cite{palee}, could all be used for this
purpose at the present time. The main role
played by such theories is providing
reliable values of parameters that feed into the basic
formulation (\ref{action0},\ref{h1}). These
include, most importantly, the pairing pseudogap $\Delta$
itself, the microscopic values of vortex core energies,
the strengths of residual interactions, etc. Once
these parameters have been supplied through such an external
input it is our task to solve for the low energy ($\ll\Delta$),
long distance physics of (\ref{action0},\ref{h1}).  At present,
these needed parameters cannot be computed reliably from a
fully microscopic approach. We therefore combine
the available theoretical arguments
with the experimentally determined phase diagram (Fig. 1)
and argue that the pseudogap is indeed pairing in
origin and that the transition from the superconductor
to the pseudogap state must proceed via thermal and
quantum unbinding of vortex-antivortex pairs.
\begin{figure}[t]
\includegraphics[width=8cm]{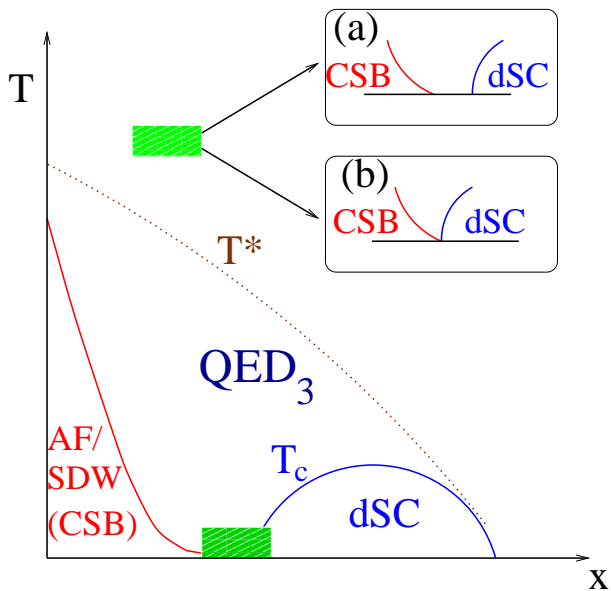}
\caption{\label{fig1}
A schematic representation of the phase
diagram of a cuprate superconductor.
Below $T^*$ ($\sim T_{\rm Nernst}$), the symmetric
QED$_3$/AFL replaces Fermi liquid as the effective
low energy theory. CSB denotes chiral symmetry
broken states, the most prominent among which is an
incommensurate antiferromagnet (SDW). Panel (a) represents
the path between dSC and an antiferromagnet via the
intervening algebraic Fermi liquid (AFL) ground state.
Panel (b) shows a direct dSC-CSB transition with AFL
describing $T\ll T^*$ behavior.
}
\end{figure}

Our ``inverted'' approach is similar in spirit
to the Landau theory
of Fermi liquids as applied to conventional metals.
In Fermi liquid (FL) theory the reference state is that
of a non-interacting gas of fermions. As the
interactions are turned on adiabatically, the Pauli
principle severely restricts the available phase
space for scattering and many of the general features
of free fermion system are preserved, albeit in a
renormalized form  \cite{pines}.
In our case, the reference state is a non-interacting
system of Bogoliubov-deGennes (BdG) fermions.
One can think of the pairing pseudogap $\Delta$
as being our ``Fermi energy'' and the highest energy scale
in the problem. While our theory cannot account for the physics
at energies (or temperatures) higher than $\Delta$, we will
endevour to show that the low energy physics can be
computed systematically and is parametrized
by a handful of material constants whose values can
be extracted either from experiments or from a more
microscopic theory.

In this context, the residual interactions among
the BdG fermions, both those arising from
the p-p channel through weak amplitude fluctuations
of $\Delta$ and the p-h interactions generated by
the effective spin fluctuations
within the pseudogap state  \cite{nodalliquid},
can be thought of as our version of
``Landau interaction parameters''
$\{F^{s,a}_{2\ell +1}\}$. These interactions
are generically short-ranged and are even less effective in
disturbing the coherence of nodal BdG fermions than
their FL counterpart. With all of the Fermi surface
gapped apart from the nodal regions, such interactions
are irrelevant in the renormalization group (RG)
sense and can be absorbed into renormalizations of
various relevant parameters (more on this in Section III).
As the system makes the transition from a phase-coherent
superconductor to a phase-incoherent pseudogap state, the
nodal BdG fermions interacting only through such
residual interactions would hardly notice -- the nodal
BdG liquid is a better Fermi liquid than the original.

There is, however, another source of interactions
among BdG quasiparticles which ruins this happy
state of affairs. Our reference state being
a superconductor, we must consider the interactions
of BdG fermions with collective
modes of $\Delta$ relevant near the superconductor-``normal''
boundary, i.e. fluctuating vortices.
As we demonstrate next, these interactions between
BdG quasiparticles and
fluctuating vortex-antivortex excitations
play the central dynamical role in our theory.

\subsection{Phase fluctuations, vortices,
quasiparticles and topological frustration}

The global U(1) gauge invariance mandates that the
partition function (\ref{action0}) must be independent
of the overall choice of phase for  $\hat\Delta$.
We should therefore aim to
eliminate the phase $\varphi ({\bf r},\tau)$
from the pairing term (\ref{h1}) in favor of
$\partial_{\mu}\varphi$ terms [$\mu = (x,y,\tau)$]
in the fermionic action. For the regular (``spin-wave'') piece of $\varphi$
this is easily accomplished by absorbing a phase factor
$\exp(i\frac{1}{2}\varphi)$ into both spin-up and spin-down
fermionic fields. This amounts to ``screening'' the
phase of $\Delta ({\bf r},\tau)$ (or an ``XY phase'', as
commonly known) by
a ``half-phase'' field (or ``half-XY phase'')
attached to $\psi_{\uparrow}$ and $\psi_{\downarrow}$.
However, as discussed in Ref.  \cite{ft,ft2}, when dealing
with the singular part of $\varphi$, such transformation
``screens'' physical singly quantized
$hc/2e$ superconducting vortices with
``half-vortices'' in the fermionic fields. Consequently,
this ``half-angle'' gauge transformation must be accompanied by
branch cuts in the fermionic fields
which originate and terminate at vortex
positions and across which the quasiparticle wavefunction
must switch its sign. These branch cuts are mathematical
manifestation of a fundamental physical effect: in presence
of vortices, which are topological defects in the phase of the
Cooper pair field and thus naturally bind the elementary flux of
$hc/2e$, the motion of quasiparticles is {\em topologically frustrated},
since their natural elementary flux is
twice as large ($hc/e$).
The physics of this topological frustration is at the origin
of all non-trivial dynamics discussed in this paper.

Dealing with branch cuts in a fluctuating vortex problem is a
rather cumbersome affair due to their
non-local character and defeats the original purpose
of reducing the problem to that of fermions interacting with {\em local}
fluctuating superflow field, i.e. with $\partial_{\mu}\varphi$.
Instead, in order to avoid the branch cuts, non-locality and
non-single valued wavefunctions, we employ
the singular gauge transformation devised in Ref.  \cite{ft}, hereafter
referred to as `FT' transformation:
\begin{equation}
\bar\psi_{\uparrow}\to\exp{(-i\varphi_A)}\bar\psi_{\uparrow},\ \ \
\bar\psi_{\downarrow}\to\exp{(-i\varphi_B)}\bar\psi_{\downarrow},
\label{ft}
\end{equation}
where $\varphi_A + \varphi_B
= \varphi$. Here $\varphi_{A(B)}$ is the singular part
of the phase due to $A(B)$ vortex defects:
\begin{equation}
\nabla\times\nabla\varphi_{A(B)}=2\pi\hat{z}\sum_i
q_i\delta\left({\bf r}-{\bf r}_i^{A(B)}(\tau)\right),
\end{equation}
with $q_i=\pm 1$ denoting the
topological charge of $i$-th vortex and
${\bf r}_i^{A(B)}(\tau)$ its  position. The labels
$A$ and $B$ represent some convenient but otherwise
arbitrary division of vortex defects (loops or
lines in $\varphi ({\bf r},\tau)$) into two sets, although
we will soon discuss  many virtues
of the {\em symmetrized} transformation (\ref{ft}) which
apportions vortex defects {\em equally} between
sets $A$ and $B$  \cite{ft,ft2}.
The transformation (\ref{ft}) ``screens'' the original
superconducting phase $\varphi$ (or ``XY phase'') with
two ordinary ``XY phases'' $\varphi_A$ and $\varphi_B$
attached to fermions. Both $\varphi_A$ and $\varphi_B$
are themselves perfectly legitimate phase configurations
of $\Delta ({\bf r},\tau)$ but simply with fewer vortex defects.
The key feature of the transformation (\ref{ft}) is that
it accomplishes ``screening''
of the physical $hc/2e$ vortices by using
only ``whole'' (i.e., not ``halved'')
vortices in fermionic fields and thus guarantees that the quasiparticle
wavefunctions remain
single-valued. The topological frustration
still remains, being the genuine physical effect of the branch cuts,
but is now incorporated directly
into the fermionic part of the action:
\begin{equation}
{\cal L}'=\bar\psi_{\uparrow}[\partial_{\tau} +
i(\partial_{\tau}\varphi_A)]\psi_{\uparrow} +
\bar\psi_{\downarrow}[\partial_{\tau} +
i(\partial_{\tau}\varphi_B)]\psi_{\downarrow}
+ \Psi^{\dag}{\cal H}^{\prime}\Psi,
\label{vmuamu}
\end{equation}
\noindent
where the transformed Hamiltonian ${\cal H}^{\prime}$ is:
\[
\left( \begin{array}{cc}
{1\over 2m}(\hat{\bf \pi}+{\bf v})^2
-\epsilon_F & \hat D \\
\hat D  &
-{1\over 2m}(\hat{\bf \pi}-{\bf v})^2
+\epsilon_F
\end{array} \right),                                                           \]
with
$\hat D =(\Delta_0/2k_F^2)(\hat \pi_x \hat \pi_y+\hat \pi_y \hat \pi_x)$ and
${\hat{\bf \pi}} ={\hat{\bf p}} + {\bf a}$.

The singular gauge transformation (\ref{ft})
generates a ``Berry'' gauge potential
\begin{equation}
a_\mu=\frac{1}{2}(\partial_\mu\varphi_A -
\partial_\mu\varphi_B)~~~,
\label{berry}
\end{equation}
which describes half-flux Aharonov-Bohm scattering of
quasiparticles on vortices. $a_\mu$ couples
to BdG fermions {\em minimally} and mimics the effect of
branch cuts in quasiparticle-vortex dynamics \cite{ft,ft2,marinelli}.
Closed fermion loops in the Feynman path-integral
representation of (\ref{action0}) acquire
the ($-1$) phase factors due to this half-flux Aharonov-Bohm effect
just as they would from a branch cut attached to a vortex defect.
The topological frustration is now
implemented through the fact that $\partial_\mu\varphi_{A(B)}$,
minimally coupled to a loop of BdG fermions in the partition
function (\ref{action0}) as it winds around the imaginary time direction,
generates the phase equal to the
advance in the XY phase $\varphi_{A(B)}({\bf r},\tau)$, which inhabits the
same spacetime, along the closed path coinciding with
the said fermion loop. Naturally, this advance must be an integer multiple of
$2\pi$ -- the overall factor of $\frac{1}{2}$  in the definition
of $a_\mu$ (\ref{berry}) reduces this further to an integer multiple of $\pi$.
In the end, the phase factors of such fermion loops, which are
the gauge invariant quantities, are all equal to $\pm 1$
and are furthermore precisely what we would have obtained
had we opted for the ``half-angle'' gauge transformation
and implemented the branch cuts.

The above $(\pm 1)$ prefactors of the BdG fermion
loops come on top of general and everpresent
U(1) phase factors $\exp(i\delta)$,
where the phase $\delta$ depends on spacetime configuration of
vortices. These U(1) phase factors are
supplied by the ``Doppler" gauge field
\begin{equation}
v_\mu=\frac{1}{2}(\partial_\mu\varphi_A +
\partial_\mu\varphi_B)\equiv\frac{1}{2}\partial_\mu\varphi ~~~,
\label{doppler}
\end{equation}
which denotes the classical part
of the quasiparticle-vortex interaction. The coupling
of $v_\mu$ to fermions is the same as that of the
usual electromagnetic gauge field $A_\mu$ and is
therefore {\em non-minimal}, due to the pairing term in the
original Hamiltonian $\cal H$ (\ref{h1}). It is
this non-minimal interaction with $v_\mu$,
which we call the Meissner coupling,
that is responsible for the U(1) phase factors $\exp(i\delta)$.
These U(1) phase factors are ``random'', in the sense that
they are not topological in nature -- their values
depend on detailed distribution of superfluid fields of all vortices
and ``spin-waves'' as well as on the internal structure of
BdG fermion loops, i.e. what is the sequence of spin-up and
spin-down portions along such loops. In this respect,
while its minimal coupling
to BdG fermions means that, within the lattice d-wave
superconductor model \cite{ft2}, one is naturally tempted
to represent the Berry gauge field $a_\mu$
as a {\em compact} U(1) gauge field,
the Doppler gauge field $v_\mu$ is decidedly {\em non-compact},
lattice or no lattice \cite{compact}.
The reader should be advised, however, that
the issue of  ``compactness'' of $a_\mu$ versus ``non-compactness''
of $v_\mu$ constitutes a moot point: $a_\mu$ and $v_\mu$
as defined by Eqs. (\ref{berry},\ref{doppler})
are {\em not independent} since the discrete spacetime
configurations of vortex defects $\{{\bf r}_i(\tau)\}$
serve as {\em sources} for {\em both}. We will belabor
this important issue in the next subsection.

	For now, note that all choices of the
sets $A$ and $B$ in transformation
(\ref{ft}) are completely
{\em equivalent} -- different choices represent different
singular gauges and $v_\mu$, and therefore $\exp(i\delta)$, are
invariant under such transformations. $a_\mu$, on the other hand,
changes but only through the introduction of ($\pm$) unit Aharonov-Bohm
fluxes at locations of those vortex defects involved in the
transformation. Consequently, the Z$_2$ style
$(\pm 1)$ phase factors associated
with $a_\mu$ that multiply the fermion loops remain unchanged.
We now symmetrize the partition function with respect
to this singular gauge by defining a generalized
transformation (\ref{ft}) as the sum over all possible choices
of $A$ and $B$, i.e., over the entire family of singular
gauge transformations. This is an Ising sum
with $2^{N_l}$ members, where $N_l$
is the total number of vortex defects in
$\varphi ({\bf r},\tau)$, and is itself
yet another choice of the singular gauge. The many benefits
of such symmetrized gauge will be discussed shortly but
we stress here that its ultimate function
is calculational convenience. What actually
matters for the physics is that
the original $\varphi$ be split into two XY
phases so that the vortex defects of every distinct topological class
are apportioned {\em equally} between $\varphi_A$ and $\varphi_B$
(\ref{ft})  \cite{ft,ft2}.

	The physics behind this last
requirement can be intuitively appreciated as
follows: imagine that we simply {\em prohibit} singly quantized
vortex excitations and replace them in
(\ref{vmuamu}) by {\em doubly quantized} $(hc/e)$ vortices. In this
case, the Berry gauge field attaches
a multiple of a {\em full} Aharonov-Bohm
flux to each vortex position and all the phase factors in
front of fermion loops equal unity -- this is equivalent to
$a_\mu =0$ and complete absence of topological frustration.
{\em It is then natural to select the gauge which eliminates
$a_\mu$ altogether}. This is straightforwardly accomplished
by screening each doubly quantized $(hc/e)$
vortex defect with a unit-flux Dirac string
in $\psi_\uparrow$ and an equivalent one in $\psi_\downarrow$.
The resulting sets of $A$ and $B$ vortices are two identical
replicas of each other and
$a_\mu ={1\over 2}(\partial_\mu\varphi_A -\partial_\mu\varphi_B)\equiv 0$.
Now, while still staying within the same gauge, we allow doubly
quantized vortices to relax into energetically more favorable
configurations -- they will immediately decay into singly
quantized vortices and our sets $A$ and $B$ will end up containing
an equal number of singly quantized vortices of each distinct
topological class. For example, in the case of 2D thermal fluctuations
$\varphi_{A(B)}$ should each contain a half of the original
vortices in $\varphi$ and a half of antivortices. This is readily
achieved by including those (anti)vortex variables
whose positions are labeled by ${\bf r}_i$ with
$i$ {\em even} into $\varphi_A$, while the {\em odd} ones
are absorbed into $\varphi_B$. The symmetrization
is just a convenient mathematical tool that
automatically guarantees this goal.

The above symmetrization leads to the new partition function:
$Z\to \tilde Z$:
\begin{equation}
\tilde Z=\int{\cal D}\tilde\Psi^{\dag}
\int{\cal D}\tilde\Psi\int{\cal D}v_\mu\int{\cal D}a_\mu
\exp{[-\int_{0}^{\beta} d\tau\int d^2r\tilde{\cal L}]}~,
\label{barez}
\end{equation}
in which the half-flux-to-minus-half-flux (Z$_2$) symmetry of
the singular gauge transformation (\ref{ft}) is now manifest:
\begin{equation}
\tilde{\cal L}=\tilde\Psi^{\dag}[(\partial_{\tau}
+ia_\tau)\sigma_0 + iv_\tau\sigma_3]
\tilde\Psi + \tilde\Psi^{\dag}{\cal H}^{\prime}\tilde\Psi
+{\cal L}_0[v_\mu,a_\mu],
\label{bareaction}
\end{equation}
\noindent
where ${\cal L}_0$ is the ``Jacobian'' of the transformation given by
\begin{eqnarray}
e^{-\int_0^\beta d\tau\int d^2r {\cal L}_0}=
2^{-N_l}\sum_{A,B}\int {\cal D}\varphi ({\bf r},\tau) \ \ \ \ \ \ \
\label{jacobian}   \\
\times
\delta[v_\mu - {\textstyle {1\over 2}}(\partial_\mu\varphi_A +
\partial_\mu\varphi_B)]
\delta[a_\mu - {\textstyle {1\over 2}}(\partial_\mu\varphi_A -
\partial_\mu\varphi_B)].\nonumber
\end{eqnarray}
Here $\sigma_\mu$ are the Pauli matrices, $\beta =1/T$,
${\cal H}^{\prime}$ is given in Eq. (\ref{vmuamu}) and,
for later convenience, ${\cal L}_0$ will also include
the energetics of vortex core overlap driven by
amplitude fluctuations and thus independent of long range
superflow (and of ${\bf A}$).
We call the transformed quasiparticles $\tilde\Psi^{\dag}=
(\bar{\tilde\psi}_{\uparrow},\tilde\psi_{\downarrow})$
appearing in (\ref{bareaction}) ``topological fermions" (TF).
TF are the natural fermionic excitations of the
pseudogapped normal state. They are electrically neutral
and are related to the original quasiparticles by the
inversion of transformation (\ref{ft}).

By recasting the original problem in terms of topological
fermions we have accomplished our original goal: the interactions
between quasiparticles and vortices are now described solely
in terms of two local superflow fields \cite{ft}:
\begin{equation}
v_{A\mu}=\partial_\mu\varphi_A~;~~~~~~~~~
v_{B\mu} =\partial_\mu\varphi_B~,
\label{vavb}
\end{equation}
which we can think of as superfluid velocities
associated with the phase configurations
$\varphi_{A(B)}({\bf r},\tau)$ of a
(2+1)-dimensional XY model with periodic boundary conditions along the
$\tau$-axis. Our Doppler and Berry gauge fields
$v_\mu$ and $a_\mu$ are linear combinations of $v_{A\mu}$ and
$v_{B\mu}$. Note that $a_\mu$ is produced {\em exclusively} by vortex
defects since the ``spin-wave'' configurations of $\varphi$ can be
fully absorbed into $v_\mu$. All that remains is to perform the
sum in (\ref{barez}) over all the ``spin-wave'' fluctuations
and all the spacetime configurations of vortex
defects $\{{\bf r}_i(\tau)\}$ of this (2+1)-dimensional XY model.

\subsection{``Coarse-grained'' Doppler and Berry U(1)
gauge fields ($v_\mu$ and $a_\mu$)
and their physical significance}

Unfortunately, the exact integration over
the phase $\varphi ({\bf r},\tau)$
is prohibitively difficult.
To proceed by analytic means we must
devise some approximate procedure to integrate over
vortex-antivortex positions $\{{\bf r}_i(\tau)\}$ in (\ref{barez}) which
will capture qualitative features of
at least the long-distance, low-energy physics
of the original problem. This is where our
recasting of the problem in terms of BdG fermions
interacting with superflow fields $v_{A\mu}$
and $v_{B\mu}$ (\ref{vavb}) will come in handy.
A hint of how to devise
such an approximation comes from examining the role
of the Doppler gauge field $v_\mu$ in the physics of this
problem. To illustrate our
reasoning and for simplicity, we consider the finite $T$ case
where we can ignore the $\tau$-dependence of $\varphi ({\bf r},\tau)$.
The results are easily generalized, with appropriate modifications,
to include quantum fluctuations.

We start by noting that
${\bf V}_s=2{\bf v} - (2e/c){\bf A}$ is just the physical superfluid
velocity  \cite{remark11}, invariant under both $A\leftrightarrow B$ singular
gauge transformations (\ref{ft}) and ordinary electromagnetic
U(1) gauge symmetry. The superfluid velocity (Doppler) field, swirling
around each (anti)vortex defect, is responsible for a
vast majority of phenomena that we associate with
vortices: long range interactions between vortex defects,
coupling to external magnetic field and the Abrikosov lattice
of the mixed state, Kosterlitz-Thouless transition, etc.
Remarkably, we will show that its role is essential even for
the physics discussed in the present paper, although
it now appears as a supporting actor
to the Berry gauge field $a_\mu$ which
ultimately occupies the center stage. To see how this comes
about, imagine that ${\bf a}~(a_\mu)$ were absent -- then,
upon integration over topological fermions
in (\ref{bareaction}), we obtain
the following term in the effective Largangian:
\begin{equation}
M^2\left({\bf \nabla}\varphi - \frac{2e}{c}{\bf A}\right)^2
+ (\cdots)~~~,
\label{meissner}
\end{equation}
where $(\cdots )$ denotes higher order powers and derivatives
of $2{\bf \nabla}\varphi - (2e/c){\bf A}$. In the above we
have replaced ${\bf v}\to (1/2){\bf\nabla}\varphi$ to emphasize
that the leading term, with the coefficient $M^2$ proportional
to the bare superfluid density, is just
the standard superfluid-velocity-squared term of the continuum
XY model -- the notation $M^2$ for the coefficient will become clear
in a moment. We can now write $\nabla\varphi =
\nabla\varphi_{\rm vortex}+\nabla\varphi_{\rm spin-wave}$
and reexpress the transverse portion of (\ref{meissner})
in terms of (anti)vortex positions $\{{\bf r}_i\}$ to obtain a
familiar form:
\begin{equation}
\to\frac{M^2}{2\pi}\sum_{\langle i,j\rangle}
\ln |{\bf r}_i - {\bf r}_j|
\label{logint}
\end{equation}
or, by using $\nabla\cdot\nabla\varphi_{\rm vortex}=0$
and $\nabla\times\nabla\varphi_{\rm vortex}= 2\pi\rho ({\bf r})$,
equivalently as:
\begin{equation}
\to\frac{M^2}{2\pi}\int d^2r\int d^2r' \rho ({\bf r})\rho ({\bf r}')
\ln |{\bf r}- {\bf r}'|~~,
\label{logint1}
\end{equation}
where $\rho ({\bf r}) = 2\pi\sum_i
q_i\delta({\bf r}-{\bf r}_i)$ is the vortex density.


The Meissner coupling of ${\bf v}$ to fermions
is very strong -- it leads to familiar long range interactions between
vortices which constrain vortex fluctuations to a remarkable
degree. To make this statement mathematically explicit we
introduce the Fourier transform of vortex density
$\rho ({\bf q})=\sum_i\exp(i{\bf q}\cdot {\bf r}_i)$ and
observe that its variance satisfies:
\begin{equation}
\langle\rho ({\bf q})\rho (-{\bf q})\rangle\propto q^2/M^2~~.
\label{compress}
\end{equation}
Vortex defects form an {\em incompressible} liquid -- the
long distance vorticity fluctuations are strongly
suppressed. This ``incompressibility constraint'' is naturally
enforced in (\ref{barez}) by replacing the integral over
{\em discrete} vortex positions $\{{\bf r}_i\}$ with the
integral over {\em continuously} distributed field
$\bar\rho ({\bf r})$ with $\langle\bar\rho ({\bf r})\rangle =0$.
The Kosterlitz-Thouless transition
and other vortex phenomenology are still maintained
in the non-trivial structure of ${\cal L}_0[\bar\rho(\br)]$. But
the long wavelength form of (\ref{meissner}) now reads:
\begin{equation}
M^2\left(2\bar{\bf v} - \frac{2e}{c}{\bf A}\right)^2
+ (\cdots)~~~,
\label{meissner2}
\end{equation}
and can be interpreted as a {\em massive} action for
a U(1) gauge field $\bar{\bf v}$. The latter is our {\em coarse grained}
Doppler gauge field defined by
\begin{equation}
\nabla\times \bar{\bf v}(\br)= \pi\bar\rho (\br)~~.
\label{meissner3}
\end{equation}
We have now gone full circle with the Doppler
gauge field ${\bf v}$. The coarse-graining procedure
has made it into a massive U(1) gauge field $\bar{\bf v}$ whose
influence on TF disappears in the long wavelength
limit. We can therefore drop it from low energy
fermiology. Hereafter we shall drop the bar and use symbol ${\bf v}$
for both the actual and the coarse grained quantity, the meaning being
obvious from the context.

The real problem also contains the Berry gauge field
${\bf a}~(a_\mu)$ which now must be restored.
However, if we are to take advantage of introducing
continuous $\rho (\br)$ and eventually dispensing with ${\bf v}$,
${\bf a}~(a_\mu)$ cannot remain
the same gauge field we started with in Eq. (\ref{bareaction})
when we embarked on our quest to derive the effective theory.
Having replaced the Doppler field ${\bf v}$ with the
{\em distributed} quantity (\ref{meissner3}), we {\em cannot}
simply continue to keep ${\bf a}~(a_\mu)$ specified by half-flux Dirac
(Aharonov-Bohm) strings located 
at {\em discrete} vortex positions $\{ {\bf r}_i\}$.
Instead, ${\bf a}~(a_\mu)$ must be replaced by some
new gauge field which reflects the ``coarse-graining'' that
has been applied to ${\bf v}$ -- simply put, the
Z$_2$-valued Berry gauge field (\ref{berry}) of the
original problem (\ref{bareaction})
must be {\em ``dressed''} in such a way as to {\em best compensate} for the
error introduced by ``coarse-graining'' ${\bf v}$. In the
language of RG, we must find such ``dressing'' of the Berry gauge field,
i.e. the form and the bare
action for ${\bf a}~(a_\mu)$, which renders any such errors
unimportant for the low energy physics.

The recipe for such a required
``dressing'' of the Berry gauge field
is straightforward in
the FT singular gauge -- it takes the form of
a non-compact U(1) gauge
field with a simple Maxwellian action.
To see how this comes about note that
if we insist on replacing ${\bf v}$ by 
its ``coarse-grained'' form (\ref{meissner3}),
the only way to achieve this is to ``coarse-grain'' {\em both}
${\bf v}_A$ and ${\bf v}_B$ in the same manner:
\begin{equation}
\nabla\times {\bf v}_A\to 2\pi\rho_A~~;~~
\nabla\times {\bf v}_B\to 2\pi\rho_B~~,
\label{rhoarhob}
\end{equation}
where $\rho_{A,B} (\br)$ are now continuously distributed
densities of $A(B)$ vortex defects (the
reader should contrast this with (\ref{berry},\ref{doppler})).
This is because the {\em elementary}
vortex variables of our problem are {\em not} the sources
of ${\bf v}$ and ${\bf a}$; rather, they are
the sources of ${\bf v}_A$ and
${\bf v}_B$. We cannot separately fluctuate
or ``coarse-grain'' the Doppler and
the Berry ``halves''
of a given vortex defect -- they are {\em permanently confined}
into a physical $(hc/2e)$ vortex. On the other hand, we {\em can}
independently fluctuate $A$ and $B$ vortices -- this is why it was
important to use the singular gauge (\ref{ft}) to
rewrite the original problem solely in terms of $A$ and $B$
vortices and associated superflow fields (\ref{vavb}).

Following this recipe we can now reassemble the coarse-grained
Doppler and Berry gauge fields as:
\begin{equation}
{\bf v} = {1\over 2}({\bf v}_A + {\bf v}_B)~~;~~
{\bf a} = {1\over 2}({\bf v}_A - {\bf v}_B)~~,
\label{va}
\end{equation}
with the straightforward generalization to (2+1)D:
\begin{equation}
v = {1\over 2}(v_A + v_B)~~;~~
a = {1\over 2}(v_A - v_B)~~.
\label{va3}
\end{equation}
The coupling of $ {\bf v}_A (v_A)$ and $ {\bf v}_B (v_B)$ to
fermions is a hybrid of Meissner and minimal coupling \cite{compact}.
They contribute a product of U(1) phase factors
$\exp (i\delta_A)\times\exp (i\delta_B)$ to the BdG fermion
loops with both $\exp (i\delta_A)$ and $\exp (i\delta_B)$
being ``random'' in the sense of previous subsection. Upon
coarse-graining $ {\bf v}_A (v_A)$ and $ {\bf v}_B (v_B)$
turn into non-compact U(1) gauge fields
and therefore $ {\bf v} (v)$ and  $ {\bf a} (a)$ must
as well.

\subsection{Further remarks on FT gauge}

The above ``coarse-grained'' theory must
have the following symmetry: it has
to be invariant under the exchange of spin-up and spin-down
labels, $\psi_\uparrow\leftrightarrow\psi_\downarrow$,
{\em without} any changes in ${\bf v}$ ($v$). This symmetry
insures that (arbitrarily preselected)
$S_z$ component of the {\em spin}, which is the
same for TF  and real electrons,
decouples from the physical superfluid velocity which
naturally must couple only to {\em charge}. When dealing with discrete
vortex defects this symmetry is guaranteed by
the singular gauge symmetry defined by the family
of transformations (\ref{ft}). However, if we
replace ${\bf v}_A$ and ${\bf v}_B$ by their distributed
``coarse-grained'' versions, the said symmetry is preserved
only in the FT gauge. This is seen by considering
the effective Lagrangian expressed in terms of
coarse-grained quantities:
\begin{equation}
{\cal L}~\rightarrow~\tilde\Psi^{\dag}\partial_{\tau}\tilde\Psi
+ \tilde\Psi^{\dag}{\cal H}^{\prime}\tilde\Psi
+{\cal L}_0[\rho_A,\rho_B]~~,
\label{bareaction1}
\end{equation}
\noindent
where ${\cal H}^{\prime}$ is given by (\ref{vmuamu}),
${\bf v}_A$, ${\bf v}_B$ are connected to
$\rho_A$, $\rho_B$ via (\ref{rhoarhob}), and
${\cal L}_0$ is independent of $\tau$.

	The problem lurks in
${\cal L}_0[\rho_A,\rho_B]$ -- this is just the entropy
functional of
fluctuating {\em free} $A(B)$ vortex-antivortex defects
and has the following symmetry: $\rho_{A(B)}\to -\rho_{A(B)}$
with $\rho_{B(A)}$ kept unchanged. This symmetry reflects the fact
that the entropic ``interactions'' do not depend on vorticity.
Above the Kosterlitz-Thouless
transition we can expand:
\begin{equation}
{\cal L}_0 \to {K_A\over 4} (\nabla\times{\bf v}_A)^2 +
{K_B\over 4}(\nabla\times{\bf v}_B)^2 + (\cdots ),
\label{lab}
\end{equation}
\noindent
where the ellipsis denote higher order terms and
the coefficients $K^{-1}_{A(B)}\to n_l^{A(B)}$
(see Appendix A for details).

The above discussed symmetry of Hamiltonian
${\cal H}^{\prime}$ (\ref{vmuamu}) demands that
${\bf v}$ and ${\bf a}$ (\ref{va},\ref{va3})
be the natural choice for independent
distributed vortex fluctuation gauge
fields which should appear in
our ultimate effective theory. ${\cal L}_0$, however,
collides with this symmetry of ${\cal H}^{\prime}$ -- if we replace
${\bf v}_{A(B)}\to {\bf v}\pm {\bf a}$ in (\ref{lab}), we
realize that ${\bf v}$ and ${\bf a}$ are {\em coupled}
through ${\cal L}_0$ in the general case $K_A\not=K_B$:
\begin{eqnarray}
{\cal L}_0&\to&\frac{K_A+K_B}{4}\left( (\nabla\times {\bf v})^2
+(\nabla\times {\bf a})^2\right) \nonumber \\
&+&
\frac{K_A-K_B}{2}(\nabla \times {\bf a}) \cdot (\nabla \times {\bf v})~.
\label{oscareq}
\end{eqnarray}
Therefore, via its coupling to ${\bf a}$, the
superfluid velocity ${\bf v}$ couples to the ``spin''
of topological fermions and ultimately to the true {\em spin}
of the real electrons. This is an
unacceptable feature for the effective theory and seriously
handicaps the general ``$A-B$'' gauge, in which the original phase
is split into $\varphi\to \varphi_A +\varphi_B$, with  $\varphi_{A(B)}$
each containing some {\em arbitrary} fraction of the original
vortex defects. In contrast,
the symmetrized transformation (\ref{ft}) which
apportions vortex defects equally between
$\varphi_A$ and $\varphi_B$  \cite{ft} leads to
$K_A =K_B$ and to decoupling of
${\bf v}$ and ${\bf a}$ at quadratic order,
thus eliminating the problem at its
root. Furthermore, even if we start with the general  ``$A-B$'' gauge,
the renormalization of  ${\cal L}_0$ arising from integration
over fermions will ultimately drive $K_A - K_B\to 0$
and make the coupling of ${\bf v}$ and ${\bf a}$
{\em irrelevant} in the
RG sense \cite{andersongauge}. This argument is actually
quite rigorous in the case of quantum fluctuations where
the symmetrized gauge (\ref{ft}) represents a fixed point in the
RG analysis (see below)  \cite{coulombgauge}.
Consequently, it appears that the symmetrized
singular transformation (\ref{ft}) employed
in (\ref{bareaction}) is the {\em preferred} gauge for
the construction of the effective low-energy
theory \cite{coulombgauge,remarkboring}.
In this respect, while all the singular
$A-B$ gauges are created equal some are ultimately more equal than others.

The above discussion provides the rationale behind using
the FT transformation in our quest for the effective
theory. At low energies, the interactions between quasiparticles
and vortices are represented by two U(1) gauge fields
$v$ and $a$ (\ref{va},\ref{va3}).
The conversion of $a$ from a Z$_2$-valued
to a non-compact U(1) field with Maxwellian action
is effected by the confinement
of the Doppler to the Berry half of a singly quantized
vortex -- in the coarse-graining process the phase factors
$\exp(i\delta)$ of the
non-compact Doppler part ``contaminate'' the original
$(\pm 1)$ factors supplied by $a$ (\ref{berry}). This
contamination diminishes as doping  $x\to 0$ since then
$v_F/v_\Delta\to 0$ and ``vortices'' are effectively liberated of
their Doppler content. In this limit, the pure Z$_2$ nature
of $a$ is recovered and one enters the realm of the
Z$_2$ gauge theory of Senthil and Fisher \cite{senthil}.
In contrast, in the pairing pseudogap regime of this paper where
$v_F/v_\Delta > 1$ and singly quantized $(hc/2e)$ vortices
appear to be the relevant excitations \cite{bonn},
we expect the effective theory to take the U(1) form
described by $v$ and $a$ (\ref{va},\ref{va3}).

\subsection{Jacobian ${\cal L}_0[v_\mu,a_\mu]$ and its
Maxwellian form}

Having elucidated the origin of the ``coarse-grained''
U(1) gauge fields $v_\mu$ and $a_\mu$  and settled on
the symmetrized FT transformation (\ref{ft})
as the natural gauge choice for this problem, there remains
one more task to be accomplished before we can conclude this Section.
We need to derive a precise expression for
the long distance, low
energy form of the ``Jacobian'' ${\cal L}_0 [v_\mu,a_\mu]$
which serves as the ``bare action'' for the gauge
fields $v_\mu$ and $a_\mu$ of our effective theory.
As shown below, this form is a non-compact Maxwellian
whose stiffness $K$ (or inverse ``charge'' $1/e^2=K$)
stands in intimate relation to the helicity modulus tensor
of a dSC and, in the pseudogap
regime of strong superconducting fluctuations, can
be expressed in terms of a {\em finite} physical superconducting
correlation length $\xi_{\rm sc}$, $K\propto\xi_{\rm sc}^2$
(2D); $K\propto\xi_{\rm sc}$ ((2+1)D). As we
enter the superconducting phase and $\xi_{\rm sc}\to\infty$,
$K\to\infty$ as well (or $e^2\to 0$), implying that $v_\mu$
and $a_\mu$ have become massive.
Our derivation, the results of which were originally quoted
and used in Ref.  \cite{ftqed}, can be accomplished
with remarkably little algebra and holds for
Ginzburg-Landau, XY model or any other representation
of superconducting fluctuations.
This is no accident -- the straightforward relationship between the
massless (or massive) character of ${\cal L}_0 [v_\mu,a_\mu]$
and the superconducting phase disorder (or order)
is a consequence of rather general physical and
symmetry principles.

To make good on the above claim consider first a simple example
of an s-wave superconductor with a large gap $\Delta$
extending over all of the
Fermi surface. We can also view this as a model
for the high energy BdG quasiparticles in dSC, those far removed
from the nodes. The action takes the form similar to (\ref{bareaction}):
\begin{equation}
\tilde{\cal L}=\tilde\Psi^{\dag}[(\partial_{\tau}
+ia_\tau)\sigma_0 + iv_\tau\sigma_3]
\tilde\Psi + \tilde\Psi^{\dag}{\cal H}^{\prime}_s\tilde\Psi
+{\cal L}_0[v_\mu,a_\mu],
\label{bareactions}
\end{equation}
\noindent
but with ${\cal H}^{\prime}_s$ defined as:
\begin{equation}
\left( \begin{array}{cc}
{1\over 2m}(\hat{\bf \pi}+{\bf v})^2
-\epsilon_F & \Delta \\
\Delta  &
-{1\over 2m}(\hat{\bf \pi}-{\bf v})^2
+\epsilon_F
\end{array} \right)~~,
\label{hhs}
\end{equation}
with ${\hat{\bf \pi}} ={\hat{\bf p}} + {\bf a}$.
$v_\mu$ in the fermionic action (but not in ${\cal L}_0$) goes into
$v_\mu - (e/c)A_\mu$ when the electromagnetic field is included.
The reader might wish to recall here that we have defined
${\cal L}_0[v_\mu,a_\mu]$
in this particular way to clearly separate the superflow mediated
interactions among vortices, which include $A_\mu$, from
the entropic effects and short range amplitude
driven core-overlap interactions,
which do not.

\subsubsection{Thermal vortex-antivortex fluctuations}

We can now reap the benefits of this convenient separation.
In the language of BdG fermions the
system (\ref{bareactions}) is a large
gap ``semiconductor'' and the Berry gauge field couples to
it minimally through ``BdG'' vector and scalar
potentials ${\bf a}$ and $a_\tau$.
Such BdG semiconductor is a poor
dielectric diamagnet with respect to $a_\mu$. We proceed to ignore its
``diamagnetic susceptibility'' and also set $a_\tau=0$ to
concentrate on thermal fluctuations. All this means is that
the Berry gauge field part of the coupling between
quasiparticles and vortices in a large gap s-wave
superconductor influences the latter only
through weak short-range interactions which are
unimportant in the region of strong vortex
fluctuations near Kosterlitz-Thouless transition.
We can therefore drop ${\bf a}$
from the fermionic part of the action and integrate
over it to obtain ${\cal L}_0$ in terms of physical
vorticity $\rho ({\bf r})=(\nabla\times {\bf v})/\pi$.
Additional integration
over the fermions produces the effective
free energy functional for vortices:
\begin{equation}
{\cal F}[\rho]=M^2(2{\bf v} - \frac{2e}{c}{\bf A}^{\rm ext})^2 + (\cdots )
+ {\cal L}_0[\rho]~~,
\label{fs}
\end{equation}
where we have used our earlier notation and have introduced
a small external transverse vector potential ${\bf A}^{\rm ext}$.
The ellipsis denotes higher order contributions to vortex
interactions.
As discussed earlier in this Section, the familiar
long range interactions between vortices lead directly to the
standard Coulomb gas representation of the vortex-antivortex
fluctuation problem and Kosterlitz-Thouless transition.

The presence of these long range interactions implies
that the vortex system is incompressible (\ref{compress})
and long distance vortex density fluctuations are suppressed.
When studying the coupling of BdG quasiparticles to these fluctuations
it therefore suffices to expand the ``entropic'' part:
\begin{equation}
{\cal L}_0[\rho] \cong \frac{1}{2}\pi^2 K \delta\rho^2 + \dots =
\frac{1}{2}K (\nabla\times{\bf v})^2 + \dots~~.
\label{maxwellv}
\end{equation}
\noindent
The above expansion is justified above $T_c$ since we
know that at $T\gg T_c$ we must match the purely
entropic form of a non-interacting particle system,
${\cal L}_0\propto \delta\rho^2$ (see Appendix A).

To uncover the physical meaning of coefficient $K$
we expand the free energy $F$ of the vortex system to
second order in ${\bf A}^{\rm ext}$. In the pseudogap
state the gauge invariance demands that $F$ depends only on
$\nabla\times{\bf A}^{\rm ext}$:
\begin{equation}
F[\nabla\times{\bf A}^{\rm ext}]= F[0] +
\frac{(2e)^2}{2c^2}\chi \int d^2r (\nabla\times{\bf A}^{\rm ext})^2
+ \dots
\label{fexp}
\end{equation}
Note that $((2e)^2/c^2)\chi$ is just the diamagnetic susceptibility
in the pseudogap state. $\chi$ determines the long wavelength
form of the helicity modulus tensor $\Upsilon_{\mu\nu}(q)$ defined as:
\begin{equation}
\Upsilon_{\mu\nu}(q) =
\Omega\frac{\delta^2F}{\delta A^{\rm ext}_\mu (q)\delta A^{\rm ext}_\nu (-q)}
\big|_{A^{\rm ext}_\mu\to 0}~~.
\label{upsilon}
\end{equation}
The above is the more general form of $\Upsilon_{\mu\nu}(q)$
applicable to uniaxially symmetric 3D and (2+1)D XY or
Ginzburg-Landau models; in 2D only $\Upsilon_{xy}({\bf q)}$ appears.
In the long wavelength limit
$\Upsilon_{\mu\nu}(q)$ vanishes as $q^2\chi$:
\begin{equation}
\frac{c^2}{(2e)^2}\Upsilon _{\mu\nu}({\bf q}) =
\chi \epsilon _{\rho\alpha\mu}\epsilon _{\rho\beta\nu}q_{\alpha}q_{\beta}
+ \dots ~~~,
\label{eaiv}
\end{equation}
for the isotropic case, while for the anisotropic situation
$\chi _{\perp}\not =\chi_{\parallel}$:
\begin{equation}
\frac{c^2}{(2e)^2}\Upsilon _{\mu\nu}({\bf q}) =
(\chi_{\parallel} -\chi _{\perp})
\epsilon _{z\alpha\mu}\epsilon _{z\beta\nu}q_{\alpha}q_{\beta}
+
\chi _{\perp} \epsilon _{\rho\alpha\mu}\epsilon _{\rho\beta\nu}q_{\alpha}q_{\beta}~.
\label{eav}
\end{equation}
$\epsilon _{\alpha\beta\gamma}$ is the Levi-Civita symbol,
summation over repeated indices is understood and index $z$
of the anisotropic 3D GL or XY model is replaced by $\tau$
for the (2+1)D case. $(2e)^2/c^2$ is factored out
for later convenience.

So, what is $K$? Let us compute the helicity modulus of the problem
explicitly.
This is done by absorbing the small transverse vector
potential ${\bf A}^{\rm ext}$ into ${\bf v}$ in
(\ref{bareactions},\ref{vmuamu}) and integrating over
the new variable ${\bf v} - (e/c){\bf A}^{\rm ext}$.
The hecility modulus tensor measures
the screening properties of the vortex system. In a
superconductor, with topological defects bound in
vortex-antivortex dipoles, there is no screening
at long distances. This translates into
Meissner effect for ${\bf A}^{\rm ext}$.
When the dipoles unbind and some
free vortex-antivortex excitations appear the
screening is now possible over all lengthscales and there is no
Meissner effect for ${\bf A}^{\rm ext}$. The
information on presence or absence of such screening
is actually stored entirely in ${\cal L}_0$, where ${\bf A}^{\rm ext}$
reemerges after the above change of variables. We
finally obtain:
\begin{equation}
4\chi = K - \frac{K^2\pi^2}{T}
\lim_{|{\bf q}|\to 0^{+}}
\int d^2r e^{i{\bf q}\cdot{\bf r}}
\langle \delta\rho ({\bf r})\delta\rho(0)\rangle~~,
\label{eaxiii}
\end{equation}
where the thermal average $\langle\cdots\rangle$ is over
the free energy (\ref{fs}) with ${\bf A}^{\rm ext}= 0$
or, equivalently, (\ref{bareactions}).
In the normal phase only the first term
contributes in the long wavelength limit,
the second being down by an extra power
of $q^2$ courtesy of long range vortex interactions.
Consequently, $K=4\chi$ (the factor of 4 is due to the fact that
the true superfluid velocity is $2{\bf v}$ \cite{remark11}).

We see that in the pseudogap phase, with free vortex
defects available to screen, ${\cal L}_0[{\bf v}]$ takes on a
massless Maxwellian form, the stiffness of which is given by the
diamagnetic susceptibility of a strongly fluctuating superconductor.
In the fluctuation region $\chi$ is given by the
superconducting correlation length $\xi_{\rm sc}$  \cite{zbtloops}:
\begin{equation}
K=4\chi = 4{\cal C}_2T\xi_{\rm sc}^2~~,
\label{susceptibility}
\end{equation}
where ${\cal C}_2$ is a numerical constant, intrinsic
to a 2D GL, XY or some other model of superconducting
fluctuations.

As we approach $T_c$, $\xi_{\rm sc}\to\infty$ and the
stiffness of Maxwellian term (\ref{maxwellv}) diverges.
This can be interpreted as the Doppler gauge field
becoming massive. Indeed, immediately bellow the
Kosterlitz-Thouless transition at $T_c$,
${\cal L}_0\cong m^2_v{\bf v}^2 + \dots$, where $m_v\ll M$
and $m_v\to 0$ as $T\to T_c^-$.
This is just a reflection of the helicity modulus
tensor now becoming {\em finite} in the long wavelength
limit,
$$\Upsilon_{xy}\to {4e^2\over c^2}{m^2_vM^2\over 4M^2 +m_v^2}.$$
Topological defects are now bound in vortex-antivortex
pairs and cannot screen resulting in the
Meissner effect for  ${\bf A}^{\rm ext}$. The
system is a superconductor and ${\bf v}$ had become massive.

Returning to a d-wave superconductor, we can
retrace the steps in  the above analysis but we must replace
${\cal H}'_s$ in (\ref{bareactions}) with ${\cal H}'$ (\ref{vmuamu}).
Now, instead of a large gap BdG ``semiconductor'', we
are dealing with a narrow gap ``semiconductor'' or BdG
``semimetal'' because of the low energy nodal quasiparticles.
This means that we must
restore the Berry gauge field ${\bf a}$ to the fermionic
action since the contribution from nodal quasiparticles
makes its  BdG ``diamagnetic susceptibility'' very large,
$\chi_{\rm BdG}\sim 1/T\gg 1/T^*$ (see the next Section).
The long distance fluctuations of ${\bf v}$
and ${\bf a}$ are now {\em both} strongly suppressed,
the former through incompressibility of the vortex
system and the latter through  $\chi_{\rm BdG}$. This allows
us to expand (\ref{lab}):
\begin{equation}
{\cal L}_0\cong
\frac{K_A}{4}(\nabla\times{\bf v}_A)^2 +
\frac{K_B}{4}(\nabla\times{\bf v}_B)^2 + (\cdots )~~,
\label{lab2}
\end{equation}
where $K_A=K_B=K$ is mandated by the FT singular gauge.
Since in our gauge the fermion spin and charge channels decouple
${\bf A}^{\rm ext}$ still couples only to ${\bf v}$
and the above arguments connecting $K$ to the helicity
modulus and diamagnetic susceptibility $\chi$ follow
through. This finally gives the Maxwellian
form of Ref.  \cite{ftqed}:
\begin{equation}
{\cal L}_0\to
\frac{K}{2}(\nabla\times{\bf v})^2 +
\frac{K}{2}(\nabla\times{\bf a})^2~~,
\label{labd}
\end{equation}
where $K$ is still given by Eq. (\ref{susceptibility}). Note
however that $\xi_{\rm sc}$ of a d-wave and of an s-wave
superconductor
are two rather {\em different} functions of $T$, $x$ and other parameters
of the problem, due to strong Berry gauge
field renormalizations of vortex interactions in the
d-wave case. Nonetheless, as $T\to T_c$, the Kosterlitz-Thouless
critical behavior
remains unaffected since $\chi_{\rm BdG}$, while
large, is still finite at all finite $T$.

Just as advertised, we have
shown that ${\cal L}_0[\rho_A,\rho_B]=
{\cal L}_0[{\bf v},{\bf a}]$ in the pseudogap state takes
on the massless Maxwellian form (\ref{labd}), with the
stiffness $K$ set by the true superconducting correlation
length $\xi_{\rm sc}$ (\ref{susceptibility}).  This result
holds as a general feature of our theory irrespective
of whether one employs a Ginzburg-Landau theory, XY model,
vortex-antivortex Coulomb plasma, or any other description
of strongly fluctuating dSC, as long as such description
properly takes into account vortex-antivortex fluctuations
and reproduces Kosterlitz-Thouless phenomenology.
In the Appendix A we show that within the continuum vortex-antivortex
Coulomb plasma model:
\begin{equation}
{\cal L}_0 \to {T\over 2\pi^2 n_l}\left[
(\nabla\times{\bf v})^2 +
(\nabla\times{\bf a})^2\right]~~,
\label{2d}
\end{equation}
where $n_l$ is the average density of {\em free} vortex
and antivortex defects. Comparison with (\ref{labd})
allows us to identify
$\xi_{\rm sc}^{-2}\leftrightarrow 4\pi^2 {\cal C}_2n_l$  \cite{ftqed}.

\subsubsection{Quantum fluctuations of (2+1)D vortex loops}

The above results can be generalized to quantum
fluctuations of spacetime vortex loops. The superflow
fields $v_{A(B)\mu}$ satisfy $(\partial\times v_{A(B)})_\mu =
2\pi j_{A(B)\mu}$, where $j_{A(B)\mu} (x)$ are the coarse-grained
vorticities associated with $A(B)$ vortex defects and
$\langle j_{A(B)\mu}\rangle =0$.
The topology of
vortex loops dictates that $j_{A(B)\mu} (x)$ be a purely transverse
field, i.e. $\partial\cdot j_{A(B)} = 0$, reflecting the fact
that loops have no starting nor ending point.
Again, we begin with a large gap s-wave superconductor and
use its poor BdG diamagnetic/dielectric nature to justify dropping
the Berry gauge field $a_\mu$ from the fermionic part
of (\ref{bareactions}) and integrating over $a_\mu$
in the ``entropic'' part containing ${\cal L}_0[v_\mu,a_\mu]$.

The integration over the fermions
contains an important novelty specific to
the (2+1)D case: the appearance
of the Berry phase terms for quantum vortices as they
wind around fermions. Such Berry phase is the consequence
of the first order time derivative in the original
fermionic action (\ref{action0}). If we think of spacetime
vortex loops as worldlines of some relativistic quantum
bosons dual to the Cooper pair
field $\Delta ({\bf r},\tau)$, as we
do in the Appendix A, then these
bosons see Cooper pairs and quasiparticles as sources
of ``magnetic'' flux \cite{nodalliquid}.
At the mean field level, this translates
into a dual ``Abrikosov lattice'' or a Wigner crystal
of holes in a dual superfluid. Accordingly, the
non-superconducting ground state in the pseudogap
regime will likely contain a weak charge modulation -- the
modulation is made weak by the same strong fluctuations
that make $T_c\ll T^* (T_{\rm Nernst})$.
The focus of the present paper being the symmetric
AFL description of the pseudogap, we postpone
the discussion of this point to Part II and will
ignore it for the rest of this paper. This is justified by the fact
that $a_\mu$ does not couple to charge directly
and is quantitatively valid in the window
$T,\Delta_f\ll (\omega, v_{F,\Delta}|{\bf q}|)\ll T^*$,
where $\Delta_f$ is any small gap in the nodal TF spectrum
produced by the said charge modulation.

Hereafter, we blissfully turn the blind eye
to the above subtleties and
assume that the transition from a dSC into a pseudogap
phase proceeds via unbinding of vortex loops of
a (2+1)D XY model or its GL counterpart or, equivalently,
an anisotropic 3D XY or GL model where the role
of imaginary time is taken on by a third spatial
axis $z$ \cite{sudbo}. Having learned all we really need from
an earlier 2D example we can now integrate the fermions to
obtain the effective Lagrangian for coarse-grained
spacetime loops ($\pi j_\mu=(\partial\times v)_\mu$):
\begin{equation}
{\cal L}[j_\mu]=M^2_\mu(2v_\mu - \frac{2e}{c}A^{\rm ext}_\mu)^2 + (\cdots )
+ {\cal L}_0[j_\mu]~~,
\label{ls}
\end{equation}
where $M_x=M_y=M$ and $M_\tau=M/c_s$, with $c_s\sim v_F$ being the
effective ``speed of light'' in the vortex loop spacetime.
The incompressibility condition
reads $\langle j_\mu (q)j_\nu(-q)\rangle
\sim q^2[\delta_{\mu\nu} -(q_\mu q_\nu/q^2)]$ and
in the pseudogap state permits the expansion:
\begin{equation}
{\cal L}_0[j_\mu]\cong \frac{1}{2}\pi^2K_\tau j_\tau^2 +
\frac{1}{2}\pi^2\sum_{i=x,y}K_i j_i^2~~,
\label{l0s}
\end{equation}
where $K_x=K_y=K\not=K_\tau$. Using the analogy with the
uniaxially symmetric anisotropic 3D XY (or GL) model we
can expand the ground state energy in the manner of (\ref{fexp}):
\begin{equation}
E[\partial\times A^{\rm ext}]= E[0] +
\frac{(2e)^2}{2c^2}\sum_{\perp,\parallel}\chi_{\perp,\parallel}
\int d^3x (\partial\times A^{\rm ext})_{\perp,\parallel}^2~~,
\label{eexp}
\end{equation}
with $\chi_\perp=\chi_x=\chi_y=\chi$ and $\chi_\parallel=\chi_\tau\not=\chi$.
Note that the form of ${\cal L}_0$ (\ref{l0s}) follows directly
from the requirement that there are infinitely large vortex loops,
resulting in vorticity fluctuations over
all distances. Combined with (\ref{eexp}) and then translated to
the language of a (2+1)D XY (GL) model it tells us something
already familiar: upon the transition to the pseudogap state
generated by vortex loops unbinding, the superconductor has
turned into an insulator \cite{nodalliquid}.
$\chi$ and $\chi_\tau$ determine
the diamagnetic and dielectric susceptibilities of this
insulating pseudogap state.

The explicit computation of
$\Upsilon_{\mu\nu}(q)$ (\ref{upsilon},\ref{eav})
leads to:
\begin{equation}
4\chi_{i,\tau} = K_{i,\tau} - K^2_{i,\tau}\pi^2
\lim_{q\to 0^{+}}
\int d^3x e^{iq\cdot x}
\langle j(x)_{i,\tau}j(0)_{i,\tau}\rangle~~,
\label{eaxiii1}
\end{equation}
where the second term is again eliminated by the incompressibility
of the vortex system. This results in:
\begin{equation}
K=4\chi=4{\cal C}_3\xi_\tau~~;~~~
K_\tau=4\chi_\tau=4{\cal C}_3\frac{\xi_{\rm sc}^2}{\xi_\tau}~~,
\label{k3d}
\end{equation}
where we used the result for the anisotropic 3D XY or
GL models: $\chi_\perp={\cal C}_3T\xi_\parallel$,
$\chi_\parallel={\cal C}_3T\xi_\perp^2/\xi_\parallel$,
${\cal C}_3$ being the intrinsic  numerical constant for
those models (${\cal C}_3\not= {\cal C}_2$).
In the case of (2+1)D vortex loops
$\xi_\tau\propto\xi_{\rm sc}$ since our adopted
model has the dynamical critical exponent $z=1$. Thus,
we again encounter a massless Maxwellian form (\ref{l0s})
whose stiffness diverges as we approach a superconductor
except now this divergence is linear in the superconducting
correlation length, $K\propto \xi_{\rm sc}$.

The application to a d-wave superconductor is straightforward:
the nodal structure of BdG quasiparticles in dSC
helps along by the way of providing the anomalous stiffness
for the Berry gauge field $a_\mu$ -- upon integration
over the nodal fermions the following term emerges
in the effective action:
$$\frac{1}{2}\chi_{\rm BdG}(q)
(q^2\delta_{\mu\nu} -q_\mu q_\nu)a_\mu(q)a_\nu(-q)\sim$$
\begin{equation}
\sim|q|[\delta_{\mu\nu} -(q_\mu q_\nu/q^2)]a_\mu(q)a_\nu(-q)~~.
\end{equation}
In the terminology of
our chimerical BdG ``semimetal'', the ``susceptibility'' $\chi_{\rm BdG}$
is not merely very large, it {\rm diverges}:
$\chi_{\rm BdG}\sim 1/q$, as $q\to 0$, and
is computed in detail in  Section IV.
We use this pleasing fact to observe
that we are fully justified in expanding ${\cal L}_0[j_{A\mu},j_{B\mu}]$
and retaining only quadratic terms as long as we keep a
safe distance from the actual phase transition:
\begin{equation}
{\cal L}_0\cong
\frac{K_{A\mu}}{4}(\partial\times v_A)^2_\mu +
\frac{K_{B\mu}}{4}(\partial\times v_B)^2_\mu + (\cdots )~~,
\label{lab3}
\end{equation}
where $K_{A\mu}=K_{B\mu}=K_\mu$ is again assured by our choice
of the FT singular gauge (\ref{ft},\ref{jacobian}).

The above reasoning merits
an amusing aside: beside the ubiquitous incompressibility
($\langle (\partial\times v)_\mu  (\partial\times v)_\nu\rangle
\sim q^2[\delta_{\mu\nu} -(q_\mu q_\nu/q^2)]$) the integration
over nodal fermions now also occasions
diverging $\chi_{\rm BdG}$
implying $\langle (\partial\times a)_\mu  (\partial\times a)_\nu\rangle
\sim q[\delta_{\mu\nu} -(q_\mu q_\nu/q^2)]$. Had we chosen
a singular gauge in which the sets $A$ and $B$ were not
equivalent, like Anderson gauge \cite{andersongauge},
and therefore $K_{A\mu}\not=K_{B\mu}$, the ensuing
$(\partial\times v)\cdot (\partial\times a)$ coupling
in ${\cal L}_0$  (\ref{lab3},\ref{oscareq})
would now be driven to {\em zero}
in the long distance limit as an
{\em irrelevant} operator in the RG sense \cite{remarkdangerous}.
The reason is simple: the coupling of gauge fields
$v_\mu$ and $a_\mu$ to topological fermions
mandates that they decouple at the quadratic (harmonic)
level due to decoupling of ``charge'' and ``spin'' channels
for TF. Such coupling of $v_\mu$ and $a_\mu$
can only arise from ${\cal L}_0$ by
our uninformed choice of a singular gauge. Since both
$(\partial\times v)^2$ and $(\partial\times a)^2$ terms
in (\ref{lab3}) are strongly relevant due to diverging
contributions they receive from fermions, the
coupling constant in front of
$(\partial\times v)\cdot (\partial\times a)$, proportional
to $K_{A\mu}-K_{B\mu}$, is driven to zero under repeated
applications of RG transformation.
Therefore, the FT gauge (\ref{ft},\ref{jacobian})
specifically designed to insure $K_{A\mu}-K_{B\mu}=0$
at the very start, is recovered as an RG fixed point \cite{coulombgauge}.

Finally, we rewrite (\ref{lab3}) in terms of $v_\mu,a_\mu =
(1/2)(v_{A\mu}\pm v_{B\mu})$:
\begin{equation}
{\cal L}_0\to
\frac{K_\mu}{2}(\partial\times v)^2_\mu +
\frac{K_\mu}{2}(\partial\times a)^2_\mu ~~,
\label{lab4}
\end{equation}
and observe that the fact that $A_\mu^{\rm ext}$ couples
only to $v_\mu$ means that the expression (\ref{k3d})
for $K_\mu$ is still valid. Of course, $\xi_{\rm sc}(x,T)$
is now truly different from its s-wave counterpart,
including a possible difference in the critical exponent,
since the coupling of d-wave quasiparticles to the Berry gauge field
is marginal at the RG engineering level and may change
the quantum critical behavior of the superconductor-pseudogap (insulator)
transition.


\section{QED$_3$ -- a low energy effective theory of the pseudogap state}

We have now elucidated the nature of the coupling of our
two gauge fields,  $v_\mu$ and $a_{\mu}$, to TF and
have specified their ``bare'' thermal and quantum dynamics encoded in
${\cal L}_0[v_\mu,a_{\mu}]$ (\ref{labd},\ref{lab4}).
To make further progress toward our ultimate goal
of describing the low energy fermiology in the pseudogap state, we
now focus our attention on the nodal quasiparticle excitations of the
Hamiltonian ${\cal H}'$ Eq.\ (\ref{vmuamu}). This will
enable us to apply the machinery of perturbative RG
to nodal (massless) TF and rid our effective theory of all
remaining excess baggage.

\subsection{Farewell to $v_\mu$ and residual interactions}

\begin{figure}
\epsfxsize=5.5cm
\hspace{1.5cm}
\hfil\epsfbox{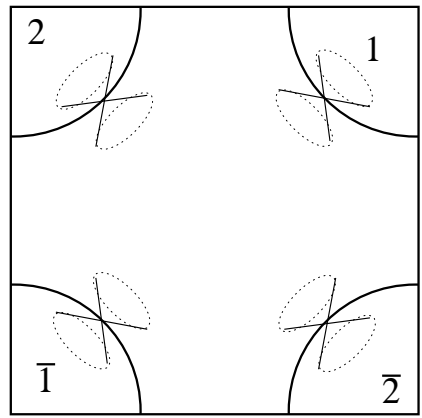}\hfill
\vspace{0.05cm}
\caption{Schematic representation of the Fermi surface of the
cuprate superconductors with the indicated nodal points of the $d_{x^2-y^2}$
gap.}
\label{fsfig}
\end{figure}

As indicated in Fig.\ \ref{fsfig},
the low energy quasiparticles are located at the four nodal
points of the $d_{xy}$ gap function:
$\bk_{1,\bar{1}}=(\pm k_F,0)$ and $\bk_{2,\bar{2}}=(0,\pm k_F)$,
hereafter denoted as $(1,\bar{1})$ and $(2,\bar{2})$ respectively.
To focus on the leading low energy behavior of the fermionic excitations
near the nodes we follow the standard procedure \cite{simon1} and linearize
the Lagrangian (\ref{bareaction}). To this end we write our TF spinor
$\tilde\Psi$ as a sum of 4 nodal fermi fields,
\begin{equation}
\tilde\Psi=e^{i\bk_1\cdot \br}\tilde\Psi_1
+e^{i\bk_{\bar{1}}\cdot \br}\sigma_2\tilde\Psi_{\bar{1}}
+e^{i\bk_2\cdot \br}\tilde\Psi_2
+e^{i\bk_{\bar{2}}\cdot \br}\sigma_2\tilde\Psi_{\bar{2}}.
\label{lin}
\end{equation}
The $\sigma_2$ matrices have been inserted here
for convenience: they insure that
we eventually recover the conventional form of the QED$_3$ Lagrangian.
(Without the $\sigma_2$ matrices the Dirac velocities at $\bar{1},\bar{2}$
nodes would have been negative.)
Inserting $\tilde\Psi$ into (\ref{bareaction}) and
systematically neglecting the higher order derivatives \cite{simon1}, we
obtain the nodal Lagrangian of the form
\begin{eqnarray}
{\cal L}_D&=&
\sum_{l=1,\bar{1}}\tilde\Psi_l^{\dag}[D_{\tau}
+ iv_FD_x\sigma_3+ iv_\Delta D_y\sigma_1]
\tilde\Psi_l
\label{qed3} \\
&+&\sum_{l=2,\bar{2}}\tilde\Psi_l^{\dag}[D_{\tau}
+ iv_FD_y\sigma_3+ iv_\Delta D_x\sigma_1]\tilde\Psi_l
+{\cal L}_0[a_\mu],
\nonumber
\end{eqnarray}
\noindent
where $D_\mu = \partial_\mu +ia_\mu$ denotes the covariant derivative
and ${\cal L}_0[a_\mu]$ isgiven by (\ref{lab4}):
\begin{equation}
{\cal L}_0[a_\mu] = {1\over 2}K_{\mu}(\partial\times a)^2_\mu
\equiv {1\over 2e^2_\mu}(\partial\times a)^2_\mu~~.
\label{finall0}
\end{equation}
$v_F={\partial \eps_{\bk}}/{\partial \bk}$
denotes the Fermi velocity at the node and
$v_{\Delta}={\partial \Delta_{\bk}}/{\partial \bk}$ denotes the
gap velocity. Note that $v_F$ and $v_{\Delta}$ already contain
renormalizations coming from high energy interactions and
are effective material parameters of our theory.
Similarly, $K_{\mu}=1/e^2_\mu$, derived in the previous
section in terms of $\xi_{\rm sc}(x,T)$, are treated
as adjustable parameters which are matched to experimentally
available information on the range of superconducting
correlations in the pseudogap state.

Doppler gauge field $v_\mu$ has disappeared from the above
expression. After informing
us on how to properly ``coarse-grain'' the theory and
dressing our Z$_2$-valued Berry gauge field in its
ultimate U(1) Maxwellian outfit, the time has come
to drop $v_\mu$, its eventual demise caused
by the Meissner coupling to BdG fermions discussed
in the previous section. After being ``screened'' by
high energy and nodal TFs it is rendered massive
both in the superconducting and pseudogap states and unimportant for
low energy physics. Its legacy lives on, however,
having given birth to the U(1) non-compact character of $a_\mu$.

In contrast,
Berry gauge field $a_\mu$ remains {\em massless} in the pseudogap
state, as it cannot acquire mass by coupling to the fermions. As seen from
Eq.\ (\ref{qed3}) $a_\mu$ couples minimally to the Dirac fermions and
therefore its massless character is protected by gauge invariance. Physically,
one can also argue that $a_\mu$ couples to the TF {\em spin} three-current
-- in a spin-singlet d-wave superconductor SU(2) spin symmetry
must remain unbroken, thereby insuring that $a_\mu$ remains massless.
Its massless Maxwellian dynamics (\ref{finall0}) in the
pseudogap state can therefore be traced
back to the topological state of spacetime vortex loops
and directly reflect the absence of true superconducting order
(Section II) or, equivalently, the presence of ``vortex loop
condensate'' and dual order (Appendix A).


We have also dispensed with the residual interactions
represented by ${\cal H}_{\rm res}$ in (\ref{h1}). These
interactions are generically short-ranged contributions from
the p-h and amplitude fluctuations part of the p-p channel
and in our new notation are exemplified by:
\begin{equation}
{\cal H}_{\rm res}\to {1\over 2}\sum_{l,l'}I_{ll'}\tilde\Psi_l^{\dag}\tilde\Psi_l
\tilde\Psi_{l'}^{\dag}\tilde\Psi_{l'}~~.
\end{equation}
The effective vertex $I_{ll'}$ has a scaling dimension $-1$
at the engineering level. This follows
from the RG analysis near the massless Dirac
points which sets the dimension of $\tilde\Psi_l$
to $[{\rm length}]^{-1}$. The implication is that
$I_{ll'}$ is {\em irrelevant} for low
energy physics in the perturbative RG sense and we are therefore
well within our rights in setting $I_{ll'}\to 0$. However,
the residual interactions will not go so quietly into the night:
such interactions are known to become important if
stronger than some critical value $I_c$ \cite{gusynin1,reenders}.
In the present theory, this is bound to happen in the severely underdoped
regime and at half-filling, as $x\to 0$  \cite{tvfqed}. In this
case, the residual interactions are becoming large
and comparable in scale to the pairing pseudogap $\Delta$,
and  are likely to cause chiral symmetry
breaking (CSB) which leads to spontaneous mass generation
for massless Dirac fermions.  The CSB and its variety of patterns
in the context of the theory (\ref{qed3},\ref{finall0}) in underdoped cuprates
were discussed in  \cite{herbut1,tvfqed} and are the subject of
Part II of this paper. Here, where we have limited ourselves
to the chirally symmetric phase, we assume $I_{ll'}< I_c$, which
we expect to be the case for moderate underdoping,
and set $I_{ll'}\to 0$.

In the end, we are left with (\ref{qed3},\ref{finall0}) as our
effective low energy theory for nodal TF. This theory,
derived previously in Ref.  \cite{ftqed}, is the
chief dynamical muscle behind the physics discussed
in this paper. It describes the problem of massless
topological fermions interacting with massless vortex
``Beryons'', i.e. the quanta of the Berry gauge field $a_\mu$, and
is formally equivalent to the
Euclidean quantum electrodynamics of massless
Dirac fermions in (2+1) dimensions (QED$_3$). It, however,
suffers from an intrinsic Dirac
anisotropy by virtue of $v_F\not = v_\Delta$.

\subsection{QED$_3$ Lagrangian for the pseudogap state}

We are now in position to do some real calculations within our theory.
Before we plunge into the algebra, however, we first apply
some cosmetics: Lagrangian (\ref{qed3}) is
not in the standard from as used in quantum
electrodynamics where the matrices associated with the components of covariant
derivatives form a Dirac algebra and mutually anticommute.
In (\ref{qed3}) the temporal derivative is associated with unit matrix and
it therefore commutes (rather than anticommutes)
with $\sigma_1$ and $\sigma_3$ matrices associated
with the spatial derivatives. These nonstandard commutation
relations, however, lead to some rather unwieldy
algebra. For this reason we manipulate the Lagrangian  (\ref{qed3})
into a slightly different form that is consistent with usual
field-theoretic notation. First, we combine each pair of antipodal
(time reversed) two-component spinors into one 4-component spinor,
\begin{equation}
\Upsilon_1=\left(\begin{array}{c} \tilde\Psi_1 \\
\tilde\Psi_{\bar{1}}\end{array}\right),
\ \ \
\Upsilon_2=\left(\begin{array}{c} \tilde\Psi_2 \\
\tilde\Psi_{\bar{2}}\end{array}\right).
\label{4comp}
\end{equation}
Second, we define a new adjoint four-component spinor
\begin{equation}
\bar\Upsilon_l=-i\Upsilon^\dagger_l\gamma_0.
\label{psib}
\end{equation}
In terms of this new spinor the Lagrangian becomes
\begin{equation}
{\cal L}_D=\sum_{l=1,2}
\bar\Upsilon_l\gamma_\mu D_\mu^{(l)}\Upsilon_l
+{1\over 2}K_{\mu}(\partial\times a)^2_\mu~~,
\label{l1}
\end{equation}
with covariant derivatives
\begin{eqnarray}
D_{\mu}^{(1)}&=&i[(\partial_\tau +ia_\tau),v_F(\partial_x +ia_x),
\vd(\partial_y +ia_y)], \nonumber \\
D_{\mu}^{(2)}&=&i[(\partial_\tau +ia_\tau),v_F(\partial_y +ia_y),
\vd(\partial_x +ia_x)].
\nonumber
\end{eqnarray}
The $4\times 4$ gamma matrices, defined as
\begin{eqnarray}
\gamma_0&=&\left(\begin{array}{cc}\sigma_2 & 0 \\ 0 & -\sigma_2\end{array}
\right), \
\gamma_1= \left(\begin{array}{cc}\sigma_1 & 0 \\ 0 & -\sigma_1\end{array}
\right),\nonumber \\
\gamma_2&=&\left(\begin{array}{cc}-\sigma_3 & 0 \\ 0 & \sigma_3\end{array}
\right)
\label{gamma}
\end{eqnarray}
now form the usual Dirac algebra,
\begin{equation}
\{\gamma_\mu,\gamma_\nu\}=2\delta_{\mu\nu}
\label{diraca}
\end{equation}
and furthermore satisfy
\begin{equation}
\tr(\gamma_\mu)=0~, \ \ \tr(\gamma_\mu\gamma_\nu)=4\delta_{\mu\nu}.
\label{traces}
\end{equation}

The use of the adjoint spinor $\bar\Upsilon$
instead of conventional $\Upsilon^\dagger$
is a purely formal device which will simplify calculations but does not
alter the physical content of the theory. At the end of the calculation
we have to remember to undo the transformation (\ref{psib}) by multiplying
the $\langle \Upsilon(x)\bar\Upsilon(x')\rangle$ correlator by $i\gamma_0$ to
obtain the physical correlator $\langle \Upsilon(x)\Upsilon^\dagger(x')
\rangle$.

Next, to make the formalism simpler
still we can eliminate the asymmetry
between the two pairs of nodes by performing an
internal SU(2) rotation at nodes 2, $\bar 2$:
\begin{equation}
\Upsilon_{2}\to e^{-i{\pi\over 4}\gamma_0}\gamma_1\Upsilon_{2},
\label{su2}
\end{equation}
leading to the anisotropic QED$_3$ Lagrangian
\begin{equation}
{\cal L}_D=\sum_{l=1,2}
\bar\Upsilon_l v_\mu^{(l)}\gamma_\mu (i\partial_\mu -a_\mu)\Upsilon_l
+{1\over 2}K_{\mu}(\partial\times a)^2_\mu~,
\label{l2}
\end{equation}
with
$v_{\mu}^{(1)}=(1,v_F,\vd)$ and $v_{\mu}^{(2)}=(1,\vd,v_F)$.


\section{Spectral properties of topological fermions and physical electrons
in QED$_3$}

We shall start by considering the isotropic case, $v_F=\vd=1$, which
although unphysical in the strictest sense,
is computationally much simpler and provides penetrating insights into
the physics embodied by the QED$_3$ Lagrangian (\ref{l2}).
After we have understood
the isotropic case we will then be ready to tackle
the calculation for the general case and will
show that Dirac cone anisotropy does
not modify the essential physics discussed here. To make contact with
standard literature on QED$_3$, we further consider a more general problem
with $N$ {\em pairs} of nodes described by Lagrangian
\begin{equation}
{\cal L}_D=\sum_{l=1}^{N}
\bar\Upsilon_l \gamma_\mu (i\partial_\mu -a_\mu)\Upsilon_l
+{1\over 2}K_{\mu}(\partial\times a)^2_\mu~~.
\label{l3}
\end{equation}
For the basic problem of a single CuO$_2$ layer $N=2$.
As we will show in the next section, $N$ itself is variable
and can be equal to four or six in bi- and multi-layer
cuprates. Our analytic results can be viewed as arising
from the formal $1/N$ expansion, although we expect them to be
qualitatively (and even quantitatively!) accurate even for
$N=2$ as long as we are within the {\em symmetric} phase
of QED$_3$  -- the quantitative accuracy stems from
a fortuitous conspiracy of small numerical prefactors \cite{gusynin1}.

\subsection{Berryon propagator}

Ultimately, we are interested in the properties of physical electrons.
To describe those we need to understand the properties of the
electron--electron interaction mediated by the gauge field $a_\mu$.
To this end we proceed to calculate the Berry gauge field propagator by
integrating out the fermion degrees
\begin{figure}[t]
\includegraphics[width=8cm]{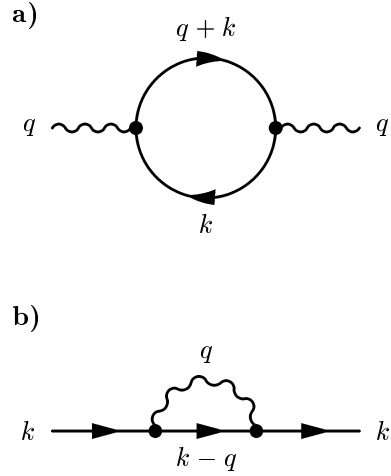}
\caption
{\label{fig_diag1}
One loop Berryon polarization (a) and TF self energy (b).
}
\end{figure}
of freedom from the Lagrangian (\ref{l3}). To one loop order this corresponds
to evaluating the vacuum polarization bubble Fig. \ref{fig_diag1}(a). Employing
the standard rules for Feynman diagrams in the momentum space \cite{peskin}
the vacuum polarization reads:
\begin{equation}
\Pi_{\mu\nu}(q)=N \int{d^3k\over (2\pi)^3}
\tr[G_0(k)\gamma_\mu G_0(k+q)\gamma_\nu].
\label{bub1}
\end{equation}
Here $G_0(k)=k_\alpha\gamma_\alpha/k^2$ is the free Dirac Green function,
$k=(k_0,\bk)$ denotes the Euclidean three-momentum, and the trace is performed
over the $\gamma$ matrices.

The integral in Eq.\ (\ref{bub1}) is a standard one (see Appendix for the
details of computation) and the result is
\begin{equation}
\Pi_{\mu\nu}(q)={N\over 8}|q|\left(\delta_{\mu\nu}-{q_\mu q_\nu\over q^2}
\right),
\label{bub2}
\end{equation}
where $|q|=\sqrt{q^2}$.
The one loop effective action for the Berry gauge field therefore becomes
$(2\pi)^{-3}\int d^3q{\cal L}_B$ with
\begin{equation}
{\cal L}_B[a]=\left({N\over 8}|q|+{1\over 2e^2} q^2\right)
\left(\delta_{\mu\nu}-{q_\mu q_\nu\over q^2}\right) a_\mu(q)a_\nu(-q)~.
\label{ber}
\end{equation}
At low energies and long wavelengths, $|q|e^2\ll N/4$,
 the fermion polarization completely overwhelms
the original Maxwell bare action term and the Berryon properties become
universal. In particular, the coupling constant $1/e^2$ drops out at low
energies and only reappears as the ultraviolet cutoff.
Physically, the medium of massless Dirac fermions screens the long range
interactions mediated by $a_\mu$. In QED$_3$ this screening is incomplete: the
gauge field becomes stiffer by one power of $q$ but still remains
{\em massless}, in accordance with our general expectations.
This anomalous stiffness of $a_\mu$
justifies the quadratic level expansion of ${\cal L}_0$
(\ref{lab4}) and renders higher order terms
{\em irrelevant} in the RG sense. The theory therefore clears
an important self-consistency check.

At low energies the fully dressed Berryon
propagator is given as an inverse of the
polarization,
\begin{equation}
D_{\mu\nu}(q)=\Pi^{-1}_{\mu\nu}(q)~~.
\label{ber1}
\end{equation}
In order to perform this inversion we have to
fix the gauge. To this end we
implement the usual gauge fixing procedure by replacing $q_\mu q_\nu/q^2
\to (1-\xi^{-1})q_\mu q_\nu/q^2$ in Eq.\ (\ref{bub2}).
$\xi\geq 0$ parametrizes the orbit of
all covariant gauges. For example, $\xi=0$ corresponds to Lorentz gauge
$k_\mu a_\mu(k)=0$ while $\xi=1$ corresponds to Feynman gauge.
Upon inversion we obtain the low energy Berryon propagator
\begin{equation}
D_{\mu\nu}(q)={8\over |q|N}\left(\delta_{\mu\nu}-{q_\mu q_\nu\over q^2}(1-\xi)
\right),
\label{ber2}
\end{equation}
in agreement with previous authors  \cite{kim}.

\subsection{TF self energy and propagator}

TF propagator is a gauge dependent entity
and one could therefore immediately object that
as such it has no direct physical content and is of no
interest. Such viewpoint,
while expressed frequently, is actually quite
naive. The reality is that in gauge theories various
gauge-variant objects can often be connected to
physical gauge-invariant quantities when computed within
a {\em particular} choice of gauge. In practice, a rather typical
occurrence is that a gauge-invariant physical propagator is given by
a hugely non-local form which is basically impossible to compute
except in a judiciously chosen gauge where it
is related to a much simpler, and therefore far easier to
compute, gauge-variant propagator. Consequently,
a gauge-variant propagator computed along a particular
gauge orbit often contains relevant information about
the true dynamics of a gauge theory -- the trick is
to know how to extract this information.

This general statement holds
in the case of QED$_3$ as well. TF propagator
evaluated in an arbitrary covariant gauge parametrized by $\xi$ contains
useful information about the nature of the fermionic
excitations of the system. We will show that its coupling to the
massless gauge field destroys the usual
Fermi liquid pole and results in the propagator
displaying a Luttinger-like behavior, characterized by
a small anomalous dimension. In the next
subsection we argue that the physical electron propagator of
our theory, which can be related to a particular
gauge invariant fermion propagator of QED$_3$,
exhibits the same Luttinger-like behavior.

The lowest order self-energy diagram is depicted in Fig.\ \ref{fig_diag1}(b)
and reads
\begin{equation}
\Sigma(k)=\int{d^3q\over (2\pi)^3}
D_{\mu\nu}(q)\gamma_\mu G_0(k+q)\gamma_\nu.
\label{self1}
\end{equation}
Again, the computation is rather straightforward
(see Appendix for details) and the most divergent
part is
\begin{equation}
\Sigma(k)={4(2-3\xi)\over 3\pi^2 N}\kslash
\ln\left({\Lambda\over |k|}\right),
\label{self2}
\end{equation}
where we have introduced the Feynman ``slash'' notation,
$\kslash=k_\mu\gamma_\mu$.

To the leading $1/N$  order, the
inverse TF propagator is given by
\begin{equation}
G^{-1}(k)=\kslash\left[1+\eta\ln\left({\Lambda\over|k|}\right)\right]
\label{g1}
\end{equation}
with
\begin{equation}
\eta=-{4(2-3\xi)\over 3\pi^2N}~~.
\label{eta}
\end{equation}
Higher order contributions in $1/N$ will necessarily affect this result.
Renormalization group arguments \cite{maris} and non-perturbative approaches
 \cite{pennington} strongly suggest that Eq.\ (\ref{g1}) represents
the start of a perturbative series that eventually resums into a power law:
\begin{equation}
G^{-1}(k)=\kslash\left({\Lambda\over|k|}\right)^{\eta}~~.
\label{g2}
\end{equation}
This implies real-space propagator of the form
\begin{equation}
G(r)=\Lambda^{-\eta}{\rslash\over r^{3+\eta}}.
\label{g3}
\end{equation}
Thus, the TF propagator exhibits a Luttinger-like
algebraic singularity at small momenta, characterized by
an anomalous exponent $\eta$. In the Lorentz gauge ($\xi=0$) we find
$\eta= -8/3\pi^2N\simeq -0.13$, for $N=2$.
This rather small numerical value for the
anomalous dimension exponent (which is even considerably
smaller for $N=4$ or $N=6$) indicates that
the unraveling of the Fermi liquid pole in the original
TF propagator brought about by its interaction with the
massless Berry gauge field is in a certain sense ``weak''.
Note also that $\eta$ is {\em negative} in the Lorentz
gauge while it becomes positive, $\eta= 4/3\pi^2N\simeq 0.06$ for $N=2$,
in the Feynman gauge ($\xi =1$). The above results provide a strong
indication that the physical, gauge-invariant fermion propagator
also has a Luttinger-like form, characterized by a small
and {\em positive} anomalous dimension  \cite{ftqed}. We now show that this
indeed is the case.

\subsection{Physical electron propagator}

Various spectroscopies on cuprates, such as ARPES and STM,
as well as numerous optical and microwave techniques, all
measure the spectral function of a real physical electron, not of TF.
Therefore, we
are ultimately interested in computing
the propagator of the physical electron in our theory,
\begin{equation}
G^{\rm elec}(x-x')=\langle\Psi(x)\Psi^\dagger(x')\rangle~~,
\label{e1}
\end{equation}
where $\Psi(x)$ is the original electron field operator
appearing in (\ref{action0}) (the reader should recall
that this operator already contains high-energy renormalizations
built in at the very beginning). If we define
a matrix
\begin{equation}
\hat\Omega(x)=\left( \begin{array}{cc}
e^{-i\phiup(x)} & 0 \\
0 & e^{i\phidown(x)}
\end{array} \right)~,
\label{om}
\end{equation}
we can write $G^{\rm elec}$ in terms of TF fields as
\begin{equation}
G^{\rm elec}(x-x')=\langle\hat\Omega(x)\tilde\Psi(x)\tilde\Psi^\dagger(x')
\hat\Omega^\dagger(x')\rangle~~.
\label{e2}
\end{equation}
Ordinary spectroscopies reflect the diagonal part of the electron propagator
\begin{equation}
[G^{\rm elec}(x-x')]_{11}=\langle e^{-i[\phiup(x)-\phiup(x')]}
[\tilde\Psi(x)\tilde\Psi^\dagger(x')]_{11}\rangle~,
\label{e3}
\end{equation}
and similar expression for $[G^{\rm elec}(x-x')]_{22}$. We may recast
this in a more convenient form by writing the phase difference in
the exponent as a line integral of a gradient along a {\em straight} line
connecting $x$ and $x'$,
\begin{equation}
\phiup(x)-\phiup(x')\to -\int_x^{x'}\partial_\mu\phiup ds_\mu,
\label{line}
\end{equation}
and then expressing the phase gradient in terms of the two gauge fields
$a_\mu$ and $v_\mu$:
\begin{equation}
[G^{\rm elec}(x-x')]_{11}=\left\langle e^{i\int_x^{x'}(v_\mu+a_\mu) ds_\mu }
[\tilde\Psi(x)\tilde\Psi^\dagger(x')]_{11}\right\rangle,
\label{e4}
\end{equation}
This expression only involves the coarse-grained Doppler and Berry
gauge fields and TF, which are precisely the fields that
enter our effective low-energy theory, and is therefore
amenable to analysis. Note that (\ref{e4}) is only a long-distance 
approximation to the exact electron propagator (\ref{e3}) which
is defined through discrete vortex variables entering via
$\phiup$ ($\phidown$). 

As discussed earlier, the Doppler gauge field $v_\mu$ is massive in both normal
and superconducting phases and therefore its fluctuations will not affect
the low energy, long wavelength properties of the electron propagator. We
may thus remove it from the line integral Eq.\ (\ref{e4}) and focus on the
quantity
\begin{equation}
{\cal G}(x-x')=\left\langle e^{i\int_x^{x'}a_\mu ds_\mu }
\Upsilon(x)\bar\Upsilon(x')\right\rangle~~.
\label{e5}
\end{equation}
By considering the
transformation properties of $\Upsilon(x)$ 
under the gauge transformations with
respect to $a_\mu$ it is easy to verify that ${\cal G}(x-x')$ is
{\em gauge invariant}. This quantity therefore 
represents a gauge invariant
propagator of QED$_3$ theory (\ref{l1})
and its knowledge allows us to reconstruct the diagonal
components of the electron propagator by means of
\begin{equation}
[G^{\rm elec}(x-x')]_{ii}=[i\gamma_0{\cal G}(x-x')]_{ii}
\label{e6}
\end{equation}
It seems natural to attempt to relate the components
of the above gauge invariant propagator
to the physical electron propagator, since
the latter by definition must be
gauge invariant under the transformations
of the internal gauge field $a_\mu$. 

The question arises how to evaluate the gauge invariant propagator
${\cal G}(x-x')$. This turns out to be a
non-trivial issue since, despite its pleasing 
manifest invariance under gauge and spacetime symmetries, (\ref{e5}) 
also exhibits a severe (linear) ultraviolet
divergence arising from the straight
line integral of the gauge field. This renders it ill-defined
in absence of some proper regularization (see below for
more details).
Here we adopt the approach discussed by Brown
\cite{brown1,boulaware} in the context of QED$_4$. Brown proves that the
following relation exists between gauge invariant 
propagator $\tilde {\cal G}$ and gauge dependent
propagator $G$  \cite{remark1} (see also Appendix C):
\begin{equation}
\tilde {\cal G}(r)=e^{F(r)} G(r),
\label{brown1}
\end{equation}
where $r=x-x'$ and
\begin{equation}
F(r)={1\over 2}\int d^3z\int d^3z' J_\mu(z) D_{\mu\nu}(z-z') J_\nu(z')~~,
\label{brown2}
\end{equation}
with
\begin{equation}
 J_\mu(z)=r_\mu\int_0^1d\alpha\delta(z-x'-\alpha r)
\label{brown3}
\end{equation}
representing the source term for the line integral in Eq.\ (\ref{e5}).
In the above $G(r)$ and $D_{\mu\nu}(r)$ 
refer to the real space gauge dependent
fermion and Berryon propagators respectively, 
obtained by Fourier transforming
the expressions (\ref{g2}) and (\ref{ber2}). 
By definition, both
are to be computed in {\em covariant} gauge. Brown's result Eq.\
(\ref{brown1}) is an
explicit statement of the fact that 
one can construct two gauge invariant propagators by using 
the line integral of the gauge field, 
${\cal G}$ and $\tilde {\cal G}$.
This is a rather general feature of Abelian gauge theories \cite{boulaware} 
and is easily generalized to QED$_3$. ${\cal G}$ and $\tilde {\cal G}$
can be formally related through a gauge transformation
(see Appendix C) and one might think of $\tilde {\cal G}$ as representing
suitably regularized ${\cal G}$ \cite{regularization}. Alternatively, we 
can simply think of  $\tilde {\cal G}$ as being another
QED$_3$ fermion propagator invariant under gauge and spacetime symmetries
just like ${\cal G}$. We discuss the details 
pertaining to Eq.\ (\ref{brown1}) in the Appendix C.

To calculate $\tilde {\cal G}(r)$ from Eq.\ (\ref{brown1}) we need to
evaluate $F(r)$. We proceed by first performing the $z$, $z'$ integrals
to obtain
\begin{equation}
F(r)={1\over 2}\int_0^1d\alpha\int_0^1d\beta
r_\mu D_{\mu\nu}(r(\alpha-\beta)) r_\nu~~,
\label{f2}
\end{equation}
By power counting
the expression for $F(r)$ suffers from linear UV divergence
reflecting the singular behavior of the gauge field line integral at
short distances. This singularity is the main reason why
direct computation of (\ref{e5}) is such a frustrating
task, despite its deceivingly compact and elegant form.
A typical scheme to regularize this linear UV divergence interferes
with gauge invariance and corrupts the effort to extract
the true physical part of (\ref{e5}) which we expect to be
scale-invariant. The advantage of Brown's
approach (\ref{brown1}) is twofold: it 
permits computation of a gauge invariant 
propagator $\tilde {\cal G}$ in the
{\em covariant} gauge where the UV divergence can be treated
with the help of {\em dimensional regularization}, which
respects the gauge invariance and preserves the long wavelength,
low energy properties
of the physical propagators, and the offending linear part
of the UV divergence cancels out between numerator and
denominator in a gauge invariant manner (Appendix C). This allows us to
extract a meaningful power law behavior for $\tilde {\cal G}$.
To take advantage of dimensional
regularization we express ${\cal P}(r)\equiv
r_\mu D_{\mu\nu}(r) r_\nu$ as a Fourier transform in $d$-dimensions,
\begin{equation}
{\cal P}(r)={8\over N}\int {d^dk\over(2\pi)^d} {e^{ik\cdot r}
\over k}\left[r^2-(1-\xi){(k\cdot r)^2\over k^2}\right]~,
\label{f3}
\end{equation}
and treat $d$ as a continuous variable.
This Fourier transform is evaluated in the Appendix, giving the result
\begin{equation}
{\cal P}(r)={4\over N}{\Gamma\left({d-1\over 2}\right)\over\pi^{d+1\over 2}}
[1+(d-2)(1-\xi)]r^{3-d}.
\label{f4}
\end{equation}
Substituting  ${\cal P}(r(\alpha-\beta))$ into Eq.\ (\ref{f2}) and performing
the remaining integrals by means of
\begin{equation}
\int_0^1d\alpha\int_0^1d\beta |\alpha-\beta|^\zeta
={2\over(\zeta+1)(\zeta+2)},
\label{zeta}
\end{equation}
we find, near $d=3$,
\begin{eqnarray}
F(r)&=&-{4(2-\xi)\over N\pi^2}\lim_{d\to 3}
\left({r^{3-d}\over 3-d}\right)
\nonumber \\
&=& -{4(2-\xi)\over N\pi^2}\left[\ln(\Lambda r)+{1\over 3-d}
\right]_{d\to 3}~~. \label{f5} \\ \nonumber
\end{eqnarray}
The UV divergence is now parametrized by the ($r$-independent) second term
in the angular brackets. The leading long-distance behavior is contained
in the log, implying a power law contribution
$e^{F(r)}\propto r^{-4(2-\xi)/N\pi^2}$ to the gauge invariant
propagator $\tilde {\cal G}(r)$. Combining (\ref{f5}) with (\ref{g3}) we
obtain  \cite{remark2}
\begin{equation}
\tilde {\cal G}(r)=\Lambda^{-\eta'}{\rslash\over r^{3+\eta'}}~,
\label{brown77}
\end{equation}
or in the momentum space
\begin{equation}
\tilde {\cal G}(k)=\Lambda^{-\eta'}{\kslash\over k^{2-\eta'}}~,
\label{brown78}
\end{equation}
where the anomalous dimension exponent $\eta'$ is given by
\begin{equation}
\eta'=\eta+{4(2-\xi)\over N\pi^2}={16\over3\pi^2N}~~.
\label{etap}
\end{equation}
The last equation informs us that in the exponent $\eta'$ the gauge
fixing parameter $\xi$ has canceled
out and  $\tilde{\cal G}(r)$ is indeed gauge invariant. We have thus
verified, by explicit calculation to leading order in $1/N$, that Eq.\
(\ref{brown1}) yields a gauge invariant TF propagator which we can connect
to the physical electron propagator by means of Eq.\ (\ref{e6}) 
with ${\cal G}\to \tilde{\cal G}$.

An interesting feature of the above result is that for $\xi=2$ we have
$F(r)=0$. In the QED literature this is
known as ``Yennie's gauge'' and its
significance is
that in this particular gauge the diagonal components of the TF
propagator are directly equal to $\tilde {\cal G}(r)$.
In Yennie's gauge one can
therefore evaluate various electron observables in terms of TF propagator
without worrying about the exponential factors.
This is just the situation we have anticipated in the previous
subsection. One could also define an ``anti-Yennie's gauge'' (or
a ``non-local gauge'' as it is known in the QED$_3$ literature),
$\xi={2\over 3}$,
in which $\eta=0$ and the effect of the gauge field fluctuations on
$\tilde {\cal G}(r)$
is contained entirely in $F(r)$. To leading order in $1/N$ this
observation further solidifies the expectation
that the leading log in the self energy
indeed exponentiates and the low energy,
long lengthscale propagator behaves as a power law.

Another important feature to observe is that $\eta'>0$ -- the
electron has acquired
a {\em positive} anomalous dimension. The positivity of
$\eta'$ is mandatory from general considerations -- once we perform
the Euclidean rotation and obtain the real time electron
propagator the conditions of unitarity and causality
of our original problem demand $\eta' >0$.
This is also a physically sensible
result implying that the 
interacting electron propagator (\ref{brown77})
decays on long lengthscales {\em faster} than the free
BdG electron propagator.
Interaction mediated by the massless
gauge field destabilizes the Fermi liquid pole in the
original propagator and leads to Luttinger liquid-like power
law fermionic correlator in the low energy,
long wavelength limit -- this is our algebraic Fermi
liquid (AFL), a non-Fermi liquid {\em symmetric} phase
which governs the physics of the pseudogap state.
The positivity of $\eta'$ means that the
interacting AFL propagator is {\em less coherent} than the free
BdG electron propagator which is just
what one expects on intuitive grounds. We therefore
propose that $\tilde {\cal G}$ be identified as the true electron
propagator in our theory.

The Luttinger-like electron propagator that follows
from (\ref{brown78}) leads to a characteristic
asymmetry between the energy and momentum distribution
curves (EDC and MDC) observed 
in ARPES experiments \cite{valla} (note that, within our
theory, this behavior is limited to the pseudogap
state) -- the MDC is a
very sharp Lorentzian close to the Fermi surface while
EDC is broad, reflecting decoherence of physical electrons
in the pseudogap state  \cite{ftqed}). 
This decoherence 
is relatively weak with
the explicit value of $\eta' \simeq 0.27$ for the case of an
individual CuO$_2$ layer where $N=2$. In bi-layer or multi-layer
systems $N$ could be four, six or even higher and $\eta'$
is even smaller. The reason for this increase in $N$
is the anisotropy of the tunneling matrix element between the
constituent CuO$_2$ planes within a multi-layer unit cell.
This matrix element effectively vanishes near the nodes, along
$(\pm\pi,\pm\pi)$ directions,
but is appreciable elsewhere in
the Brillouin zone \cite{andersen}.
The result is that low energy BdG fermions on different
constituent CuO$_2$ layers remain {\em decoupled} while
the vortex excitations on these same layers are strongly {\em coupled}
within the unit cell of a multi-layer, since their coupling
comes from an integral over the full Brillouin zone. This
translates to a larger effective $N$ in our QED$_3$ (\ref{l3})
and to a corresponding reduction in the anomalous dimension:
$\eta'\simeq 0.13$ for $N=4$ (bi-layer like YBCO)
or $\eta'\simeq 0.09$ for $N=6$ (tri-layer like
HgBa$_2$Ca$_2$Cu$_3$O$_8$). The actual
numerical value of the anomalous dimension exponent
$\eta'$ is the ``fingerprint'', a unique
mathematical signature of the symmetric phase
of QED$_3$ and therefore of the algebraic Fermi liquid (AFL)
state within the pseudogap regime of underdoped
cuprates. Determining $\eta'$ directly from experiments,
either through various spectroscopies or transport
measurements, would be a major step toward testing
theoretical ideas expressed in this paper.

The exponent $\eta'$ of the physical electron
propagator had come
under much scrutiny as of late since
several effective theories related to QED$_3$
emerged recently in condensed matter physics, in problems
like Heisenberg antiferromagnets or spin liquids.
While in each case the physical content of these multiple reincarnations
of QED$_3$ differs completely from the one discussed
in the present paper and from each other, the issue of the gauge invariant
QED$_3$ fermion propagator and the value of $\eta'$ loom
large in all these different contexts, for obvious reasons.
In particular, $\eta'$ has been calculated
recently by Rantner and Wen  \cite{rantner1,rantner2}
and also by Khveshchenko  \cite{khvesh1}.
The former authors obtain 
$\eta'=-32/3\pi^2 N$  \cite{remark4} by performing a
calculation of ${\cal G}$ in the so-called axial gauge \cite{leibrandt},
in which the line integral of the
gauge field in Eq.\ (\ref{e5}) is taken to vanish for a particular
direction in the
real space. Negative anomalous dimension $\eta'<0$ would imply that
the interacting electron propagator 
is {\em more coherent} at long distances
than the propagator of a free electron and this is
prohibited on general grounds, as discussed above. For example,
negative $\eta'$ produces
divergent electronic density of states and leads to unphysical singular
behavior in various thermodynamic and transport quantities.
Thus, negative anomalous dimension for the physical 
electron should be, in our view,
rejected out of hand. For reader's benefit, we should stress that
we believe that the calculations
carried out in Refs.  \cite{rantner1,rantner2,khvesh1} are
perfectly correct, in the sense that $\eta'=-32/3\pi^2 N$ indeed
follows from the algebra once we adopt the axial gauge regularization
of (\ref{e5}) as implemented in  \cite{rantner1,rantner2}
and perform the calculation with logarithmic accuracy;
we have done such a calculation ourselves and
have obtained the same result. Operationally, the problem
with computation of  ${\cal G}$ (\ref{e5})
is not in the algebra but resides in the physical
interpretation of the obtained results. In the axial
gauge regularization followed by the momentum space computation
employed by the authors of Refs.  \cite{rantner1,rantner2},
the negative value for
$\eta'$ arises from the treatment of spurious gauge singularities
that are invariably introduced by
writing the gauge boson propagator in the
axial gauge  \cite{leibrandt}.
These singularities then must be regularized in
some way and this is done using an
ad hoc principal value prescription for momentum space
integrals. The problem with this prescription is that
the momentum space propagator in the axial gauge
{\em does not exist} \cite{zwanziger}.
Axial gauge is an example of a singularly
non-covariant gauge (like Coulomb, temporal or similar
gauges) and as
such does not fully fix the gauge. The gauge transformations
which are independent of the spacetime variable $x_1$ singled out by
the axial gauge but have arbitrary dependence on the remaining
$d-1$ variables $(x_2,x_3,\dots)$ are still allowed and cost no energy.
Consequently, the gauge-variant two-point fermion propagator
$G({\bf x},{\bf x}')$ must vanish whenever $x_2\not= x_2'$,
$x_3\not= x_3'$, $\dots$ \cite{zwanziger}. The same problem resurfaces
in a different form when one computes (\ref{e5}) directly
in the covariant gauge \cite{melikyan}: the expectation value of the 
transverse part ($\xi=0$) of the line integral, 
$\langle\exp (i\int_x^{x'}a_\mu ds_\mu )\rangle$,
decouples from the rest of the expression. In
fact, the result $\eta'=-32/3\pi^2 N$ is easily understood
as arising from the TF propagator in the Lorentz gauge $G_{\xi=0}$
($\eta_L = -8/3\pi^2 N$)
being made more coherent through simply being multiplied
by the said expectation value
of the transverse part of line integral ($\eta'=
\eta_L + \eta_t$ where $\eta_t = -8/\pi^2 N$
from the above computation of $F$, with $\xi=0$).
Since $\langle\exp (i\int_x^{x'}a_\mu ds_\mu )\rangle_{\xi =0}$ 
is an expectation value of
a phase factor, this is clearly a troubling
result -- such a multiplicative factor can make the
full propagator only less
coherent than $G_{\xi=0}$ as $|x-x'|\to\infty$. 
The problem is that the transverse part of the line
integral is more divergent than just a simple $\log r$
appearing in the exponent.
When the dominant (linear) divergence is included,
the full propagator ${\cal G}$ is exponentially suppressed 
($\eta\to +\infty$) and cannot be computed without some 
physically motivated UV
regularization scheme \cite{regularization}. 

We have also carried out our own
calculation in the axial gauge using a
different regularization scheme and
obtained  different (negative)
exponent $\eta'=-16/3\pi^2 N$. We are thus
forced to conclude that the axial gauge calculations yield values
of $\eta'$ that are regularization scheme dependent, and are therefore
inherently unreliable.
By contrast no such ambiguities arise when employing Eq.\ (\ref{brown1}) and
the computation of $\tilde {\cal G}$ yields 
physically reasonable positive anomalous dimension
given by Eq.\ (\ref{etap}). 

We now discuss the off-diagonal (anomalous) components of the electron
propagator. According to Eq.\ (\ref{om}) we have
\begin{equation}
[G^{\rm elec}(x-x')]_{12}=\langle e^{-i[\phiup(x)+\phidown(x')]}
[\tilde\Psi(x)\tilde\Psi^\dagger(x')]_{12}\rangle,
\label{oo1}
\end{equation}
and similar expression for $[G^{\rm elec}(x-x')]_{21}$. It is now less
straightforward to interpret the phase factor in terms of our gauge fields
$a_\mu$ and $v_\mu$. One way to do this is to add and subtract $\phiup(x')$
and write
\begin{eqnarray}
\phiup(x)+\phidown(x')&=&[\phiup(x)-\phiup(x')]+[\phiup(x')+\phidown(x')]
\nonumber \\
&=& -\int_x^{x'}(v_\mu+a_\mu)ds_\mu+2\int_0^{x'}v_\mu ds_\mu.
\label{oo2}
\end{eqnarray}
The first term on the right hand side is just like the one we encountered in
our discussion of the diagonal propagator. The second term, however, involves
a line integral of $v_\mu$ originating at an {\em arbitrary} fixed reference
point in space-time. In the non-superconducting phase fluctuations in
$v_\mu$ will clearly drive any such term to zero, making the off-diagonal terms
of the electron propagator vanish. This is consistent with our general
expectation that electron propagator does not exhibit anomalous off-diagonal
correlations in the normal state.

Finally, pulling different strands together, we can write down the full
electron Green function in the AFL phase of the pseudogap state. To make
connection with the notation prevalent in the condensed matter physics we
perform a Euclidean rotation $k_0\to i\omega$ in Eq.\ (\ref{brown78}) and
with help of Eq.\ (\ref{e6}) we obtain
\begin{equation}
G^{\rm elec}(\bk,\omega)= \Lambda^{-\eta'}{\omega+\sigma_3\epsilon_\bk
\over[\epsilon_\bk^2+\Delta_\bk^2-\omega^2]^{1-\eta'/2}},
\label{eprop}
\end{equation}
where we have also restored the full electron dispersion $\epsilon_\bk$ and
the gap function $\Delta_\bk$ with the understanding that the above form for
the propagator is strictly valid only in the vicinity of the nodal point
and close to the isotropic limit. We note that the Eq.\ (\ref{eprop}) implies
an anomalous electron density of states
\begin{equation}
N(\omega)\sim \omega^{1+\eta'}
\label{dos}
\end{equation}
at low energies. It would be very interesting if
such anomalous electron density of states could be
measured in tunneling experiments.
Similarly, the Luttinger-like behavior of the propagator
Eq.\ (\ref{eprop}) will be reflected in other physical observables.


\section{Effects of Dirac anisotropy in symmetric QED$_3$}

It is natural to examine to what extend is the theory modified
by the inclusion of the Dirac anisotropy,
i.e. the finite difference in the Fermi velocity
$v_F$ and the gap velocity $v_{\Delta}$.
In the actual materials the Dirac anisotropy
$\alpha_D =\frac{v_F}{v_{\Delta}}$ decreases with
decreasing doping from $\sim15$ in the optimally doped to $\sim3$ in the
heavily underdoped samples.

There are two key issues: first, for a large
enough number of Dirac fermion species $N$, how is the chirally symmetric
infrared (IR) fixed point modified by the fact that
$\alpha_D \neq 1$, and second, as we
decrease $N$, does the chiral symmetry breaking
occur at the same value of $N$ as in the isotropic theory.
In this section we address in detail
the first issue and defer the discussion of
the second one to Part II.

We determine the effect of the Dirac anisotropy,
marginal by power counting,
by the perturbative renormalization group
(RG) to first order in the large $N$ expansion.
To the leading order $\delta$ in the small anisotropy
$\alpha_D=1+\delta$, we obtain the analytic value of
the RG $\beta_{\alpha_D}$-function
and find that it is proportional to $\delta$, i.e. in the infra red
the $\alpha_D$ decreases when $\delta > 0$ and the
anisotropic theory flows to the isotropic fixed
point. On the other hand, when $\delta < 0$, $\alpha_D$
increases in the IR and again the theory flows into
the isotropic fixed point. These results hold even
when anisotropy is not small as shown
by numerical evaluation of the $\beta$-function.
Therefore, we conclude that the isotropic
fixed point is stable against small anisotropy.

Furthermore,
we show that in any covariant gauge renormalization of
$\Sigma$ due to the unphysical longitudinal
degrees of freedom is {\em exactly} the same along any space-time direction.
Therefore the only contribution to the RG flow of anisotropy comes from the
physical degrees of freedom and our results for $\beta_{\alpha_D}$ stated above
are in fact gauge invariant.

%
\subsection{Anisotropic QED$_3$}
In the realm of condensed matter physics there is no Lorentz
symmetry to safeguard the space-time isotropy of the theory. Rather,
the intrinsic Dirac anisotropy is always present since it ultimately
arises from complicated
microscopic interactions in the solid which eventually renormalize
to band and pairing amplitude dispersion.
Thus there is nothing to protect the difference in the
Fermi velocity $v_F=\frac{\partial \eps_{\bk}}{\partial \bk}$
and the gap velocity $v_{\Delta}=\frac{\partial \Delta_{\bk}}{\partial \bk}$
from vanishing and, in fact, all HTS materials are anisotropic.

The value of $\alpha_D$ can be directly measured by the
angle resolved photo-emission spectroscopy (ARPES), which is ultimately
a "high"
energy local probe of $v_F$ and $v_{\Delta}$. Since QED$_3$ is
free on short distances, we can take the experimental values as the starting
bare parameters of the field theory.

The pairing amplitude of the HTS cuprates has $d_{x^2-y^2}$ symmetry,
and consequently there are four nodal points on the Fermi surface with Dirac
dispersion around which we can linearize the theory.
Note that the roles of $x$ and $y$ directions are
interchanged between adjacent nodes. As before, we combine the four
two-component Dirac spinors for the opposite (time reversed) nodes
into two four-components spinors and label them as
$(1,\bar1)$ and $(2,\bar{2})$ (see Fig. \ref{fsfig}).

Thus, the two-point vertex function of the non-interacting theory for, say,
$1\bar{1}$ fermions is
\begin{equation}\label{2ptfreevertex}
\Gamma^{(2)free}_{1\bar{1}}= \gamma_0 k_0 +
v_F \gamma_1 k_1 + v_{\Delta} \gamma_2 k_2.
\end{equation}
Therefore the corresponding non-interacting ``nodal'' Green functions are
\begin{equation}
G^n_0(k)=\frac{\sqrt{g^n}_{\mu\nu}\gamma_{\mu}k_{\nu}}{k_{\mu}g_{\mu\nu}k_{\nu}} \equiv
\frac{\gamma^n_{\mu}k_{\mu}}{k_{\mu}g_{\mu\nu}k_{\nu}}.
\end{equation}
Here we introduced the (diagonal) ``nodal'' metric
$g^{(n)}_{\mu\nu}$:
$g^{(1)}_{00}\!=\!g^{(2)}_{00}\!=\!1$,
$g^{(1)}_{11}\!=\!g^{(2)}_{22}\!=\!v^2_F$,
$g^{(1)}_{22}\!=\!g^{(2)}_{11}\!=\!v^2_{\Delta}$,
as well as the ``nodal'' $\gamma$ matrices $\gamma^n$. In what follows
we assume that both $v_F$ and $v_{\Delta}$ are dimensionless and that
eventually one of them can be chosen to be unity by an appropriate choice
of the "speed of light".

Since $\alpha_D \neq 1$ breaks Lorentz invariance of the theory
and since it is the Lorentz invariance that protects the spacetime isotropy,
we expect the $\beta$ functions for $\alpha_D$ to acquire finite values.
However, the theory still respects time-reversal and parity and
for $N$ large enough the system is in the chirally symmetric phase.
These symmetries force the fermion self-energy of the interacting
theory to have the form
\begin{equation}
\Sigma_{1\bar{1}} =A(k_{1\bar1},k_{2\bar2})
\left(\gamma_0 k_0 + v_F\zeta_1 \gamma_1 k_1 +
v_{\Delta} \zeta_2 \gamma_2 k_2 \right).
\end{equation}
where $k_{1\bar1}\equiv k_{\mu}g^{(1)}_{\mu\nu}k_{\nu}$ and
$k_{2\bar2}\equiv k_{\mu}g^{(2)}_{\mu\nu}k_{\nu}$.
The coefficients $\zeta_i$ are in general different from unity.
Furthermore, there is a discrete symmetry which relates flavors $1,\bar{1}$
and $2,\bar{2}$
and the $x$ and $y$ directions in such a way that
\begin{equation}
\Sigma_{2\bar{2}} =A(k_{2\bar2},k_{1\bar1}) \left(\gamma_0 k_0 + v_{\Delta} \zeta_2 \gamma_1 k_1
+ v_F \zeta_1 \gamma_2 k_2 \right).
\end{equation}
In the computation of the fermion self-energy,
this discrete symmetry allows us to concentrate on a particular pair of nodes
without any loss of generality.

\subsection{Gauge field propagator}

As discussed above in the isotropic case,
the effect of vortex-antivortex fluctuations
at T=0 on the fermions can, at large distances, be included by coupling
the nodal fermions minimally to a fluctuating $U(1)$ gauge field with
a standard Maxwell action. Upon integrating out the fermions, the gauge
field acquires a stiffness proportional to $k$, which is another way
of saying that at the charged, chirally symmetric fixed point the gauge
field has an anomalous dimension $\eta_A=1$
(for discussion of this point in the bosonic QED see  \cite{herbutzbt}).

We first proceed in the transverse gauge ($k_{\mu}a_{\mu}=0$)
which is in some sense the most physical
one considering that the $\nabla \times \bf{a}$ is physically related to
the vorticity, i.e. an intrinsically transverse quantity. We later extend
our results to a general covariant gauge.
To one-loop order the screening effects of the fermions on the gauge field
are given by the polarization function
\begin{equation}
\Pi_{\mu\nu}(k)=\frac{N}{2}\sum_{n=1,2} \int \frac{d^3q}{(2\pi)^3}Tr[G_0^n(q)\gamma_{\mu}
^n G_0^n(q+k)\gamma_{\nu}^n]
\end{equation}
where the index $n$ denotes the fermion ``nodal'' flavor.
The above expression can be evaluated straightforwardly
by noting that it reduces to the isotropic $\Pi_{\mu\nu}(k)$
once the integrals are properly rescaled  \cite{ftqed}.
The result can be conveniently presented by taking advantage of the
``nodal'' metric $g^n_{\mu\nu}$ as
\begin{equation}\label{polarization}
\Pi_{\mu\nu}(k)=\sum_n
\frac{N}{16v_Fv_{\Delta}}\sqrt{k_{\alpha}g^n_{\alpha \beta}k_{\beta}}
\left(
g^n_{\mu\nu}-\frac{g^n_{\mu\rho}k_{\rho}g^n_{\nu\lambda}k_{\lambda}}
{k_{\alpha}g^n_{\alpha \beta}k_{\beta}}
\right).
\end{equation}
Note that this expression is explicitly transverse,
$k_{\mu}\Pi_{\mu\nu}(k)=\Pi_{\mu\nu}(k)k_{\nu}=0$, and symmetric in its
space-time indices. It also properly reduces to the isotropic expression
when $v_F=v_{\Delta}=1$.

However, as opposed to the isotropic case, it is not quite as straightforward
to determine the gauge field propagator $D_{\mu\nu}$.
For example, as it stands the
polarization matrix (\ref{polarization})
is not invertible, which makes it necessary to introduce
some gauge-fixing conditions. In our case the direct inversion of the
$3\times3$ matrix would obscure the analysis and
therefore, we choose to follow a more physical and notationally
transparent line of reasoning which
eventually leads to the correct expression for the gauge field propagator.
Upon integrating out the fermions and expanding the effective action to the
one-loop order, we find that
\begin{equation}
{\cal L}_{eff}[a_{\mu}]=(\Pi^{(0)}_{\mu\nu}+\Pi_{\mu\nu})a_{\mu}a_{\nu}
\end{equation}
where the bare gauge field stiffness is
\begin{equation}
\Pi^{(0)}_{\mu\nu}=\frac{1}{2e^2} k^2 \left(\delta_{\mu\nu}-
\frac{k_{\mu}k_{\nu}}{k^2} \right).
\end{equation}
At this point we introduce the dual field $b_{\mu}$ which is related to
$a_{\mu}$ as
\begin{equation}
b_{\mu}=\epsilon_{\mu\nu\lambda}q_{\nu}a_{\lambda}.
\end{equation}
Physically, the $b_{\mu}$ field represents vorticity and we integrate over
all possible vorticity configurations with the restriction that $b_{\mu}$
is transverse. We note that
\begin{equation}\label{lb}
{\cal L}[b_{\mu}]=\chi_0 b_0^2 +\chi_1 b_1^2 +\chi_2 b_2^2
\end{equation}
where $\chi_{\mu}$'s are functions of $k_{\mu}$ and upon a straightforward
calculation they can be found to read
\begin{equation}
\chi_{\mu}=\frac{1}{2e^2}+
\frac{N}{16v_Fv_{\Delta}}\sum_{n=1,2}
\frac{g^n_{\nu\nu}g^n_{\la\la}}{\sqrt{k_{\alpha}g^n_{\alpha\beta}k_{\beta}}}~~,
\end{equation}
where
$\mu\neq \nu\neq \lambda\in \{0,1,2\}$.
At low energies we can neglect the non-divergent bare stiffness and thus
set $1/e^2=0$ in the above expression.

The expression (\ref{lb}) is manifestly gauge invariant and
has the merit of not only being
quadratic but also diagonal in the individual components of $b_{\mu}$.
Thus, integration over the vorticity (even with the restriction of
transverse $b_{\mu}$) is simple and we can easily determine the
$b_{\mu}$ field correlation function
\begin{equation}
\langle b_{\mu} b_{\nu} \rangle=\frac{\delta_{\mu\nu}}{\chi_{\mu}}-
\frac{k_{\mu}k_{\nu}}{\chi_{\mu}\chi_{\nu}}\left( \sum_i \frac{k_i^2}{\chi_i}
\right)^{-1}.
\end{equation}
The repeated indices are not summed in the above expression.
Note that, in addition to being transverse,
$\langle b_{\mu} b_{\nu} \rangle$ is also symmetric in its space time indices.

It is now quite simple to determine the correlation function for the $a_{\mu}$
field and in the transverse gauge we obtain
\begin{equation}
D_{\mu\nu}(q)=\langle a_{\mu} a_{\nu} \rangle=
\epsilon_{\mu i j}\epsilon_{\nu k l}
\frac{q_i q_k}{q^4}\langle b_j b_l \rangle.
\end{equation}
Using the transverse character of $\langle b_{\mu}b_{\nu} \rangle$
(which is independent of the gauge)
the above expression can be further reduced to
\begin{equation}\label{bp}
D_{\mu\nu}(q)=
\frac{1}{q^2}\left(\left(\delta_{\mu\nu}-\frac{q_{\mu}q_{\nu}}{q^2}\right)
\langle b^2 \rangle - \langle b_{\mu} b_{\nu} \rangle \right).
\end{equation}
It can be easily checked that in the isotropic limit
the expression (\ref{bp})
properly reduces to the results obtained in a different way.

We can further extend this result to include a general gauge by writing
\begin{equation}\label{bpgi}
D_{\mu\nu}(q)=
\frac{1}{q^2}\left(\left(\delta_{\mu\nu}-(1-\frac{\xi}{2})\frac{q_{\mu}q_{\nu}}{q^2}\right)
\langle b^2 \rangle - \langle b_{\mu} b_{\nu} \rangle \right).
\end{equation}
where $\xi$ is our continuous parameterization of the gauge fixing.
This expression can be justified by the Fadeev-Popov type of procedure
starting from the Lagrangian
\begin{equation}
{\cal L}=\left(\Pi_{\mu \nu}+\frac{1}{\xi}\frac{2k^2}{\langle b^2 \rangle}
\frac{k_{\mu}k_{\nu}}{k^2}\right)a_{\mu}a_{\nu},
\end{equation}
where the stiffness for the unphysical modes was judiciously chosen to scale
as $k$ in a particular combination of the physical scalars of the theory.
Note that $\langle b^2 \rangle$ can be determined without ever considering
gauge fixing terms.
In this way, the extension of a transverse gauge $\xi=0$ to a general covariant
gauge is accomplished by a simple substitution
$k_{\mu}k_{\nu} \rightarrow (1-\frac{\xi}{2})k_{\mu}k_{\nu}$. The expression
(\ref{bpgi}) is our final result for the gauge field propagator in a
covariant gauge.

\subsection{TF self energy}
%
As discussed in Sec. IV,
$\Sigma$ is not gauge invariant in that it has an explicit dependence
on the gauge fixing parameter. As we will show in this Section (and
more generally in the Appendix D), the
renormalization of $\Sigma$ by the unphysical longitudinal degrees of freedom
does not depend on the space-time direction: the term in $\Sigma$ which
is proportional to $\g_0$ is renormalized the same way by the gauge dependent
part of the action as the terms proportional
to $\g_1$ and $\g_2$. Therefore, the only contribution to the RG flow of
$\alpha_D$ comes from the physical degrees of freedom.

We denote the topological fermion self-energy at the node $n$ by
$\Sigma_{n}(q)$.
Hence, to the leading order in large $N$ expansion we have
\begin{equation}
\Sigma_{n}(q)=\int \frac{d^3k}{(2\pi)^3}\gamma^{n}_{\mu}
G_0^{n}(q-k) \gamma^{n}_{\nu}D_{\mu\nu}(k).
\end{equation}
or explicitly
\begin{equation}
\Sigma_{n}(q)=\int \frac{d^3k}{(2\pi)^3}\gamma^{n}_{\mu}
\frac{(q-k)_{\lambda}\gamma^n_{\lambda}}{(q-k)_{\mu}g_{\mu\nu}(q-k)_{\nu}} \gamma^{n}_{\nu}D_{\mu\nu}(k),
\end{equation}
where the gauge field propagator $D_{\mu\nu}$ is already screened by the nodal
fermions (\ref{bpgi}).
Using the fact that
\begin{equation}
\gm\gl\gn=i\eps_{\mu\lambda\nu}\gamma_{5}\gamma_{3}
+\delta_{\mu\lambda}\gn-\delta_{\mu\nu}\gl+\delta_{\lambda\nu}\gm,
\end{equation}
where $\mu,\nu,\lambda \in \{0,1,2\}$ and $\gamma_5 \equiv -i\g_0\g_1\g_2\g_3$,
we can easily see that
\begin{equation}
\gnm\gnl\gnn D_{\mu\nu}=(2g^n_{\lambda\mu}\g^n_{\nu}-\g^n_{\lambda}g^n_{\mu\nu})D_{\mu\nu},
\end{equation}
where we used the symmetry of the gauge field propagator tensor $D_{\mu\nu}$.
Thus,
\begin{equation}
\Sigma_{n}(q)=\int \frac{d^3k}{(2\pi)^3}
\frac{(q-k)_{\lambda}(2g^n_{\lambda\mu}\g^n_{\nu} -\g^n_{\lambda}g^n_{\mu\nu})D_{\mu\nu}(k)}
{(q-k)_{\mu}g^n_{\mu\nu}(q-k)_{\nu}}
\end{equation}
and as shown in the Appendix D at low energies this can be written as
\begin{equation}\label{selfenergy}
\Sigma_{n}(q)=-\sum_{\mu} \eta^n_{\mu}(\g^n_{\mu}q_{\mu})
\ln{\left(\frac{\Lambda}{\sqrt{q_{\alpha}g^n_{\alpha\beta}q_{\beta}}}\right)}.
\end{equation}
Here $\Lambda$ is an upper cutoff and
the coefficients $\eta$ are functions of the bare anisotropy, which have
been reduced to a quadrature (see Appendix D).
It is straightforward, even if somewhat tedious, to show that in case of weak
anisotropy ($v_F=1+\delta,\;v_{\Delta}=1$), to order $\delta^2$,
\begin{equation}\label{eta0}
\eta_0^{1\bar1}=-\frac{8}{3\pi^2N}\left(1-\frac{3}{2}\xi-\frac{1}{35}
\left(40-7\xi\right)\delta^2\right)
\end{equation}
\begin{equation}\label{eta1}
\eta_1^{1\bar1}=-\frac{8}{3\pi^2N}\left(1-\frac{3}{2}\xi+\frac{6}{5}\delta-\frac{1}{35}
\left(43-7\xi\right)\delta^2\right)
\end{equation}
\begin{equation}\label{eta2}
\eta_2^{1\bar1}=-\frac{8}{3\pi^2N}\left(1-\frac{3}{2}\xi-\frac{6}{5}\delta-\frac{1}{35}
\left(1-7\xi\right)\delta^2\right).
\end{equation}
In the isotropic limit ($v_F=v_{\Delta}=1$)
we regain $\eta_{\mu}^n=-8(1-\frac{3}{2}\xi)/3\pi^2N$ as previously found by others.
%
\subsection{Dirac anisotropy and its $\beta$ function}
%
Before plunging into any formal analysis, we wish to discuss some immediate
observations regarding the RG flow of the anisotropy.
Examining the Eq. (\ref{selfenergy}) it is 
clear that if $\eta^n_{1}=\eta^n_{2}$
then the anisotropy {\em does not} flow and remains equal to its bare value.
That would mean that anisotropy is marginal and the theory
flows into the anisotropic fixed point. In fact, such a theory would
have a {\em critical line} of $\alpha_D$. For this to happen, however, there
would have to be a symmetry which would protect the equality
$\eta^n_1=\eta^n_{2}$. For example, in the isotropic QED$_3$
the symmetry which protects the equality of $\eta$'s is the Lorentz invariance.
In the case at hand,  this symmetry is broken and therefore we expect that
$\eta^n_1$ will be different from $\eta^n_2$, suggesting that the anisotropy flows away
from its bare value. If we start with $\alpha_D>1$ and find that $\eta_2^{1\bar1}>\eta_1^{1\bar1}$
at some scale $p<\Lambda$, we would conclude that the anisotropy
is marginally irrelevant and decreases towards $1$.
On the other hand if $\eta_2^{1\bar1} < \eta_1^{1\bar1}$, then anisotropy
continues increasing beyond its bare value and the theory flows into
a {\em critical point} with (in)finite anisotropy.

The issue is further complicated by the
fact that $\eta^n_{\mu}$ is not
a gauge invariant quantity, i.e. it
depends on the gauge fixing parameter $\xi$.
The statement that, say $\eta_1^{n} > \eta_2^{n}$,
makes sense only if the $\xi$ dependence of
$\eta_1^n$ and $\eta_2^n$ is exactly the same,
otherwise we could choose a gauge
in which the difference $\eta_2^n-\eta_1^n$
can have either sign. However, we see from the
equations (\ref{eta0}-\ref{eta2}) that in
fact the $\xi$ dependence of all $\eta$'s
is indeed the same. Although it was
explicitly demonstrated only to the
${\cal O}(\delta^2)$, in the Appendix D
we show that it is in fact true to all orders
of anisotropy for any choice of covariant gauge fixing.
This fact provides the justification for our procedure. Now we supply the
formal analysis reflecting the above discussion.

The renormalized 2-point vertex function is related to the
``bare'' 2-point vertex function via a fermion field rescaling
$Z_{\psi}$ as
\begin{equation}
\Gamma^{(2)}_R=Z_{\psi}\Gamma^{(2)}.
\end{equation}
It is natural to demand that for example at nodes $1$ and $\bar1$ at
some renormalization scale $p$,
$\Gamma^{(2)}_R(p)$ will have the form
\begin{equation}\label{rgcond}
\Gamma^{(2)}_R(p)=\g_0p_0+v_F^R\g_1p_1 + v_{\Delta}^R\g_2p_2.
\end{equation}
Thus, the equation (\ref{rgcond})
corresponds to our renormalization condition through
which we can eliminate the cutoff dependence and calculate the RG flows.

\begin{figure}
\epsfxsize=8.0cm
\hfil\epsfbox{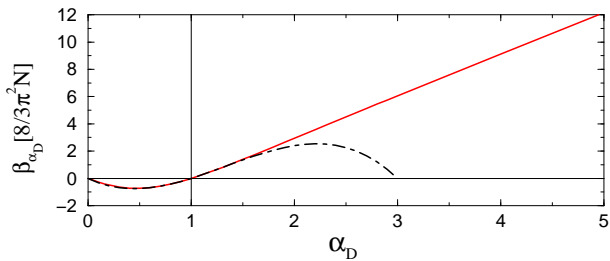}\hfill
\vspace{0.25cm}
\caption{The RG $\beta$-function for the Dirac anisotropy in units of
$8/3\pi^2N$. The solid line is the numerical integration
of the quadrature in the Eq. (\ref{etas}) while the dash-dotted
line is the analytical expansion around the small anisotropy
(see Eq. (\ref{eta0}-\ref{eta2})).
At $\alpha_D=1$, $\beta_{\alpha_D}$ crosses zero with positive slope,
and therefore at large
length-scales the anisotropic QED$_3$ scales to an isotropic theory.}
\label{beta}
\end{figure}

To the order of $1/N$ we can write
\begin{equation}
\Gamma^{(2)}_R(p)=Z_{\psi}
\g^n_{\mu}p_{\mu}\left(1+\eta^n_{\mu}\ln{\frac{\Lambda}{p}}\right)
\end{equation}
where we used the fermionic self-energy (\ref{selfenergy}).
Multiplying both sides by $\g_0$ and tracing the resulting expression determines
the field strength renormalization
\begin{equation}
Z_{\psi}=\frac{1}{1+\eta^n_0\ln{\frac{\Lambda}{p}}}\approx
1-\eta^n_0\ln{\frac{\Lambda}{p}}.
\end{equation}
We can now determine the renormalized Fermi and gap velocities
\begin{equation}
\frac{v_F^R}{v_F}\approx
(1-\eta^{1\bar{1}}_0\ln{\frac{\Lambda}{p}})(1+\eta^{1\bar{1}}_1
\ln{\frac{\Lambda}{p}})\approx
1-(\eta^{1\bar{1}}_0-\eta^{1\bar{1}}_1)\ln{\frac{\Lambda}{p}}
\end{equation}
and
\begin{equation}
\frac{v_{\Delta}^R}{v_{\Delta}}\approx
(1-\eta^{1\bar{1}}_0\ln{\frac{\Lambda}{p}})(1+\eta^{1\bar{1}}_2
\ln{\frac{\Lambda}{p}})\approx
1-(\eta^{1\bar{1}}_0-\eta^{1\bar{1}}_2)\ln{\frac{\Lambda}{p}}.
\end{equation}
The corresponding renormalized Dirac anisotropy is therefore
\begin{equation}
\alpha_D^R\equiv\frac{v_F^R}{v_{\Delta}^R}\approx
\alpha_D(1-(\eta^{1\bar{1}}_2-\eta^{1\bar{1}}_1)\ln{\frac{\Lambda}{p}}).
\end{equation}
The RG beta function can now be determined
\begin{equation}
\beta_{\alpha_D}=\frac{d \alpha^R_D}{d\ln{p}}=
\alpha_D(\eta^{1\bar{1}}_2-\eta^{1\bar{1}}_1).
\end{equation}
In the case of weak anisotropy ($v_F=1+\delta,\;v_{\Delta}=1$) the above
expression can be determined analytically as an expansion in $\delta.$
Using  Eqs.(\ref{eta1}-\ref{eta2}) we obtain
\begin{equation}
\beta_{\alpha_D}=\frac{8}{3\pi^2N}\left(\frac{6}{5}\delta(1+\delta)(2-\delta)+{\cal O}(\delta^3)
\right).
\end{equation}
Note that this expression is independent of the gauge fixing parameter $\xi$.
For $0<\delta\ll 1$ the $\beta$ function is positive which means that anisotropy
decreases in the IR and thus
the anisotropic QED$_3$ scales to an isotropic QED$_3$.
For $-1\ll \delta<0$ the $\beta$ function is negative and in this case
the anisotropy increases towards the fixed point $\alpha_D=1$, i.e. again
towards the isotropic QED$_3$. Note that for $\delta>2$, $\beta<0$  which may
naively indicate that there is a fixed point at $\delta=2$; this however
cannot be trusted as it is outside
of the range of validity of the power expansion of $\eta_{\mu}$. The numerical
evaluation of the quadrature in Eq. (\ref{etas}) shows that, apart from the
isotropic fixed point and the unstable fixed point at $\alpha_D=0$,
$\beta_{\alpha_D}$ does not vanish
(see Fig. \ref{beta}).
This indicates that to the leading order in the $1/N$ expansion,
the theory flows into the isotropic fixed point.

%

\section{Summary and conclusions}
By appealing to the unique features of high-$T_c$ cuprates -- strong electron
correlations, unconventional order parameter symmetry and 
pronounced fluctuation
effects -- we argued in favor of the inverted approach to the problem of
describing
various thermodynamic phases appearing in the underdoped region. This
inverted approach can be thought of as a ``Fermi liquid" theory of the phase
fluctuating d-wave superconductor where the role of the
Fermi energy as the large
energy scale of the problem is played by $\Delta$, the amplitude of the
pseudogap, which we assume to be predominantly
of the pairing nature. Under the umbrella of
this pseudogap -- inside the pairing protectorate -- we identify the BdG
quasiparticles as the relevant low lying fermionic
excitations of the theory and study their evolution under the effects of
interactions mediated by vortex-antivortex fluctuations. By carefully
treating these interactions we find that the low energy effective
theory for the quasiparticles inside the pairing protectorate is the (2+1)
dimensional quantum electrodynamics (QED$_3$)
with inherent spatial anisotropy,
described by Lagrangian ${\cal L}_D$ specified by Eqs.\ (\ref{qed3},\ref{l2}).

Within the superconducting state the gauge fields of the theory are massive
by virtue of vortex defects being bound into finite loops or
vortex-antivortex pairs. Such massive
gauge fields produce only short ranged interactions between our BdG
quasiparticles and are therefore irrelevant: in the superconductor
quasiparticles remain sharp in agreement with prevailing experimental
data  \cite{norman1}. Loss of the long range superconducting
order is brought about by unbinding the topological defects -- vortex loops
or vortex-antivortex
pairs -- via Kosterlitz-Thouless type transition and its quantum
cousin. Remarkably, this is accompanied by
the Berry gauge field becoming massless.
Such massless gauge field mediates long
range interactions between the fermions and becomes a relevant perturbation.
Exactly what is the consequence of this relevant perturbation depends on
the number of fermion species $N$ in the problem. For cuprates we argued
that $N=2n_{\rm CuO}$ where $n_{\rm CuO}$
is the number of CuO$_2$ layers per
unit cell. If $N<N_c\simeq 3$ \cite{appelquist},
the interactions cause spontaneous opening of a gap
for the fermionic excitations at $T=0$, via
the mechanism of chiral
symmetry breaking in QED$_3$  \cite{csb}.
Formation of the gap corresponds to the onset of AF SDW
instability \cite{herbut1,tvfqed} which must be considered as a
progenitor of the Mott-Hubbard-Neel antiferromagnet at half-filling.
If, on the other hand, $N>N_c$ as will be the case in bilayer or trilayer
materials, the theory remains in
its chirally symmetric nonsuperconducting
phase even as $T\to 0$ and AF order arises from within such state
only upon further underdoping (Fig. 1).
We call this symmetric state of  QED$_3$ an algebraic Fermi liquid (AFL).
In both cases AFL controls the low temperature, low energy
behavior of the pseudogap state and in this sense
assumes the role played by Fermi liquid theory in
conventional metals and superconductors.
In AFL the quasiparticle
pole is replaced by a branch cut -- the quasiparticle is no longer sharp --
and the gauge invariant electron propagator acquires a Luttinger-like form
Eq.\ (\ref{brown78}) with positive anomalous dimension $\eta'=16/3\pi^2N$.
To our knowledge this is one of the very few cases where a non-Fermi liquid
nature of the excitations has been demonstrated in dimension greater than one
in the absence of disorder or magnetic field. This Luttinger-like behavior
of AFL will manifest itself in anomalous power law functional form
of many physical properties of the system.

Dirac anisotropy, i.e. the fact that $v_F\neq\vd$, plays important role in
the cuprates where the ratio $\alpha_D=v_F/\vd$ in most
materials ranges between 3 and 15$-$20.
Such anisotropy is nontrivial as it cannot be rescaled and it significantly
complicates any calculation within the theory. Using the perturbative
renormalization group theory we have shown that anisotropic QED$_3$
flows back into isotropic stable fixed point. This means that for weak
anisotropy at long lengthscales the universal properties of the
theory are identical to those of the simple isotropic case. It remains to
be seen what are the properties of the theory at intermediate lengthscales
when anisotropy is strong.

Our theory of the pseudogap state gets its inspiration and builds on the
ideas originally articulated by Emery and Kivelson
in Ref.\  \cite{emerykivelson} and by Randeria and
collaborators \cite{randeria}. These ideas were later
explored and extended in various directions by others
\cite{nodalliquid,franzmillis,dorsey,loktev,levin,giovannini}.
These approaches share a common
philosophical platform of assuming that the pseudogap is primarily due to
pairing in the p-p channel and in the underdoped regime superconducting long
range coherence is destroyed by the phase fluctuations. The
essential differences lie
in the implementation of these ideas.
Balents {\em et al.} \cite{nodalliquid}, for instance,
argue for true separation of spin and charge \cite{senthil}
in their "nodal liquid" phase
which then gives way to an unconventional antiferromagnet ``AF*" with
deconfined spin 1/2 excitations. Our theory also shares considerable formal
similarities with the SU(2) gauge theories of Wen, Lee and co-workers
\cite{kim} and works by Aitchinson and Mavromatos   \cite{csb},
in that their low energy effective theory is also related to QED$_3$.
In those cases, however, the physical content of the theory -- the identity
of its excitations -- are totally different.

Finally there is a class of theories where the pseudogap is assumed to be
due to some competing order, usually in the p-h channel. This includes
SO(5)  \cite{zhang}, d-density wave  \cite{chakravarty} and various other
competing orders \cite{sachdev}. At present time we believe that experimental
evidence favors the pairing origin of the pseudogap, however the evidence
is far from conclusive and experiments can be found to support virtually
any aspect of the above mentioned theories. We believe, therefore, that the
road ahead necessitates making specific predictions based on controlled
and well defined approaches. The QED$_3$ theory of the pairing pseudogap,
as presented in this paper, starts from a remarkably simple set of
assumptions, and via manipulations that are controlled in the sense of
$1/N$ expansion, arrives at nontrivial consequences, including the
algebraic Fermi liquid and the antiferromagnet. Another notable
feature of this
theory is that it is fully calculable:
having the explicit form of
the low energy effective action, free of constraints and uncertainties,
physical observables can be computed through a systematic procedure
\cite{superrenormalizable}.


\acknowledgments

The authors are grateful to P. W. Anderson, A. Auerbach,
T. Appelquist, G. Baskaran, A. Cohen, \O. Fischer,
M. P. A. Fisher, B. Giovannini, S. M. Girvin, L. P. Gor'kov, V. P. Gusynin,
I. F. Herbut, C. Kallin, D. V. Khveshchenko,
S. A. Kivelson, R. B. Laughlin,
A. J. Millis, H. Monien, N. P. Ong,
W. Rantner, M. Reenders, S. Sachdev, T. Senthil,
D. E. Sheehy, A. Sudb\o, A.-M. Tremblay,
P. B. Wiegmann, A. Zee and S. C. Zhang
for helpful discussions and correspondence. This research was
supported in part by NSERC (MF) and  the NSF Grants
PHY99-07949 (ITP) and DMR00-94981 (ZT,OV).


\appendix

\section{Jacobian ${\cal L}_0$}

Here we derive the explicit form of the ``Jacobian'' ${\cal L}_0$
for two cases of interest:
{\em i}) the thermal vortex-antivortex fluctuations in 2D layers and
{\em ii}) the spacetime vortex
loop excitations relevant for low temperatures ($T\ll T^*$) deep in
the underdoped regime.

\subsection{2D thermal vortex-antivortex fluctuations}

In order to perform specific computations we have to
adopt a model for vortex-antivortex excitations. We will
use a 2D Coulomb gas picture of vortex-antivortex plasma.
In this model (anti)vortices are either point-like objects
or are assumed to have a small hard-disk radius of size
of the coherence length $\xi_0$ which emulates the core region.
As long as $\xi_0\ll n^{-\textstyle{1\over 2}}$,
where $n=n_v+n_a$ is the average
density of vortex defects, the two models lead to very
similar results and both undergo a vortex-antivortex pair
unbinding transition of the Kosterlitz-Thouless variety.

Above the transition we have
\begin{eqnarray}
\exp{[-\beta\int d^2r {\cal L}_0]}=
2^{-N_l}\sum_{A,B}\int {\cal D}\varphi ({\bf r})
\label{o1}     \\
\times
\delta[\nabla\times{\bf v} - {\textstyle {1\over 2}}
\nabla\times (\nabla\varphi_A +
\nabla\varphi_B)]~~~ \nonumber  \\
\times
\delta[\nabla\times{\bf a} - {\textstyle {1\over 2}}
\nabla\times (\nabla\varphi_A -
\nabla\varphi_B)]~~.\nonumber
\end{eqnarray}
The phase $\varphi ({\bf r})$ is due solely to vortices
and we can rewrite (\ref{o1}) as:
\begin{eqnarray}
\sum_{N_v,N_a}
{2^{-N_l}\over N_v!N_a!}
\sum_{A,B}\prod_i^{N_v}\int d^2r_i\prod_j^{N_a}\int d^2r_j
e^{-\beta E_c(N_v+N_a)}
\label{o2}     \\
\times
\delta\bigl[\rho_v ({\bf r})- \sum_i^{N_v}\delta ({\bf r} - {\bf r}_i)\bigr]
\delta\bigl[\rho_a ({\bf r})- \sum_j^{N_a}\delta ({\bf r} - {\bf r}_j)\bigr]\nonumber  \\
\times
\delta\bigl[b({\bf r})- \pi\sum_i^{N_v^A}\delta ({\bf r} - {\bf r}_i^A)+
\pi\sum_i^{N_v^B}\delta ({\bf r} - {\bf r}_i^B) \nonumber \\
+\pi\sum_j^{N_a^A}\delta ({\bf r} - {\bf r}_j^A) -
\pi\sum_i^{N_a^B}\delta ({\bf r} - {\bf r}_j^B)\bigr]~~.\nonumber
\end{eqnarray}
Here $N_v (N_a)$ is the number of free vortices (antivortices),
$N_l = N_v +N_a$, ${\bf r}_i$ (${\bf r}_j$) are vortex (antivortex)
coordinates and $\rho_v ({\bf r})$ ($\rho_a ({\bf r})$) are
the corresponding densities. $b ({\bf r}) = (\nabla\times a ({\bf r}))_z
= \pi (\rho_v^A -\rho_v^B - \rho_a^A + \rho_a^B)$ and
$E_c$ is the core energy which we have absorbed into ${\cal L}_0$
for convenience. We now express the above $\delta$-functions
as functional integrals over three new fields:
$d_v({\bf r})$, $d_a({\bf r})$ and $\kappa ({\bf r})$:
\begin{eqnarray}
\sum_{N_v,N_a}
{2^{-N_l}\over N_v!N_a!}
\sum_{A,B}\prod_i^{N_v}\int \!d^2r_i\prod_j^{N_a}\int\! d^2r_j
e^{-\beta E_c(N_v\!+\!N_a)}
\label{o3}     \\
\times\int {\cal D}d_v{\cal D}d_a{\cal D}\kappa
\exp\bigl\{
i\int d^2rd_v(\rho_v ({\bf r})- \sum_i^{N_v}\delta ({\bf r} - {\bf r}_i))\nonumber \\
+i\int d^2rd_a(\rho_a ({\bf r})- \sum_j^{N_a}\delta ({\bf r} - {\bf r}_j))\nonumber \\
 +
i\int d^2r\kappa \bigl[b({\bf r})- \pi\sum_i^{N_v^A}\delta ({\bf r} - {\bf r}_i^A)+
\pi\sum_i^{N_v^B}\delta ({\bf r} - {\bf r}_i^B) \nonumber \\
+\pi\sum_j^{N_a^A}\delta ({\bf r} - {\bf r}_j^A) -
\pi\sum_i^{N_a^B}\delta ({\bf r} - {\bf r}_j^B)\bigr]\bigr\}~~.\nonumber
\end{eqnarray}
The integration over $\delta$-functions in the exponential
is easily performed and the summation over $A(B)$ labels can be
carried out explicitly to obtain:
\begin{eqnarray}
\int {\cal D}d_v{\cal D}d_a{\cal D}\kappa
\exp\left[i\int d^2r\bigl(d_v\rho_v+d_a\rho_a+\kappa b\bigr)\right]
\label{o4}     \\
\times
\sum_{N_v,N_a}
{e^{-\beta E_cN_v}\over N_v!}
\prod_i^{N_v}\int d^2r_i\exp(-id_v({\bf r}_i))
\cos(\pi\kappa ({\bf r}_i)) \nonumber \\
\times {e^{-\beta E_cN_a}\over N_a!}
\prod_j^{N_a}\int d^2r_j\exp(-id_a({\bf r}_j))
\cos(\pi\kappa ({\bf r}_j))\nonumber.
\end{eqnarray}
In the thermodynamic limit the sum (\ref{o4}) is dominated by
$N_{v(a)} = \langle N_{v(a)}\rangle$, where $\langle N_{v(a)}\rangle$
is the average number of free (anti)vortices determined by
solving the full problem. Furthermore, as   $\langle N_{v(a)}\rangle\to\infty$
in the thermodynamic limit the integration over
$d_v({\bf r})$, $d_a({\bf r})$ and $\kappa ({\bf r})$ can
be performed in the saddle-point approximation leading to the
following saddle-point equations:
\begin{eqnarray}
- \rho_v ({\bf r})
+ \langle N_{v}\rangle \Omega_v(\br)=0
\label{o5}     \\
- \rho_a ({\bf r})
+ \langle N_{a}\rangle\Omega_a(\br)  =0
\label{o6} \\
- b(\br) + [\langle N_{v}\rangle \Omega_v(\br)+
\langle N_{a}\rangle \Omega_a(\br)] \ \ \ \ \   \label{o7} \\
\times\pi\tanh(\pi\kappa ({\bf r}))=0
\nonumber
\end{eqnarray}
with
$$
\Omega_m(\br)={e^{d_m (\br)}\cosh(\pi\kappa ({\bf r}))\over
\int d^2r' e^{d_m (\br')}\cosh(\pi\kappa (\br'))},\ \ \ \  m=a,v.
$$
Eqs. (\ref{o5}-\ref{o7}) follow from functional derivatives of
(\ref{o4}) with respect to $d_v({\bf r})$,
$d_a({\bf r})$ and $\kappa ({\bf r})$, respectively. We have
also built in the fact that the saddle-point solutions
occur at $d_v\to id_v$, $d_a\to id_a$, $\kappa\to i\kappa$.

The saddle-point equations (\ref{o5}-\ref{o7})
can be solved exactly leading to:
\begin{eqnarray}
d_v (\br) = \ln\rho_v ({\bf r}) - \ln\cosh(\pi\kappa ({\bf r}))
\label{o8}\\
d_a (\br) = \ln\rho_a ({\bf r}) - \ln\cosh(\pi\kappa ({\bf r}))
\label{o9}
\end{eqnarray}
where
\begin{equation}
\kappa ({\bf r}) = {1\over \pi}\tanh^{-1}\left[{b(\br)\over
\pi(\rho_v ({\bf r}) + \rho_a ({\bf r}))}\right].
\label{o10}
\end{equation}
Inserting (\ref{o8}-\ref{o10}) back into (\ref{o4}) finally
gives the entropic part of ${\cal L}_0/T$:
\begin{eqnarray}
&\rho_v&\ln \rho_v +
\rho_a\ln \rho_a
- \frac{1}{\pi}(\nabla\times{\bf a})_z\tanh^{-1}\left[ {(\nabla
\times{\bf a})_z \over \pi(\rho_v + \rho_a)} \right] \nonumber \\
&+& (\rho_v + \rho_a)\ln\cosh\tanh^{-1}\left[ {(\nabla
\times{\bf a})_z \over \pi(\rho_v + \rho_a)} \right],
\label{o11}
\end{eqnarray}
where $\rho_{v(a)}({\bf r})$ are densities of {\em free}
(anti)vortices. We display ${\cal L}_0$ in this form
to make contact with familiar physics: the first two
terms in (\ref{o11}) are the entropic contribution of free (anti)vortices
and the Doppler gauge field
$\nabla\times{\bf v}\to\pi(\rho_v - \rho_a)$
($\langle\nabla\times{\bf v}\rangle = 0$).
The last two terms encode
the ``Berry phase" physics of topological frustration.
Note that the ``Berry" magnetic field $b=(\nabla\times{\bf a})_z$
couples directly only to the {\em total density}
of vortex defects $\rho_v + \rho_a$ and is insensitive
to the vortex charge. This is a reflection of the
Z$_2$ symmetry of the original problem defined
on discrete (i.e., not
coarse-grained) vortices.
We write $\rho_{v(a)}({\bf r}) = \langle\rho_{v(a)}\rangle
+ \delta\rho_{v(a)}({\bf r})$ and expand (\ref{o11}) to leading order
in $\delta\rho_{v(a)}$ and $\nabla\times{\bf a}$:
\begin{equation}
{\cal L}_0/T \to
(\nabla\times{\bf v})^2/(2\pi^2 n_l) +
(\nabla\times{\bf a})^2/(2\pi^2 n_l),
\label{2dd}
\end{equation}
where $n_l = \langle\rho_v\rangle+
\langle\rho_a\rangle$ is the average density of free vortex defects.
Both ${\bf v}$ and ${\bf a}$ have a Maxwellian {\em bare} stiffness
and are {\em massless} in the normal
state. As one approaches $T_{c}$, $n_l\sim \xi_{\rm sc}^{-2}\to 0$,
where $\xi_{\rm sc}(x,T)$ is the superconducting correlation
length, and ${\bf v}$ and ${\bf a}$ become {\em massive} (see
the main text).

\subsection{Quantum fluctuations of (2+1)D vortex loops}

The expression for ${\cal  L}_0[j_\mu]$ given
by Eq. (\ref{l0s}) follows directly once the system
contains unbound vortex loops in its ground state
and thus can respond to the external perturbation $A^{\rm ext}_\mu$
over arbitrary large distances in (2+1)-dimensional spacetime.
This is already clear at intuitive level if we
just think of the geometry of infinite versus finite
loops and the fact that only unbound loops allow
vorticity fluctuations, described by $\langle j_\mu (q)j_\nu(-q)\rangle$,
to proceed unhindered. Still, it is useful to derive
(\ref{l0s}) and its consequences belabored in Section II
within an explicit model for vortex loop fluctuations.

Here we consider fluctuating vortex loops in continuous
(2+1)D spacetime and compute ${\cal  L}_0[j_\mu]$ using
duality map to the relativistic Bose superfluid \cite{kleinert}. We
again start by using our model of a large gap s-wave superconductor
which enables us to neglect $a_\mu$ in the fermion action
and integrate over it in the expression
for ${\cal  L}_0[v_\mu,a_\mu]$ (\ref{jacobian}). This
gives:
\begin{equation}\label{p1}
e^{-\int d^3x{\cal  L}_0}
=\sum_{N=0}^{\infty}\frac{1}{N!}
\prod_{l=1}^{N} \oint {\cal D}x_l[s_{l}]
\delta\bigl(j_\mu (x)- n_\mu (x)\bigr)e^{-\tilde{S}}
\end{equation}
where
\begin{equation}
\tilde{S}=\sum_{l=1}^NS_0\oint ds_l +{1\over 2}\sum_{l,l'=1}^N
\oint ds_l\oint ds_{l'}g(|x_l[s_l] - x_{l'}[s_{l'}]|)
\end{equation}
and
\begin{equation}
\bigl(\partial\times\partial\varphi (x)\bigr)_\mu
= 2\pi n_\mu (x) = 2\pi\sum_l^{N}\oint_L dx_{l\mu}\delta(x-x_{l}[s_l])~.
\label{p2}
\end{equation}
In the above equations $N$ is the number of loops,
$s_l$ is the Schwinger proper time (or ``proper length'')
of loop $l$, $S_0$ is the action per unit length associated
with motion of vortex cores
in (2+1)-dimensional spacetime (in an analogous 3D model this would be
$\varepsilon_c/T$, where $\varepsilon_c$ is the core line energy),
$g(|x_l[s_l] - x_{l'}[s_{l'}]|)$ is the short range penalty for
core overlap and $L$ denotes a line integral. We kept our
practice of including core terms independent of vorticity into
${\cal  L}_0$. Note that vortex loops must be periodic along
$\tau$ reflecting the original periodicity of $\varphi (\br,\tau)$.

We can think of vortex loops as worldline trajectories of
some relativistic charged (complex) bosons (charged since the loops
have two orientations) in two spatial dimensions.
${\cal  L}_0$ without the $\delta$-function describes
the vacuum Lagrangian of such a theory, with vortex loops
representing particle-antiparticle virtual creation and
annihilation process. The duality map is based on the
following relation between the Green function of free
charged bosons and a gas of free oriented loops:
$$
G(x) =\langle\phi(0)\phi^*(x)\rangle = \int {d^3k\over (2\pi)^3}
{e^{ik\cdot x}\over k^2 + m^2_d} =\int_0^\infty dse^{-sm^2_d}$$
\begin{equation}
\times\int {d^3k\over (2\pi)^3}
e^{ik\cdot x - sk^2} = \int_0^\infty dse^{-sm^2_d}
\left({1\over 4\pi s}\right)^{3\over 2}e^{-{x^2\over 4s}},
\label{p3}
\end{equation}
where $m_d$ is the mass of the complex dual field $\phi (x)$.
Within the Feynman path integral representation we can write:
\begin{equation}
\left(\frac{1}{4\pi s}\right)^{3/2}e^{-\frac{1}{4}x^2/s}=
\int_{x(0)=0}^{x(s)=x}{\cal D}x(s')\exp\left[-\frac{1}{4}\int_0^s ds'
\dot{x}^2(s')\right]~,
\label{p4}
\end{equation}
where $\dot{x}\equiv dx/ds$.
Furthermore, by simple integration (\ref{p3}) can
be manipulated into
\begin{multline}
{\rm Tr}\left[  \ln(-\partial^2+m_d^2)\right]=
-\int_0^{\infty} \frac{ds}{s}e^{-s
m_d^2}
\int\frac{d^3k}{(2\pi)^3}e^{-s k^2}\\
=-\int_0^{\infty} \frac{ds}{s}e^{-s m_d^2}\oint
{\cal D}x(s')\exp\left[-\frac{1}{4}\int_0^s ds' \dot{x}^2(s')\right]
\label{p5}
\end{multline}
where the path integral now runs over closed loops
($x(s)=x(0)$).
In the dual theory this can be reexpressed as
\begin{equation}
{\rm Tr} \left[\ln(-\partial^2+m_d^2)\right]=\int\frac{d^3k}{(2\pi)^3}
\ln(k^2+m_d^2)~.
\label{p6}
\end{equation}
Combining (\ref{p5}) and (\ref{p6}) and using
\begin{equation}
{\rm Tr} \left[\ln(-\partial^2+m_d^2)\right]=
\ln {\rm Det}(-\partial^2+m_d^2) \equiv {\cal W}
\label{p7}
\end{equation}
finally leads to:
\begin{multline}
e^{-{\cal W}}=
\sum_{N=0}^{\infty}\frac{1}{N!}\prod_{l=1}^{N}
\left[\int_0^{\infty}\frac{ds_l}{s_l}e^{-s_l m_d^2}
\oint {\cal D}x(s'_l)\right]\times\\
\exp\left[-\frac{1}{4}\sum_{l=1}^{N}\int_0^{s_l} ds'_l
\dot{x}^2(s'_l)\right]~~.
\label{p8}
\end{multline}
This is nothing else but the partition function of the free
loop gas. The size of loops is regulated by $m_d$. As $m_d\to 0$
the average loop size diverges. On the other hand,
through ${\cal W}$ (\ref{p7}), we can also think of (\ref{p8}) as the
partition function of the free bosonic theory.

To exploit this equivalence further we write
\begin{equation}
Z_d =\int {\cal D}\phi^{\ast} {\cal D}\phi e^{-\int d^3 x{\cal L}_d}~~,
\label{p9}
\end{equation}
where
\begin{equation}
{\cal L}_d = \abs{\partial \phi}^2 + m^2_d \abs{\phi}^2+
{g\over 2} \abs{\phi}^4~~,
\label{p10}
\end{equation}
and argue that $Z_d$ describes vortex loops with short range
interactions in (\ref{p1}). This can be easily demonstrated by decoupling
$\abs{\phi}^4$ through Hubbard-Stratonovich transformation and
retracing the above steps. The reader is referred to the book
by Kleinert for further details
of the above duality mapping \cite{kleinert}.

We can now rewrite the $\delta$-function in (\ref{p1}) as
\begin{equation}
\delta\bigl(j_\mu (x)- n_\mu (x)\bigr)\to
\int {\cal D}\kappa_\mu\exp\bigl(i\int d^3x \kappa_\mu(j_\mu - n_\mu)\bigr)
\label{p11}
\end{equation}
and observe that in the above language of Feynman path
integrals in proper time $ i\int d^3x \kappa_\mu n_\mu$ (\ref{p2})
assumes the meaning of a particle current three-vector $  n_\mu$  coupled
to a three-vector potential $\kappa_\mu$. Employing the same
arguments that led to (\ref{p10}) we now have:
\begin{equation}
{\cal L}_d [\kappa_\mu ]= \abs{(\partial -i\kappa)\phi}^2
+ m^2_d \abs{\phi}^2+
{g\over 2} \abs{\phi}^4~~,
\label{p12}
\end{equation}
which leads to
\begin{equation}
e^{-\int d^3 x{\cal L}_0[j_\mu]}
\to \int {\cal D}\phi^{\ast} {\cal D}\phi{\cal D}\kappa_\mu
e^{-\int d^3 x\left(-i\kappa_\mu j_\mu + {\cal L}_d[\kappa_\mu]\right)}~~.
\label{p13}
\end{equation}
In the pseudogap state vortex loop unbinding causes loss
of superconducting phase coherence. In the dual language,
the appearance of such infinite loops as $m_d\to 0$ implies
superfluidity in the system of charged bosons described by
${\cal L}_d$ (\ref{p10}). The dual off-diagonal long range
order (ODLRO) in $\langle\phi (x)\phi ^*(x')\rangle$ means that there
are worldline trajectories that
connect points $x$ and $x'$ over infinite spacetime distances -- these
infinite worldlines are nothing else but unbound vortex
paths in this ``vortex loop condensate''.
Thus, the dual and the true superconducting ODLRO are two
opposite sides of the same coin.

${\cal L}_d [\kappa_\mu ]$ is the Lagrangian of this dual superfluid
in presence of the external vector potential $\kappa_\mu $.
In the ordered phase, the response of the system is just the
dual version of the Meissner effect. Consequently, upon
functional integration over $\phi$ we are allowed to write:
\begin{equation}
e^{-\int d^3 x{\cal L}_0[j_\mu]}
=\int {\cal D}\kappa_\mu
e^{-\int d^3 x\left(-i\kappa_\mu j_\mu + M_d^2\kappa_\mu\kappa_\mu\right)}~~,
\label{p14}
\end{equation}
where $M_d^2 = |\langle\phi\rangle |^2$, with $\langle\phi\rangle$
being the dual order parameter. The remaining functional integration
over $\kappa_\mu$ finally results in:
\begin{equation}
{\cal L}_0[j_\mu] \to {j_\mu j_\mu\over 4|\langle\phi\rangle |^2}~~.
\label{p15}
\end{equation}
We have tacitly assumed that the system of loops is isotropic.
The intrinsic anisotropy of the (2+1)D theory
is easily reinstated and (\ref{p15}) becomes
Eq. (\ref{l0s}) of the main text.

It is now time to recall that we are interested in a d-wave
superconductor. This means we must restore $a_\mu$ to the
problem. To accomplish this we engage in a bit of thievery:
imagine now that it was $v_\mu$ whose coupling to fermions
was negligible and we could integrate over it in
(\ref{jacobian}). We would then be left with only
the $\delta$-function containing $a_\mu$. Actually,
we can compute such ${\cal L}_0[b_\mu/\pi]$, where
$\partial\times\partial a = b$, without any additional
work. Note that ${\cal L}_0$ contains only vorticity
independent terms. We can equally well proclaim that
it is the $A(B)$ labels that determine the true orientation
of our loops while the actual  vorticity is simply a
gauge label -- in essence, $v_\mu$ and $a_\mu$ trade
places. After the same algebra as before we obtain:
\begin{equation}
{\cal L}_0[b_\mu/\pi] \to {b_\mu b_\mu\over 4\pi^2 |\langle\phi\rangle |^2}~~,
\label{p16}
\end{equation}
which is just the Maxwell action for $a_\mu$.

Of course, this simple argument that led to (\ref{p16}) is illegal.
We cannot just forget $v_\mu$. If we did we would have no right
to coarse-grain $a_\mu$ to begin with and would have to face up
to its purely Z$_2$ character (see Section II). Still, the above
reasoning does illustrate that the Maxwellian stiffness of $a_\mu$
follows the same pattern as that of $v_\mu$: both are determined
by the  order parameter $\langle\phi\rangle$ of condensed
{\em dual} superfluid. Thus, we can write the correct form
of ${\cal L}_0$, with both $v_\mu$ and  $a_\mu$ fully included
into our accounting, as
\begin{equation}
{\cal L}_0[v_\mu,a_\mu]\to {(\partial\times v)_\mu(\partial\times v)_\mu
\over 4\pi^2|\langle\phi\rangle |^2}+
{(\partial\times a)_\mu(\partial\times a)_\mu
\over 4\pi^2|\langle\phi\rangle |^2}~~,
\label{p17}
\end{equation}
where our ignorance is now stored in computing the actual
value of $\langle\phi\rangle$ from the original parameters
of the d-wave superconductor
problem. With anisotropy restored this is precisely
Eq. (\ref{lab4}) of Section II.


\section{Feynman integrals in QED$_3$}

Many of the integrals encountered here are considered
standard in particle physics. Since the techniques involved are not as
common in the condensed matter physics we provide some of the technical
details in this Appendix. A more in-depth discussion can be found in many
field theory textbooks \cite{peskin}.

\subsection{Vacuum polarization bubble}
The vacuum polarization  Eq.\ (\ref{bub1}) can be written more explicitly as
\begin{equation}
\Pi_{\mu\nu}(q)=2N\tr[\gamma_\alpha\gamma_\mu\gamma_\beta\gamma_\nu]
I_{\alpha\beta}(q)
\label{a1}
\end{equation}
with
\begin{equation}
I_{\alpha\beta}(q)=\int{d^3k\over (2\pi)^3}
{k_\alpha(k_\beta+q_\beta)\over k^2(k+q)^2}.
\label{a2}
\end{equation}
The integrals of this type are most easily evaluated by employing the Feynman
parametrization \cite{peskin}. This consists in combining the denominators
using the formula
\begin{equation}
{1\over A^aB^b}={\Gamma(a+b)\over\Gamma(a)\Gamma(b)}
\int_0^1dx{x^{a-1}(1-x)^{b-1}\over [xA+(1-x)B]^{a+b}},
\label{a3}
\end{equation}
valid for any positive
real numbers $a$, $b$, $A$, $B$. Setting $A=k^2$ and $B=(k+q)^2$ allows us
to rewrite (\ref{a2}) as
\begin{equation}
I_{\alpha\beta}(q)=\int_0^1 dx\int{d^3k\over (2\pi)^3}
{k_\alpha(k_\beta+q_\beta)\over [(k+(1-x)q)^2+x(1-x)q^2]^2}.
\label{a4}
\end{equation}
We now shift the integration variable $k\to k-(1-x)q$ to obtain
\begin{equation}
I_{\alpha\beta}(q)=\int_0^1 dx\int{d^3k\over (2\pi)^3}
{k_\alpha k_\beta-x(1-x)q_\alpha q_\beta\over [k^2+x(1-x)q^2]^2},
\label{a5}
\end{equation}
where we have dropped terms odd in $k$ which vanish by symmetry upon
integration. We now
notice that $k_\alpha k_\beta$ term will only contribute if $\alpha=\beta$
and we can therefore replace it in Eq.\ (\ref{a5}) by ${1\over 3}
\delta_{\alpha\beta}k^2$. With this replacement the angular integrals are
trivial and we have
\begin{equation}
I_{\alpha\beta}(q)={1\over 2\pi^2}\int_0^1 dx\int_0^\infty dk k^2
{{1\over 3}
\delta_{\alpha\beta}k^2-x(1-x)q_\alpha q_\beta\over [k^2+x(1-x)q^2]^2}.
\label{a6}
\end{equation}
The only remaining difficulty stems from the fact that the integral
formally diverges at the upper bound.
This divergence is an artifact of our linearization of the fermionic spectrum
which breaks down for energies approaching the superconducting gap value.
It is therefore completely legitimate to introduce an ultraviolet cutoff.
Such UV cutoffs however tend to interfere with gauge invariance preservation
of which is crucial in this computation.
A more physical way of regularizing the integral Eq.\ (\ref{a6}) is to
recall that the gauge field must remain massless, i.e.\
$\Pi_{\mu\nu}(q\to 0)=0$. To enforce this property we write Eq.\ (\ref{a1})
as
\begin{equation}
\Pi_{\mu\nu}(q)=2N\tr[\gamma_\alpha\gamma_\mu\gamma_\beta\gamma_\nu]
[I_{\alpha\beta}(q)-I_{\alpha\beta}(0)],
\label{a7}
\end{equation}
and we see that proper regularization of Eq.\ (\ref{a6}) involves subtracting
the value of the integral at $q=0$. The remaining integral is convergent
and elementary; explicit evaluation gives
\begin{equation}
I_{\alpha\beta}(q)-I_{\alpha\beta}(0)= -{|q|\over 64}
\left(\delta_{\alpha\beta} +{q_\alpha q_\beta\over q^2}\right).
\label{a8}
\end{equation}
Inserting this in (\ref{a7}) and working out the trace using
Eqs.\ (\ref{diraca}) and (\ref{traces}) we find the result (\ref{bub2}).
Identical result can be obtained using dimensional regularization.

\subsection{TF self energy: Lorentz gauge}
For simplicity we evaluate the self energy (\ref{self1}) in the Lorentz gauge
($\xi=0$). Extension to arbitrary covariant gauge is trivial.
Eq. (\ref{self1}) can be written as
\begin{equation}
\Sigma_L(k)=-{2\over \pi^3N}\gamma_\mu I_\mu(k)
\label{a9}
\end{equation}
with
\begin{equation}
I_{\mu}(k)=\int d^3q q_\mu
{q\cdot(k+q)\over |q|^3(k+q)^2},
\label{a10}
\end{equation}
where we used an identity
\[
\gamma_\mu\gamma_\alpha\gamma_\nu\left(\delta_{\mu\nu} -
{q_\mu q_\nu\over q^2}\right)=-2\qslash{q_\alpha\over q^2}.
\]
Since the only 3-vector available is $k$, clearly the vector integral
$I_{\mu}(k)$ can only have components in the $k_\mu$ direction,
$I_{\mu}(k)={\cal C}(k)k_\mu/k^2$. By forming a scalar product
$k_\mu I_{\mu}(k)$ we obtain
\begin{equation}
{\cal C}(k)=k_\mu I_{\mu}(k)=\int d^3q
{(q\cdot k)(q\cdot k+q^2)\over |q|^3(k+q)^2}.
\label{a100}
\end{equation}
Combining the denominators using Eq.\ (\ref{a3}) and following the same
steps as in the computation of polarization bubble above we obtain
\begin{eqnarray}
{\cal C}(k)&=&{3\over 2}\int_0^1 dx\sqrt{x}\int d^3q \label{a11} \\
&\times&
{(2x-1)(k\cdot q)^2-(1-x)k^2q^2+x(1-x)^2k^4 \over [q^2+x(1-x)k^2]^{5/2}}.
\nonumber
\end{eqnarray}
The angular integrals are trivial and after introducing an ultraviolet
cutoff $\Lambda$ and rescaling the integration variable by $|k|$ the integral
becomes
\begin{equation}
{\cal C}(k)=2\pi k^2\int_0^1 dx\sqrt{x}\int_0^{\Lambda\over|k|} dq
{(5x-4)q^4+3x(1-x)^2q^2 \over [q^2+x(1-x)]^{5/2}}.
\label{a12}
\end{equation}
The remaining integrals are elementary. Isolating the leading infrared
divergent term we obtain
\begin{equation}
{\cal C}(k)\to -{4\pi\over 3}k^2\ln{\Lambda\over|k|}.
\label{a13}
\end{equation}
Substituting this to Eq.\ (\ref{a9}) we get the self energy (\ref{self2}).

\subsection{Dimensionally regularized gauge field line integral}
To evaluate ${\cal P}(r)$ specified by Eq.\ (\ref{f3}) we write it as a sum
of two contributions,
\begin{equation}
{\cal P}(r)={8r^2\over (2\pi)^dN}[I_1(r)-(1-\xi)I_2(r)],
\label{b1}
\end{equation}
where
\begin{eqnarray}
I_1(r)&=&\int d^dk{e^{ik\cdot r}\over k},
\label{b2:a}\\
I_2(r)&=&\int d^dk{e^{ik\cdot r}\over k}{(k\cdot r)^2\over k^2r^2}.
\label{b2:b}
\end{eqnarray}
We first consider $I_1(r)$. It is convenient to exponentiate the denominator
by use of the formula
\begin{equation}
{1\over A^a}={1\over\Gamma(a)}\int_0^\infty ds s^{a-1}e^{-sA}.
\label{b3}
\end{equation}
Taking $A=k^2$, $a={1\over 2}$ we have
\begin{equation}
I_1(r)={1\over\sqrt{\pi}}\int_0^\infty {ds\over\sqrt{s}}
  \int d^dke^{-sk^2+ik\cdot r}.
\label{b4}
\end{equation}
The $k$ integral is now easy to evaluate by completing the square,
shifting the integration variable, and making use of $\int d^dk \exp(-sk^2)
=(\pi/s)^{d/2}$. We thus obtain
\begin{equation}
I_1(r)=\pi^{d-1\over 2}\int_0^\infty ds s^{-{d+1\over 2}} e^{-{r^2\over 4s}}.
\label{b5}
\end{equation}
Substitution $t=r^2/4s$ transforms (\ref{b5}) to an integral of the type
shown in Eq.\ (\ref{b3}) and the result reads,
\begin{equation}
I_1(r)=(4\pi)^{d-1\over 2}\Gamma\left({\scriptstyle{d-1\over 2}}\right) r^{1-d}.
\label{b6}
\end{equation}
Using the same procedure we obtain
\begin{equation}
I_2(r)=-(4\pi)^{d-1\over 2}(d-2)\Gamma\left({\scriptstyle{d-1\over 2}}\right)
r^{1-d}.
\label{b7}
\end{equation}
Substituting (\ref{b6}) and (\ref{b7}) back into (\ref{b1}) we obtain the
result quoted in the text, Eq.\ (\ref{f4}).

\section{Proof of Brown's relation for gauge invariant propagator}

To prove Brown's relation
Eq.\ (\ref{brown1}) it is useful to introduce fermion propagator
for a {\em fixed} configuration of gauge field $a_\mu$,
\begin{equation}
G[x,x';a]={1\over z[a]}\int{\cal D}[\bar\Upsilon,\Upsilon]\
\Upsilon(x)\bar\Upsilon(x') e^{-S[\bar\Upsilon,\Upsilon,a]}
\label{c1}
\end{equation}
where $S[\bar\Upsilon,\Upsilon,a]=\int d^3r {\cal L}_D$ and
${\cal L}_D$ is the QED$_3$ Lagrangian
given by Eq.\ (\ref{l3}), $z[a]=\int{\cal D}[\bar\Upsilon,\Upsilon]\
e^{-S[\bar\Upsilon,\Upsilon,a]}$.
It is easy to see that $G[x,x';a]$ solves the Dirac equation
\begin{equation}
\gamma_\mu(i\partial_\mu-a_\mu)G[x,x';a]=\delta(x-x').
\label{c2}
\end{equation}
Furthermore, in terms of $G[x,x';a]$ the TF propagator can be written as
\begin{equation}
G(x-x')={1\over Z}\int{\cal D}a\ G[x,x';a] e^{-S_B[a]}
\label{c3}
\end{equation}
where $S_B=\int d^3x {\cal L}_B$ is the effective action [for our case,
to one-loop ${\cal L}_B$ is given by Eq.\ (\ref{ber})] and
$Z=\int{\cal D}a\ e^{-S_B[a]}$.
We now introduce a {\em gauge invariant} analog of $G[x,x';a]$,
\begin{equation}
{\cal G}[x,x';a] =e^{-i\int^{x'}_x a\cdot dr} G[x,x';a]
\label{c4}
\end{equation}
where the integral in the exponent is taken along the straight line
connecting space-time points $x'$ and $x$.
The gauge invariant TF propagator defined by Eq.\ (\ref{e5}) is then given
by an expression analogous to (\ref{c3}),
\begin{equation}
{\cal G}(x-x')={1\over Z}\int{\cal D}a\ {\cal G}[x,x';a] e^{-S_B[a]}.
\label{c5}
\end{equation}

Our task is to relate Brown propagator $\tilde{\cal G}(x-x')$ to 
$G(x-x')$ (\ref{brown1}). To this end we rewrite
Eq.\ (\ref{c3}) as follows
\begin{eqnarray}
G(x-x')&=&{1\over Z}\int{\cal D}a\ {\cal G}[x,x';a] e^{iJ\cdot a-S_B[a]}
\nonumber \\
&=& \left({{\tilde Z}\over Z}\right)\left[{1\over{\tilde Z}}
\int{\cal D}a\ {\cal G}[x,x';a] e^{-\tilde{S}_B[a]}\right] \nonumber \\
&=& \left({{\tilde Z}\over Z}\right) \tilde{\cal G}(x-x'),
\label{c6}
\end{eqnarray}
where $\tilde{S}_B[a]={S}_B[a]-iJ\cdot a$, $\tilde Z=\int{\cal D}a\
e^{-\tilde S_B[a]}$ and the last equality in Eq.\ (\ref{c6}) should be taken
as a definition of $\tilde{\cal G}(x-x')$. The source term $J$ is defined by
Eq.\  (\ref{brown3}) and $J\cdot a$ is a shorthand for $\int d^3 r J(r)\cdot
a(r)\equiv \int^{x'}_x a\cdot dr$. Note that the linear
UV divergence of phase factor 
in $\tilde Z$ cancels out between numerator and
denominator (\ref{c6}).

We now observe that
\begin{equation}
{\tilde Z\over Z}={1\over Z}\int{\cal D}a\ e^{iJ\cdot a} e^{-S_B[a]}
=\langle e^{iJ\cdot a}\rangle \equiv e^{-F(x-x')},
\label{c7}
\end{equation}
with $F$ defined by Eq.\ (\ref{brown2}). Substituting this into (\ref{c6})
we have (\ref{brown1}):
\begin{equation}
G(x-x')=e^{-F(x-x')}\tilde{\cal G}(x-x').
\label{c8}
\end{equation}
To complete this part it remains to address the relation
between $\tilde{\cal G}(x-x')$ and
${\cal G}(x-x')$. To this end we notice that both ${\cal G}$ and
$\tilde{\cal G}$ are {\em gauge invariant}: 
the former by construction and the
latter by the following simple observation. 
Since both ${\cal G}[x,x';a]$ and
$S_B[a]$ depend on the
transverse part of $a$, any dependence on longitudinal part of $a$ comes
through the $iJ\cdot a$ term and therefore identically cancels between the
numerator and $\tilde Z$ in the denominator. 
Therefore, once defined by Eq. (\ref{c6}) as a ratio of two gauge-variant
objects computed in the same  
covariant gauge, $\tilde{\cal G}$
takes on life of its own and is a fully gauge invariant quantity.
We may rewrite 
$\tilde{\cal G}$ as such ratio in arbitrary linear gauge; 
specifically we may chose the axial
gauge in which the $iJ\cdot a$ vanishes.
In this gauge $\tilde{\cal G}$ is exactly equal to ${\cal G}$. Since they
are both gauge invariant they must be equal in arbitrary gauge. 
The reader should be cautioned, however, that this equality
is a formal one since it involves manipulations of gauge-dependent
quantities which are ill-defined in 
absence of some specific regularization.

\section{Feynman integrals for anisotropic case}
\subsection{Self energy}
As shown in the Section V, to first order in $1/N$ expansion,
the topological fermion self energy can be reduced to
\begin{equation}
\Sigma_{n}(q)=\int \frac{d^3k}{(2\pi)^3}
\frac{(q-k)_{\lambda}(2g^n_{\lambda\mu}\g^n_{\nu} -\g^n_{\lambda}g^n_{\mu\nu})Dx_{\mu\nu}(k)}
{(q-k)_{\mu}g^n_{\mu\nu}(q-k)_{\nu}}.
\end{equation}
where $D_{\mu\nu}(q)$ is the screened gauge field propagator evaluated in the Section 3.
In order to perform the radial integral, we rescale the momenta as
\begin{equation}\label{rescaling}
K_{\mu}= \sqrt{g^n}_{\mu\nu}k_{\nu};\;\;\;
Q_{\mu}= \sqrt{g^n}_{\mu\nu}q_{\nu}
\end{equation}
and obtain
\begin{multline}
\!\!\!\!\!\!\Sigma_{n}(q)\!\!=\!\!\int \!\! \frac{d^3K}{(2\pi)^3}\!
\frac{(Q\!-\!K)_{\lambda}(2\sqrt{g^n}_{\lambda\mu}\g^n_{\nu}\!-\!\g_{\lambda}g^n_{\mu\nu})D_{\mu\nu}
\!\left(\!\!\frac{K_{\nu}}{\sqrt{g^n}_{\mu\nu}}\!\!\right)}{v_F v_{\Delta}(Q\!-\!K)^2}.
\end{multline}
At low energies we can neglect the contribution from the bare field stiffness $\rho$ in the gauge
propagator (see Section 3) and the resulting $D_{\mu\nu}(q)$ exhibits $\frac{1}{q}$ scaling
\begin{equation}
D_{\mu\nu}
\left(\frac{K_{\nu}}{\sqrt{g^n}_{\mu\nu}}\right)=\frac{F_{\mu\nu}(\theta,\phi)}{\abs{K}}
\end{equation}
Now we can explicitly integrate over the magnitude of the rescaled momentum $K$
by introducing an
upper cut-off $\Lambda$ and in the leading order we find that
\begin{equation}
\int_0^{\Lambda}\!\! dK \frac{K^2(Q\!-\!K)_{\lambda}}{K(Q\!-\!K)^2}\!=\!
-\Lambda \hat{K}_{\lambda}+ \ln{\left(\frac{\Lambda}{Q}\right)}\left(Q_{\lambda}
\!-\!2\hat{K}_{\lambda}\hat{K}\cdot Q \right)
\end{equation}
where $\hat{K}=(\cos{\theta},\; \sin{\theta}\cos{\phi},\; \sin{\theta}\sin{\phi})$.
Since $\hat{K}_{\mu}$ is odd under inversion while $F_{\mu\nu}$ is even,
it is not difficult to see that the term proportional to $\Lambda$
vanishes upon the angular integration.
Thus
\begin{multline}
\Sigma_{n}(q)=\int \frac{d\Omega }{(2\pi)^3v_F v_{\Delta}}
\left(Q_{\lambda}-2\hat{K}_{\lambda}\hat{K}\cdot Q \right)(\times)\\
(\times)(2\sqrt{g^n}_{\lambda\mu}\g^n_{\nu} -\g_{\lambda}g^n_{\mu\nu})F_{\mu\nu}(\theta,\phi)
\ln{\left(\frac{\Lambda}{Q}\right)}.
\end{multline}
Using the fact that the diagonal elements of $F_{\mu\nu}(\theta,\phi)$
are even under parity, while the off-diagonal elements are odd
under parity, the above expression can be further simplified to
\begin{equation}
\Sigma_{n}(q)=-\sum_{\mu}(\g^n_{\mu}q_{\mu} \eta^n_{\mu})
\ln{\left(\frac{\Lambda}{\sqrt{q_{\alpha}g^n_{\alpha\beta}q_{\beta}}}\right)}
\end{equation}
where the coefficients $\eta^n_{\mu}$ are pure numbers depending on the
bare anisotropy and are thereby reduced to a quadrature
\begin{multline}\label{etas}
\eta_{\mu}^n\!=\!\!\int\!\! \frac{d\Omega}{v_F v_{\Delta}(2\pi)^3}\! \biggl(\!
(2\hat{K}_{\mu}\hat{K}_{\mu}\!-\!1)(2g^n_{\mu\mu}F_{\mu\mu}\!-\!
\sum_{\nu}g^n_{\nu\nu}F_{\nu\nu})\\
\!+\!\sum_{\nu \neq \mu}4\hat{K}_{\mu}\hat{K}_{\nu}
 \sqrt{g^n}_{\mu\mu}\sqrt{g^n}_{\nu\nu}F_{\mu\nu}\biggr).
\end{multline}
Repeated indices are not summed in the above expression, unless explicitly indicated.
In the case of weak anisotropy ($v_F=1+\eps, v_{\Delta}=1$)
we can show that the Eq.(\ref{etas}) reduces
to Eqs.(\ref{eta0}-\ref{eta2}).

Consider now the effect of the (covariant) gauge fixing term on $\eta_{\mu}$.
Let us define the part of $F_{\mu \nu}$ which depends on the gauge fixing parameter $\xi$
as $F^{(\xi)}_{\mu \nu}$. The general form of this term is
$F^{(\xi)}_{\mu \nu}=\xi\; k_{\mu}k_{\nu} f(k)$ where
$f(k)$ is a scalar function of all three components of $k_{\mu}$;
$f(k)$ does in general depend on the
anisotropy. Upon rescaling with the nodal metric, see Eq.(\ref{rescaling}), we have
$F^{(\xi)}_{\mu \nu}=\xi \frac{1}{\sqrt{g_{\mu\mu}}}K_{\mu} \frac{1}{\sqrt{g_{\nu\nu}}}K_{\nu}
\tilde{f}(K)$, where
$\tilde{f}(K)$ is the corresponding scalar function of $K_{\mu}$. Substituting
$F^{(\xi)}_{\mu \nu}$ into the Eq. (\ref{etas}) we find
\begin{multline}
\eta^{\xi}_{\mu}\!=\!\xi\!\int\!\! \frac{d\Omega\;\tilde{f}(K)}
{v_F v_{\Delta}(2\pi)^3}\! \biggl(\!
(2\hat{K}_{\mu}\hat{K}_{\mu}\!-\!1)(2\hat{K}_{\mu}\hat{K}_{\mu}\!-\!
\sum_{\nu}\hat{K}_{\nu}\hat{K}_{\nu})
\\
\!+\!\sum_{\nu \neq \mu}4\hat{K}_{\mu}\hat{K}_{\nu}
 \hat{K}_{\mu}\hat{K}_{\nu}\biggr),
\end{multline}
where $\eta^{\xi}_{\mu}$ is the part of $\eta_{\mu}$ which comes entirely from the
gauge fixing term.
Using the fact that $\sum_{\nu}\hat{K}_{\nu}\hat{K}_{\nu}=1$, it is a matter
of simple algebra to show that
\begin{equation}
\eta^{\xi}_{\mu}\!=\!\xi\!\int\!\! \frac{d\Omega\;\tilde{f}(K)}
{v_F v_{\Delta}(2\pi)^3}
\end{equation}
i.e., the dependence on the index $\mu$ drops out! That means that the renormalization
of $\eta_{\mu}$ due to the unphysical longitudinal modes is exactly the same for all of its
components. Therefore, the difference in $\eta_1$ and $\eta_2$, 
which is related to the
RG flow of the Dirac anisotropy comes entirely from the physical modes and
is a gauge independent quantity. Note that this statement does not depend on
the choice of the covariant gauge, i.e. on the exact form of the function $f$,
only on the fact that the gauge is covariant.



\begin{references}
\bibitem{coulombgauge} For the aficionados of gauge theories, this situation
is reminiscent of the Landau gauge ($\partial\cdot A=0$)
being the fixed point among all
covariant gauges of the Higgs-Abelian theory.
\bibitem{remarkboring} At the risk of boring the reader, we
reiterate that all the different singular gauges lead to precisely
the same results if the partition function is evaluated
{\em exactly}, by integrating over the {\em discrete} vortex
variables  $\{{\bf r}_i(\tau)\}$ -- it is only when we seek
to perform this integration approximately, in terms of
{\em distributed} quantities like $\rho ({\bf r})$ or $j_\mu (x)$, that
the trouble arises.
\bibitem{regularization} It is not immediately obvious what
such regularization might be within QED$_3$ itself -- it might involve
replacing the straight line integral with a more ramified path
or a family of paths. Within our theory, however, it is clear that
such regularization ultimately reduces (\ref{e5}) to the original
expression (\ref{e3}), with discrete vortex variables.
\bibitem{remark4} References \cite{rantner1,rantner2} quote positive value
for $\eta'$. We find that this is a result of an overall sign error and
$\eta'$ should be negative, in agreement with Ref.\ \cite{khvesh1}. We are
indebted to W. Rantner and D. V. Khveshchenko
for communications regarding this issue
and for their many insightful comments concerning gauge invariance.
\end{references}
\end{document}